\documentclass[prd,preprintnumbers,nofootinbib,tightenlines,superscriptaddress]{revtex4}
\usepackage[utf8]{inputenc}
\usepackage{natbib}
\usepackage{units}
\usepackage{graphicx}
\usepackage{xcolor}
\usepackage{dcolumn}
\usepackage{bm}
\usepackage{amssymb,amsmath}
\usepackage{verbatim}
\usepackage{tikz}
\usepackage{hyperref}
\usepackage{dsfont}
\usepackage{animate}
\usepackage{media9}
\usepackage{slashed}
\usepackage{braket}
\usepackage{subfigure}
\usepackage[section]{placeins}
\begin{document}

\title{Testing the parametric model for self-interacting dark matter using matched halos in cosmological simulations} 

\author{Daneng Yang}
\email{yangdn@pmo.ac.cn}
\affiliation{Purple Mountain Observatory, Chinese Academy of Sciences, Nanjing 210023, China}
\affiliation{Department of Physics and Astronomy, University of California, Riverside, California 92521, USA}

\author{Ethan O.~Nadler}
\email{enadler@carnegiescience.edu}
\affiliation{Department of Astronomy \& Astrophysics, University of California, San Diego, La Jolla, CA 92093, USA}
\affiliation{Carnegie Observatories, 813 Santa Barbara Street, Pasadena, CA 91101, USA}
\affiliation{Department of Physics $\&$ Astronomy, University of Southern California, Los Angeles, CA, 90007, USA}

\author{Hai-Bo Yu}
\email{haiboyu@ucr.edu}
\affiliation{Department of Physics and Astronomy, University of California, Riverside, California 92521, USA}

\date{\today}

\begin{abstract}
We systemically evaluate the performance of the self-interacting dark matter (SIDM) halo model proposed in Ref.~\cite{Yang:2023jwn} with matched halos from high-resolution cosmological CDM and SIDM simulations. 
The model incorporates SIDM effects along mass evolution histories of CDM halos and it is applicable to both isolated halos and suhbhalos.  
We focus on the accuracy of the model in predicting halo density profiles at $z=0$ and the evolution of maximum circular velocity. 
We find the model predictions agree with the simulations within $10\%\textup{--}50\%$ for most of the simulated (sub)halos, $50\%\textup{--}100\%$ for extreme cases. This indicates that the model effectively captures the gravothermal evolution of the halos with very strong, velocity-dependent self-interactions.
For an example application, we apply the model to study the impact of various SIDM scenarios on strong lensing perturber systems, demonstrating its utility in predicting SIDM effects for small-scale structure analyses. 
Our findings confirm that the model is an effective tool for mapping CDM halos into their SIDM counterparts. 
\end{abstract}

\maketitle

\section{Introduction}

Elastic self-interactions of dark matter particles can dynamically affect dark matter halos, leading to phenomena such as halo core formation and collapse~\cite{1994ApJ...427L...1F,Spergel9909386,deBlok:2001hbg,Gentile:2004tb,2009MNRAS.397.1169D,deNaray09123518,2011AJ....141..193O,Oman150401437,Salucci:2018hqu,2019MNRAS.490.5451D,DeLaurentis:2022nrv}. 
This characteristic of self-interacting dark matter (SIDM) provides a compelling framework for explaining the diverse internal structures observed across a broad mass range of galaxies~\cite{Tulin170502358,Adhikari220710638}, from satellite galaxies of the Milky Way~\cite{Nishikawa:2019lsc,Sameie:2019zfo,Zavala:2019sjk,Kaplinghat:2019svz,Kahlhoefer:2019oyt,Correa:2020qam,Turner:2020vlf, Slone:2021nqd,Silverman:2022bhs,Correa:2022dey,Yang:2022mxl,Zhang:2024ggu} to galaxy clusters~\cite{2011MNRAS.415..448W,Robertson:2018anx,Banerjee:2019bjp,Andrade2020,Yang:2021kdf,Girmohanta:2022izb,Yang:2023stn,Sabarish:2023ija}, including ultra-diffuse galaxies and dwarfs~\cite{Hayashi:2020syu,Yang:2022mxl,Kong220405981,Zhang:2024ggu}, as well as spiral galaxies in the field~\cite{Kamada:2016euw,Creasey:2017qxc,Ren180805695} and elliptical galaxies~\cite{Gnedin:2000ea,McDaniel:2021kpq,Kong:2024zyw}.  
The interplay between SIDM halos and baryons can further amplify the diversity in these systems~\cite{Kaplinghat:2013xca,Vogelsberger:2014pda,Robertson:2018anx,Robertson:2020pxj,Yang:2021kdf,Rose:2022mqj,Mastromarino:2022hwx,Rahimi:2022ymu,Zhong:2023yzk,Creasey:2017qxc,Fry:2015rta,Fischer:2023eet,Kong:2024zyw,Correa:2024vgl,Yang:2024tba}.
To robustly constrain or detect SIDM signatures, dedicated efforts to search for these effects are crucial. 
For instance, Ref.~\cite{Nadler:2023nrd} investigated the signatures of a perturbed strong-lens image~\cite{Vegetti:2009cz,Vegetti:2012mc,Minor:2020hic}, suggesting it could be attributed to a core-collapsing subhalo. 
At the opposite end of the spectrum, Ref.~\cite{Zhang:2024ggu} demonstrated that the unique characteristics of Crater-II, one of the faintest satellites observed in the Milky Way~\cite{torrealba2016feeble-CraterII-discover,caldwell2017crater-CraterII-obs,Fu:2019,vivas2020decam,Ji:2021,2022MNRAS.512.5247B}, align with predictions from SIDM models but are challenging to explain in CDM.
Moreover, velocity-dependent SIDM models are increasingly recognized for their potential to reconcile small-scale observational discrepancies~\cite{Colin:2002nk,Nadler200108754,Yang220503392,Outmezguine:2022bhq,Yang:2022zkd,Yang:2023jwn,Fischer:2023lvl,Sabarish:2023ija,Vogelsberger:2014pda,Robertson:2018anx,Correa:2020qam,Robertson:2020pxj,Rose:2022mqj,Mastromarino:2022hwx,Rahimi:2022ymu}.
The core collapsed SIDM halos could also provide seeds for supermassive black holes~\cite{Balberg:2001qg,Pollack:2014rja,Choquette:2018lvq,Feng:2020kxv,Feng:2021rst,Xiao:2021ftk,Meshveliani:2022rih}.

The evolution of SIDM halos has been studied using various methods.  
N-body simulations can provide detailed and realistic representations of SIDM halos, but they are computationally expensive.
A high resolution is typically required to accurately model the inner halo regions, where the SIDM effects are expected to be most prominent~\cite{Vogelsberger:2012ku,Rocha:2012jg,Peter12083026,Zavala:2012us,Vogelsberger:2015gpr,Fitts:2018ycl,Robles:2019mfq,Banerjee:2019bjp,Sameie:2021ang,Nadler:2020ulu,Ebisu:2021bjh,Elbert:2016dbb,Creasey:2017qxc,Sameie:2018chj,Huo:2019yhk,Zeng:2021ldo,Burger:2022cjo,Yang220503392,Ray:2022ydr,Fischer:2023lvl,Du:2024sbt,Zhang:2024qem,Fischer:2024dte}. 
In comparison, the conducting fluid model allows for a more efficient alternative for simulating SIDM halos~\cite{Balberg:2002ue,Koda11013097,Essig:2018pzq,Nishikawa:2019lsc,Feng:2020kxv,Yang:2022zkd,Yang:2023stn,Zhong:2023yzk,Gad-Nasr:2023gvf}. 
It employs a set of fluid-like equations to model the dynamical properties of an isolated SIDM halo, enabling a theoretical understanding within the fluid framework referred to as gravothermal evolution. 
As a further simplification, a semi-analytic method, based on self-consistently adding an isothermal core to the Navarro-Frenk-White (NFW) profile~\cite{1997ApJ...490..493N}, has been proposed to model the thermalization of inner halo regions, which is most suitable for core-forming halos~\cite{Kaplinghat150803339,Elbert:2016dbb,Sameie:2018chj,Robertson:2020pxj,Jiang:2022aqw,Kamada:2016euw,Ren180805695,2020MNRAS.495...58S,Zentner:2022xux,Yang:2023stn}.

Recently, Ref.~\cite{Yang:2023jwn} introduced a parametric model that offers a universal solution to modeling gravothermal evolution of SIDM halos.
The model employs a parametric density profile and its time evolution is calibrated using high-resolution N-body simulations.
The parametric model builds on two key ingredients. 
First, the gravothermal evolution of isolated halos under constant cross sections reveals a universal pattern, allowing halo evolution to be parameterized through a single set of equations~\cite{Outmezguine:2022bhq,Yang:2023jwn}. 
Second, differential cross sections that have angular and velocity dependencies can be effectively approximated by an equivalent constant cross section for a given halo. 
This is achieved by integrating out the velocity ($v$) and angular ($\theta$ for the polar angle) dependencies using a $v^5\sin^2\theta$ kernel, tailored to the velocity distribution of the halo~\cite{Yang220503392,Yang:2022zkd}. 

\begin{figure}[htbp]
  \centering
  \includegraphics[width=8.cm]{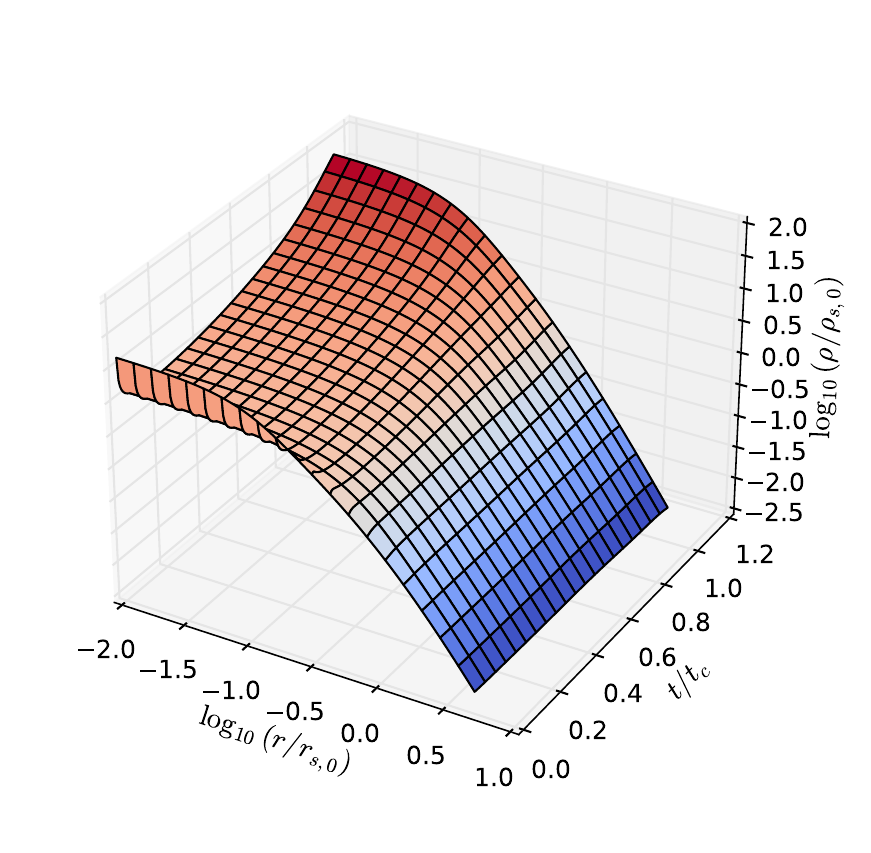}
  \caption{\label{fig:demo1}
A visual representation of the parametric model illustrating gravothermal evolution through a single parametric density profile, which is normalized using the initial NFW scale parameters ($\rho_{s,0}$, $r_{s,0}$) and the core collapse time ($t_c$).
The surface is colored based on the values of the density: high (low) densities correspond to red (blue) colors. }
\end{figure}

In Fig.~\ref{fig:demo1}, we present the evolution of the parametric density profile, which is normalized using the initial NFW scale parameters ($\rho_{s,0}$, $r_{s,0}$) and the core collapse time ($t_c$). 
The gravothermal evolution of the SIDM halo starts with an NFW initial condition at $t=0$ Gyr, and its inner density decreases first, reaching a minimum at approximately $0.2 t_c$, and subsequently increase rapidly, particularly around $t_c$.
It illustrates the universal behavior in the gravothermal evolution of SIDM halos, a crucial feature that enables parametric modeling of the evolution~\cite{Outmezguine:2022bhq,Yang:2023jwn,Zhong:2023yzk,Yang:2024tba}. 

Ref.~\cite{Yang:2023jwn} proposed two approaches to map CDM halos into their SIDM counterparts for a given particle physics realization of SIDM. 
The basic approach takes only the maximum circular velocity ($V_{\rm max}$) and the corresponding radius at which $V_{\mathrm{max}}$ is achieved ($R_{\rm max}$) of a CDM halo at $z=0$ as input and provides a smooth evolution history of its SIDM counterpart. 
It is accurate for isolated halos that have yet to undergo strong accretion and tidal events since their formation. 
The integral approach utilizes the entire accretion history of a CDM halo and integrates SIDM effects throughout its evolution. This method successfully accounts for both mass loss and accretion, proving to be effective for modeling the evolution of both isolated halos and subhalos.
More recently, the parametric model has been implemented into the {\sc SASHIMI} semi-analytic subhalo modeling program~\cite{Ando:2024kpk}. 
It has also been extended to incorporate the effect of baryons~\cite{Yang:2024tba}.

In this study, we comprehensively test the parametric model introduced in Ref.~\cite{Yang:2023jwn} by applying it to a series of halos matched in both CDM and SIDM simulations.
While the original study focused on statistical tests with the Milky Way simulation in Ref.~\cite{Yang:2022mxl}, this work tests the model on a halo-by-halo basis.
We evaluate the model's accuracy for both basic and integral approaches, considering a diverse range of masses and effective cross sections.
Additionally, we apply the model to group-scale simulations featuring extreme cross sections, as discussed in Ref.~\cite{Nadler:2023nrd}. This allows us to explore the limits of the model's accuracy and identify numerical challenges in N-body simulations. 
We further demonstrate the model's utility in generating predictions for strong lensing perturber systems under various SIDM scenarios.
Building upon the initial findings of Ref.~\cite{Yang:2023jwn}, our work offers a more systematic and extensive assessment of the parametric model using a larger sample of resolved halos within the simulations.

The rest of the paper is organized as follows. In Sec.~\ref{sec:model}, we revisit the parametric model, discussing its application to simulated halos. Sec.~\ref{sec:sim} provides a detailed description of the simulation data used in this study, including the method for matching simulated CDM halos to their SIDM counterparts. 
The accuracy of the parametric model is evaluated in Sec.~\ref{sec:valid}, where both the basic and integral approaches are examined. In Sec.~\ref{sec:group}, 
we apply the model to a simulated group system, considering an exceptionally large SIDM cross section.
An example application concerning lensing perturber systems within the group simulation is presented in Sec.~\ref{sec:application}. Finally, the paper concludes with a summary and discussion in Sec.~\ref{sec:sum}.

\section{The parametric model}
\label{sec:model}

The parametric model comprises an SIDM halo that evolves from its initial NFW profile to a later time. 
The parameters of the profile, after appropriate normalization, have no explicit dependence on the initial halo parameters and the SIDM cross section.
Ref.~\cite{Yang:2023jwn} considered two analytical forms for the evolving SIDM density profile, i.e., the $\beta4$ and Read profiles. The $\beta4$ profile takes the following form
\begin{eqnarray}
\label{eq:cnfw}
\rho_{\rm \beta4}(r) = \frac{\rho_s}{\frac{\left(r^{4}+r_c^{4} \right)^{1/{4}}}{r_s} \left(1 + \frac{r}{r_s} \right)^2}, 
\end{eqnarray}
where $\rho_s$, $r_s$, and $r_c$ are three parameters that evolve in time. 
The Read profile, which was originally proposed in Refs.~\cite{2016MNRAS.462.3628R,2016MNRAS.459.2573R}, takes a more complicated form but has the advantage that the mass profile can be obtained analytically. 
For clarity, we mainly focus on the $\beta4$ profile and provide numerical results for this profile in the main text. 
We provide in Appendix~\ref{app:compProfs} details of the Read profile and show that the two profiles yield almost the same results when used in the parametric model. 

The evolution of the profile parameters can be put in a universal form and extracted from high-resolution N-body simulation, see Ref.~\cite{Yang:2023jwn} for details. 
For the $\beta4$ profile, their evolution trajectories are
\begin{widetext}
\begin{eqnarray}
\label{eq:m0}
\frac{\rho_s(\tau)}{\rho_{s,0}} &=& 2.033 + 0.7381 \tau + 7.264 \tau^5 -12.73 \tau^7  + 9.915 \tau^9 + (1-2.033) (\ln 0.001)^{-1} \ln \left( \tau + 0.001 \right), \nonumber \\
\frac{r_s(\tau)}{r_{s,0}} &=& 0.7178 - 0.1026 \tau +  0.2474 \tau^2 -0.4079 \tau^3 + (1-0.7178) (\ln 0.001)^{-1} \ln \left( \tau + 0.001 \right), \nonumber \\
\frac{r_c(\tau)}{r_{s,0}} &=& 2.555 \sqrt{\tau} -3.632 \tau + 2.131 \tau^2 -1.415 \tau^3 + 0.4683 \tau^4,  
\end{eqnarray}
\end{widetext}
where $\tau\equiv t/t_c$ is a normalized evolution time that incorporates the SIDM dependence, and the subscript ``$0$'' denotes the corresponding value of the initial NFW profile.
We have chosen the functional forms such that $\rho_s/\rho_{s,0}=1$, $r_s/r_{s,0}=1$, and $r_c/r_{s,0}=0$ at $\tau=0$.
The core collapse time 
\begin{eqnarray}
\label{eq:tc0}
t_{\rm c}  &=& \frac{150}{C} \frac{1}{(\sigma/m) \rho_s r_s} \frac{1}{\sqrt{4\pi G \rho_s}},
\end{eqnarray}
where $C=0.75$ is a constant that can be calibrated with N-body simulations~\cite{Koda11013097,Essig:2018pzq,Nishikawa:2019lsc,Yang:2022zkd}.
The $\rho_s$ and $r_s$ are scale density and radius of NFW halos computed using more conveniently and accurately determined quantities $V_{\rm max}$ and $R_{\rm max}$ as
$\rho_s = (V_{\rm max}/(1.648 r_s))^2/G$ and $r_s = R_{\rm max}/2.1626$.
In Eq.~\ref{eq:tc0},  $\sigma/m$ refers to a constant SIDM cross section per particle mass.

\begin{figure*}[htbp]
  \centering
  \includegraphics[height=7.2cm]{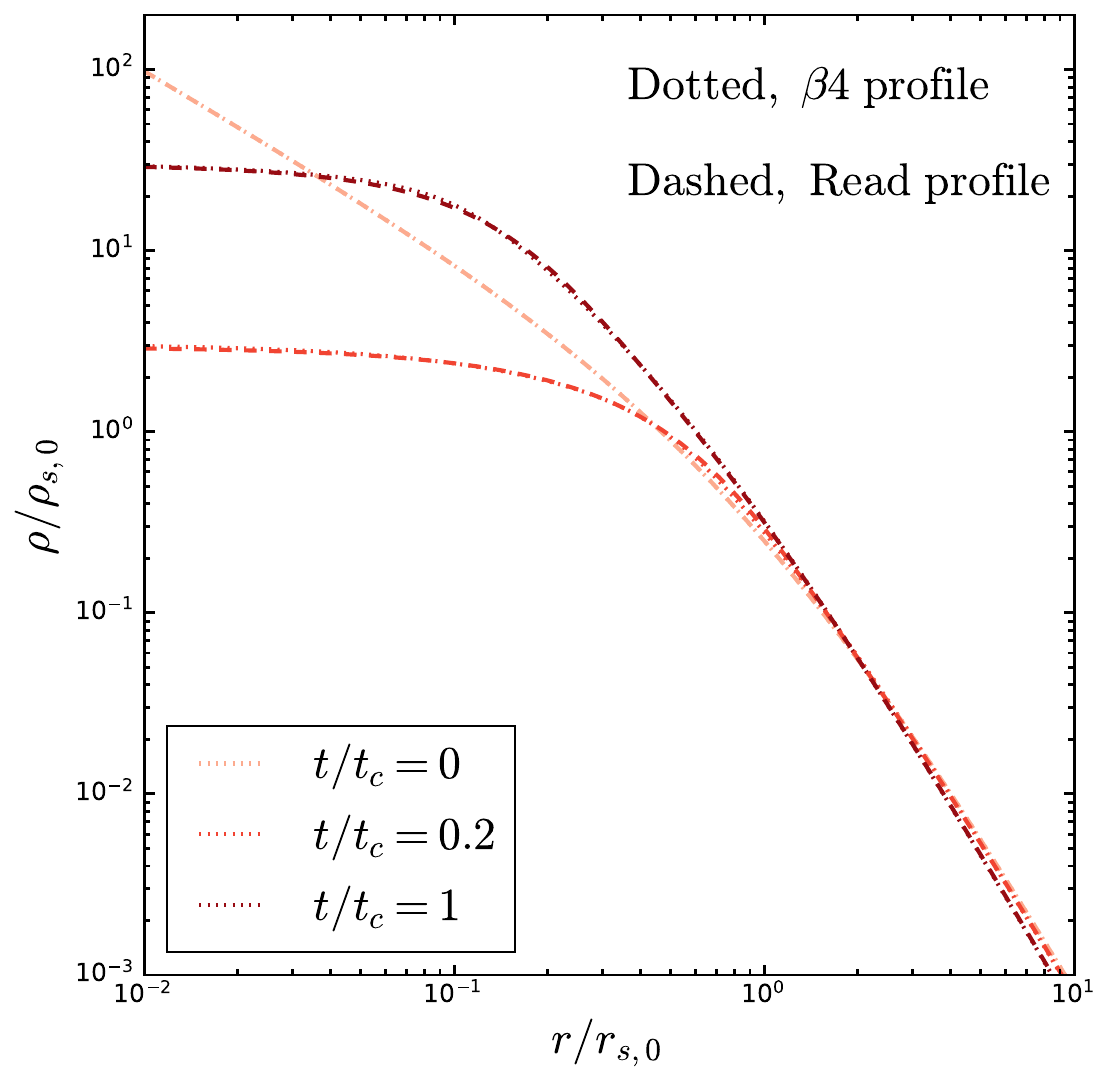}
  \includegraphics[height=7.2cm]{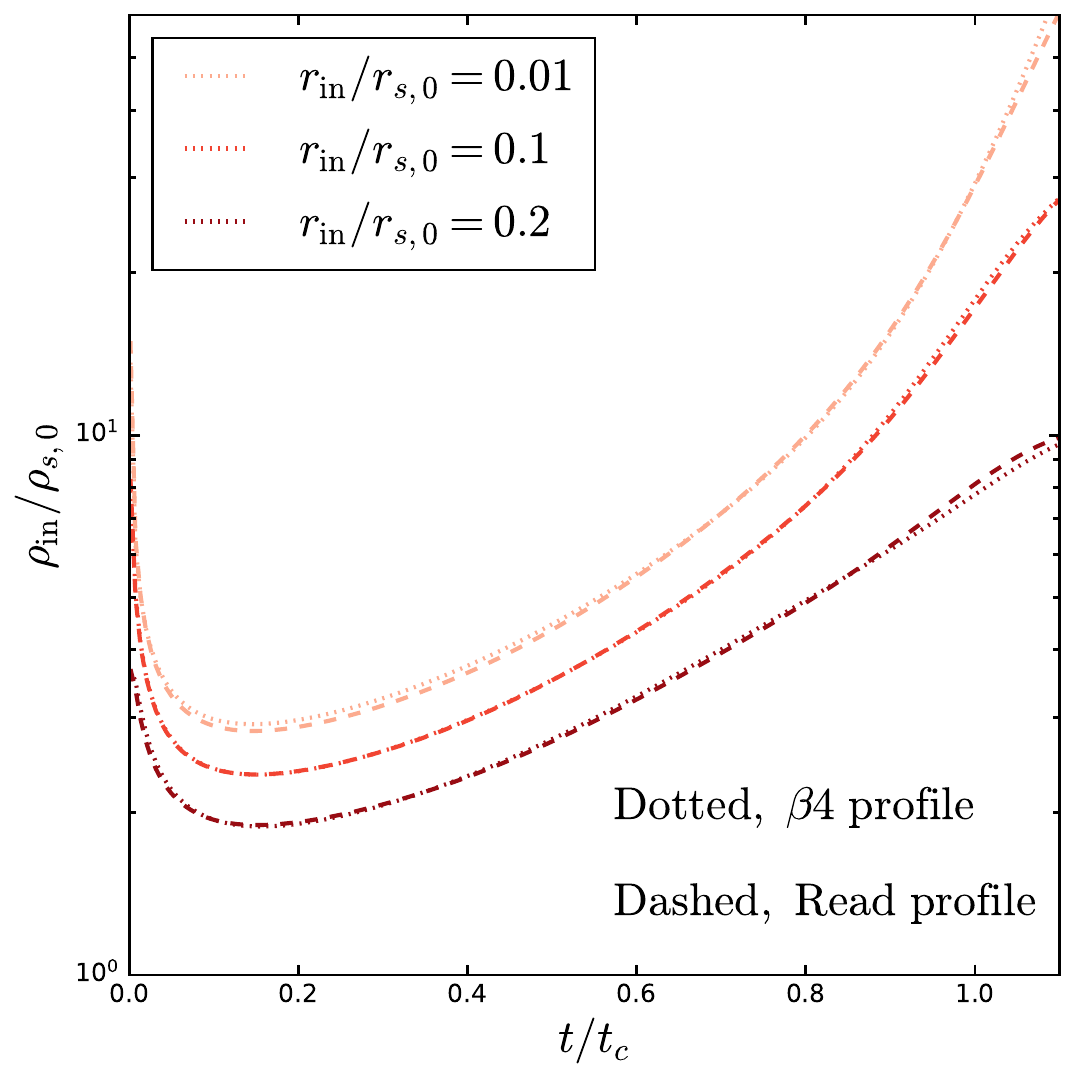}
  \caption{\label{fig:demo2} {\it Left:} Normalized $\beta4$ (dotted) and Read (dashed) density profiles at $t/t_c=0,0.2,$ and $1$.
{\it Right:} Densities at $r_{\rm in}/r_s=0.01,0.1,$ and $0.2$ as a function of $t/t_c$ for the normalized $\beta4$ (dotted) and Read (dashed) density profiles. }
\end{figure*}

Fig.~\ref{fig:demo2} (left) depicts the normalized density profiles at $t/t_c=0$, $0.2$, and $1$, based on the $\beta4$  (dotted) and Read (dashed) profiles. Fig.~\ref{fig:demo2} (right) illustrates the evolution of the inner densities at three specific profile radii: $r_{\rm in}/r_s=0.01, 0.1$, and $0.2$. 
We derive these results using analytic equations, such as Eqs.~\ref{eq:cnfw} and \ref{eq:m0} for the $\beta4$ profile.
We see the agreement between $\beta4$ and Read profiles are excellent, within the level of a few percent. 

The parametric model presented so far (Eqs. \ref{eq:cnfw} and \ref{eq:m0}) works only for SIDM models with constant cross sections and isotropic scatterings.
For dark matter velocity- and angular-dependent scatterings, we use the effective cross cross section in estimating the collapse time in equation~\cite{Yang220503392,Yang:2022zkd}, which is calculated as
\begin{eqnarray}
\label{eq:eff}
\sigma_{\rm eff} &=& \frac{2 \int d v d \cos\theta \frac{d \sigma}{d \cos\theta} \sin^2\theta v^5 f_{\rm MB}(v,\nu_{\rm eff}) }{\int d v d \cos\theta \sin^2\theta v^5 f_{\rm MB}(v,\nu_{\rm eff}) } \\ \nonumber
\end{eqnarray}
where $v$ denotes the relative velocity between the two incoming particles, $\theta$ refers to the polar angle that takes values in $[0,\pi]$,
and
\begin{eqnarray}
f_{\rm MB} (v,\nu_{\rm eff}) \propto v^2 \exp \left[-\frac{v^2}{4\nu_{\rm eff}^2}\right], 
\end{eqnarray}
is a Maxwell-Boltzmann velocity distribution that approximates the dark matter velocity distribution.
We have used $\nu_{\rm eff} =0.64 V_{\rm max, NFW} \approx 1.05 r_s \sqrt{G\rho_s}$ as a characteristic velocity dispersion for the inner halo region,
which has been found to perform well in several studies~\cite{Outmezguine:2022bhq,Yang220503392}.
The above integration kernel coincides with that of the heat conductivity of a fluid in the short-mean-free-path (SMFP) regime.
This is an intriguing result that can be interpreted as follows: the scattering probability is microscopically determined, depending on an SIDM model, while the subsequent evolution of energy transfer after scattering is solely governed by gravity and remains independent of the SIDM model, as discussed in~\cite{Yang220503392}.
Note that the SMFP regime is only relevant to central regions of a halo and at very late times in the gravothermal evolution, where the run-away collapse would result in the formation of a black hole~\cite{Lynden-Bell:1980xip,Pollack:2014rja,Choquette:2018lvq,Feng:2020kxv,Feng210811967,Xiao:2021ftk,Meshveliani:2022rih}. 
The inner region profiles in the SMFP regime, if needed, could be approximated through extrapolation assuming a power-law behavior $r^{-u}$ with $u\approx 2.21$~\cite{Lynden-Bell:1980xip}. 
For applying the parametric model, we focus on the density profile in the long-mean-free-path (LMFP) regime. 

\subsection{The basic approach}

The basic approach allows for efficient estimation of the SIDM effect on isolated halos. 
For a given halo, one computes the halo mass and NFW scale parameters $\rho_{s,0}$, $r_{s,0}$, using $V_{\rm max,0}$ and $R_{\rm max,0}$ at $z=0$.
With these parameters, the core collapse time, $t_c$, is calculated using Eq.~\ref{eq:tc0} and an estimated halo formation time. The halo formation time $t_f$ is defined as the current age of the universe (approximately 14 Gyr) minus the lookback time to its formation. 
Here we use a simplified equation provided in Ref.~\cite{Yang:2023jwn} for an estimate.
We first compute the halo formation redshift as
$z_f = -0.0064 \left(\log_{10}\left(\frac{M_{\rm vir, 0}}{{10^{10}~\rm M_{\odot}}}\right) \right)^2 -0.1043 \log_{10} \left(\frac{M_{\rm vir, 0}}{{10^{10}~ \rm M_{\odot}}}\right) + 1.4807$,
where $M_{\rm vir, 0}$ is the halo mass at $z=0$. 
Then we compute the halo formation time ($t_f$) as the current age of the universe (about 14 Gyr) minus the lookback time at $z_f$.

To obtain the density profile at a lookback time $t_L\leq t_L(z_f)$, one evaluates Eq.~\ref{eq:m0} at $\tau = (t_L(z_f)-t_L)/t_c$ and plugs in $\rho_{s,0}$, $r_{s,0}$ to obtain the $\rho_s$, $r_s$ and $r_c$ at that time.
If $t_L/t_c > 1$, the halo rapidly transitions into SMFP and Eq.~\ref{eq:m0} may not be applicable.   
In such extreme cases, the calculations are truncated at $\tau = 1$.

\subsection{The integral approach}
\label{sec:integral}

The integral approach aims to obtain more accurate predictions for halos with realistic growth histories. 
One first obtains the evolution of $V_{\rm max}$ and $R_{\rm max}$ in the parametric model using Eq.~\ref{eq:m0}. 
For the $\beta4$ profile, it reads 
\begin{widetext}
\begin{eqnarray}
\label{eq:m1}
\nonumber
\frac{V_{\rm max}}{V_{\rm max,0}} &=& 1 + 0.1777 \tau -4.399 \tau^3 + 16.66 \tau^4 - 18.87 \tau^5 + 9.077 \tau^7 - 2.436 \tau^9  \\ 
\frac{R_{\rm max}}{R_{\rm max,0}} &=& 1 + 0.007623 \tau - 0.7200 \tau^2 + 0.3376 \tau^3 -0.1375 \tau^4. 
\end{eqnarray}
\end{widetext}
Next, we integrate the SIDM effect along the evolution histories of $V_{\rm max,CDM}$ and $R_{\rm max,CDM}$ in CDM to obtain the SIDM predictions at a time $t$
\begin{widetext}
\begin{eqnarray}
\label{eq:int}
\nonumber 
V_{\rm max}(t)    &=&  V_{\rm max, CDM}(t_h) + \int_{t_h}^{t} d t'  \frac{d V_{\rm max,CDM}(t')}{d t'} +  \int_{t_h}^{t} \frac{dt'}{t_c(t')} \frac{d V_{\rm max, Model} (\tau') }{d \tau'} \\  
R_{\rm max}(t)    &=&  R_{\rm max, CDM}(t_h) + \int_{t_h}^{t} d t'  \frac{d R_{\rm max,CDM}(t')}{d t'} +  \int_{t_h}^{t} \frac{dt'}{t_c(t')} \frac{d R_{\rm max, Model} (\tau')}{d \tau'}, 
\end{eqnarray}
\end{widetext}
where 
$t_h = (14~{\rm Gyr} - t_{L}(z_f))/2 = t_f/2$
 is a halo formation time earlier than $t_f$, which is chosen sufficiently early while being after the exponential accretion phase to reduce numerical uncertainties in modeling the accretion history.
Only the last terms on the right-hand sides are relevant for capturing the SIDM effect, and the CDM evolution is recovered by removing these terms.
The $d V_{\rm max, Model} (\tau) /d \tau$ and $d R_{\rm max, Model} (\tau) /d \tau$ terms are obtained by taking derivatives of the corresponding functions of Eq.~\ref{eq:m1}, with the $V_{\rm max,0}$ and $R_{\rm max,0}$ being replaced by $V_{\rm max,CDM}(t')$ and $R_{\rm max,CDM}(t')$, respectively.

After obtaining $V_{\rm max}(t)$ and $R_{\rm max}(t)$ from Eq.~\ref{eq:int}, we need to establish the corresponding SIDM halo profile at time $t$. 
This can be done by determining $V_{\rm max,0}(t)$ and $R_{\rm max,0}(t)$ for an NFW profile that reproduces the given $V_{\rm max}(t)$ and $R_{\rm max}(t)$ at $\tau(t)$. Once these are found, we can directly compute the corresponding NFW scale parameters, $\rho_{s,0}(t)$ and $r_{s,0}(t)$.
In practice, we obtain $V_{\rm max,0}(t)$ and $R_{\rm max,0}(t)$ by solving Eq.\ref{eq:m1} with the known values of $V_{\rm max}(t)$, $R_{\rm max}(t)$, and $\tau(t)=(t-t_h)/t_c(t)$.
Then, the SIDM profile at $t$ is obtained using Eq.~\ref{eq:m0} with the $\rho_{s,0}(t)$, $r_{s,0}(t)$ and $\tau(t)$.  
Conceptually, it is important to note the distinction between the quantities $\rho_{s,0}(t)$ and $r_{s,0}(t)$ reconstructed from $V_{\rm max}(t)$, $R_{\rm max}(t)$, and the ones coming from the CDM evolution history.
In practice, such a distinction has a minor effect in dark matter-only cases.
We perform a quantitative comparative study in Appendix~\ref{app:phase}.
In the presence of baryons, the distinction can become more noticeable, see Appendix B of Ref.~\cite{Yang:2024tba} for details. 

\begin{figure*}[htbp]
  \centering
  \includegraphics[width=16cm]{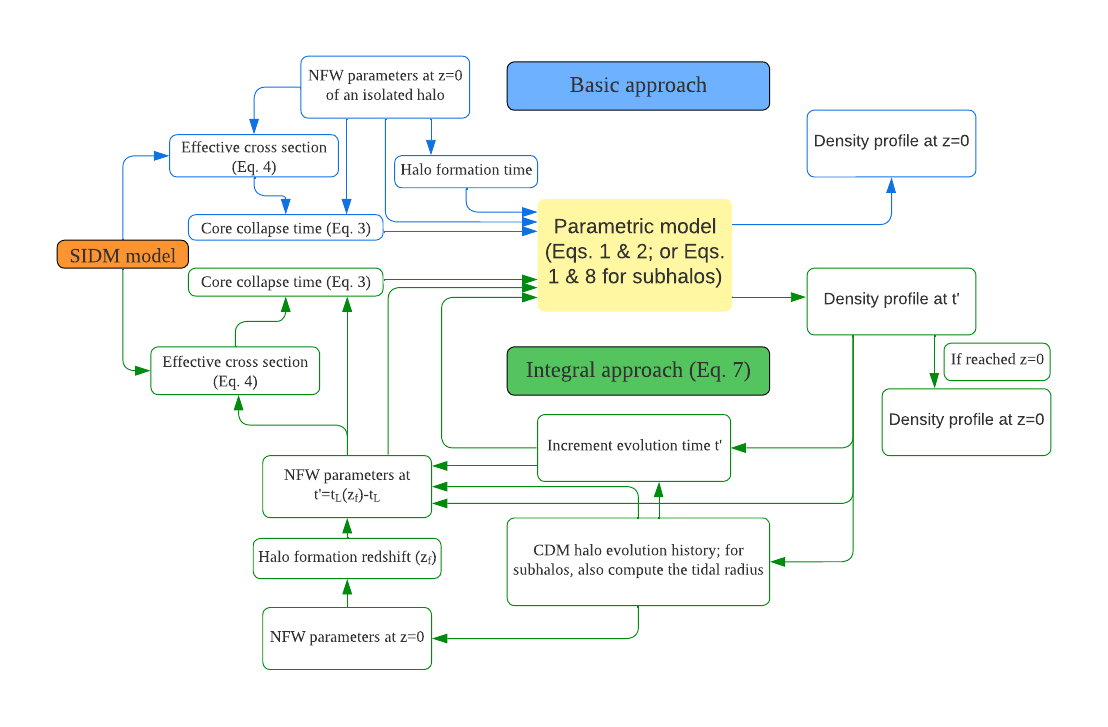}
  \caption{\label{fig:fc}
This flow chart delineates the two approaches of the parametric model for SIDM halos: the upper section for the basic approach and the lower section for the integral approach. It traces each step from initial parameters to final results, with references to key equations in the main text at relevant stages. The arrows for the basic and integral approaches are distinctly color-coded in blue and green, respectively. Detailed explanations of each step are provided in Sec.~\ref{sec:model}.
}
\end{figure*}

In our study, we reconstruct the evolution histories of $V_{\rm max,CDM}(t)$ and $R_{\rm max,CDM}(t)$ using the 
\textsc{Rockstar}~\cite{Behroozi11104372} and \textsc{consistent-trees}~\citep{Behroozi11104370} algorithms, and tabulate them at discrete snapshots. To effectively obtain the model predictions at all times, we take advantage of the fact that the integrals in Eq.~\ref{eq:int} can be obtained by summing over discrete contributions from divided time intervals.

The number of timesteps should be chosen such that the change in gravothermal evolution is always small during any increment of the evolution.  
The choice depends on the complexity of accretion histories, and typically, a few hundred steps should be sufficient for reaching the convergence over a smooth evolution history. 
For spiky accretion histories, some of the small wiggles are numerical fluctuations that arise from reconstruction, while the larger ones could be physical. Smoothing out the small wiggles would enable the use of fewer timesteps. 
In this work, we take a high number of timesteps of $10^4$ to retain all details from the reconstructed accretion history and ensure converged results.
Even with the high timestep count, our computational framework allows the completion of these calculations within a few seconds.

For halos that have evolved to $\tau>1$, we truncate their $V_{\rm max}$ and $R_{\rm max}$ evolution under SIDM, but still allow for their evolution under CDM, i.e., we set the last two terms in the right-hand sides in Eq.~\ref{eq:int} to zero.~\footnote{In practice, this truncation can be extended to $\tau=1.1$, as we have validated the performance of the parametric model through a prolonged simulation up to that point.}
For the sake of numerical stability, we also stop the $V_{\rm max}$ and $R_{\rm max}$ evolution under SIDM when their values drop below their resolution thresholds.
For the studies in this work, we consider $V_{\rm max}>2 ~\rm km/s$ and $R_{\rm max}>0.1~\rm kpc$, which are chosen to be slightly below the resolution limits of these variables.

\subsection{Subhalos}

Subhalos have density profiles distinct from the NFW profile at large radii. 
We introduce a tidal radius to smoothly truncate the density profile as 
\begin{eqnarray}
\label{eq:tcnfw}
\rho_{\rm tSIDM}(r) = \frac{\rho_{\beta4}(r,\rho_s,r_s,r_c) }{\left(1 + \left(\frac{r}{r_t}\right)^{2-u} \right)^{1+3u}}, 
\end{eqnarray}
where $u=0.25$ is a fixed constant.
Ref.~\cite{Yang:2023jwn} found that $u$ has a minor dependence on the halo concentration. 
In this work, with more matched halos, we have checked that the dependence cannot be distinguished from other modeling uncertainties and we take the fixed value for simplicity. 
The chosen form of Eq.~\ref{eq:tcnfw} implies that the same parametric model in the integral approach can be applied to subhalos, provided that $r_t$ exceeds $r_s$.
However, in extreme cases where $r_t$ is comparable to $r_s$ or $r_c$, SIDM-induced core formation may enhance tidal mass loss, thereby reducing the precision of our model predictions.
We will illustrate such an example later (Fig.~\ref{fig:sepcial}).
Additionally, subhalos with small pericentric passages are expected to develop an anisotropic velocity dispersion profile, as particles on more radial orbits are more likely to extend beyond the tidal radius and be stripped~\cite{Chiang:2024dlw}. For modeling this effect in CDM, see, e.g., Ref.~\cite{Ogiya:2019del}.
 In SIDM, the self-interactions tend to erase this anisotropy, pushing the system toward a more isotropic velocity dispersion~\cite{Vogelsberger:2012sa}. However, tidal heating and stripping can counteract this process, reintroducing anisotropy. These two competing effects differ from what occurs in CDM. We leave modeling this secondary effect for future work.

In Fig.~\ref{fig:fc}, we illustrate these procedures of the applications.
It outlines the basic and integral approaches in the upper and lower sections, respectively.
In the basic approach, it starts with the NFW parameters at $z=0$ for an isolated halo, then calculates the effective cross section, halo formation time, and core collapse time. This leads to the application of the parametric density model equations.
For the integral approach, it begins with the NFW parameters at a given time $t'=t_L(z_f)-t_L$. It then effectively captures the SIDM effect on the halo for a small timestep forward, providing the SIDM halo density profile at the incremented timestep.
If the incremented time reaches $z=0$, the present-day density profile is obtained.

\section{Simulation data}
\label{sec:sim}

We use the CDM and SIDM simulation data from Refs.~\cite{Yang:2022mxl,Nadler:2023nrd}
In Ref.~\cite{Yang:2022mxl}, we performed high-resolution cosmological zoom-in SIDM and CDM simulations with a particle mass of $5\times 10^4~\rm M_{\odot}$,
a Plummer-equivalent gravitational softening length $\epsilon=114~\rm pc$,
 and the main halo is $10^{12} M_{\odot}$, similar to the Milky Way halo. 
The SIDM effect is modeled considering a Rutherford-like scattering cross section~\cite{Feng09110422,Ibe:2009mk}
\begin{eqnarray}
\label{eq:xsr}
\frac{d\sigma}{d \cos\theta} = \frac{\sigma_{0}w^4}{2\left[w^2+{v^{2}}\sin^2(\theta/2)\right]^2 },
\end{eqnarray}
where $\sigma_0/m=147.1~\rm cm^2/g$ and $w=24.33~\rm km/s$. 
The cross section, hereafter ``MilkyWaySIDM,'' has a high value in halos hosting satellite galaxies, while being significantly suppressed at the mass scale of the Milky Way halo.
In Ref.~\cite{Nadler:2023nrd}, we performed a zoom-in simulation for a host halo on group mass scales, $10^{13}~\rm M_{\odot}$, with a minimum simulation particle mass of $4\times 10^5~\rm M_{\odot}$ and a Plummer-equivalent gravitational softening length $\epsilon=243~\rm pc$.
The ``GroupSIDM'' cross section used there has $\sigma_0/m=147.1~\rm cm^2/g$ and $w=120~\rm km/s$, and the simulation is based on a viscosity cross section, with the angular dependence averaged using a $\sin^2\theta$ kernel~\cite{Tulin:2013teo}.
Compared to the MilkyWaySIDM model, the GroupSIDM model has a much higher cross section towards the higher halo mass, and hence more halos would be in the collapse phase. 
For example, for a halo with $V_{\rm max}\approx 100~ \rm km/s$, $\sigma_{\rm eff}/m \approx 0.3~ \rm cm^2/g$ and $20~{\rm cm^2/g}$ for the former and latter, respectively; 
see Fig.~1 of Ref.~\cite{Nadler:2023nrd} for details. 
The GroupSIDM cross section represents an extreme case, resulting in a large population of core collapsed halos. 
The cosmological parameters used in both the simulations are $\Omega_{M}=0.286$, $\Omega_{\Lambda}=0.714$, $n_s=0.96$, $h=0.7$, and $\sigma_8=0.82$~\cite{WMAP:2012nax}.

\begin{figure*}[htbp]
  \centering
  \includegraphics[width=5.6cm]{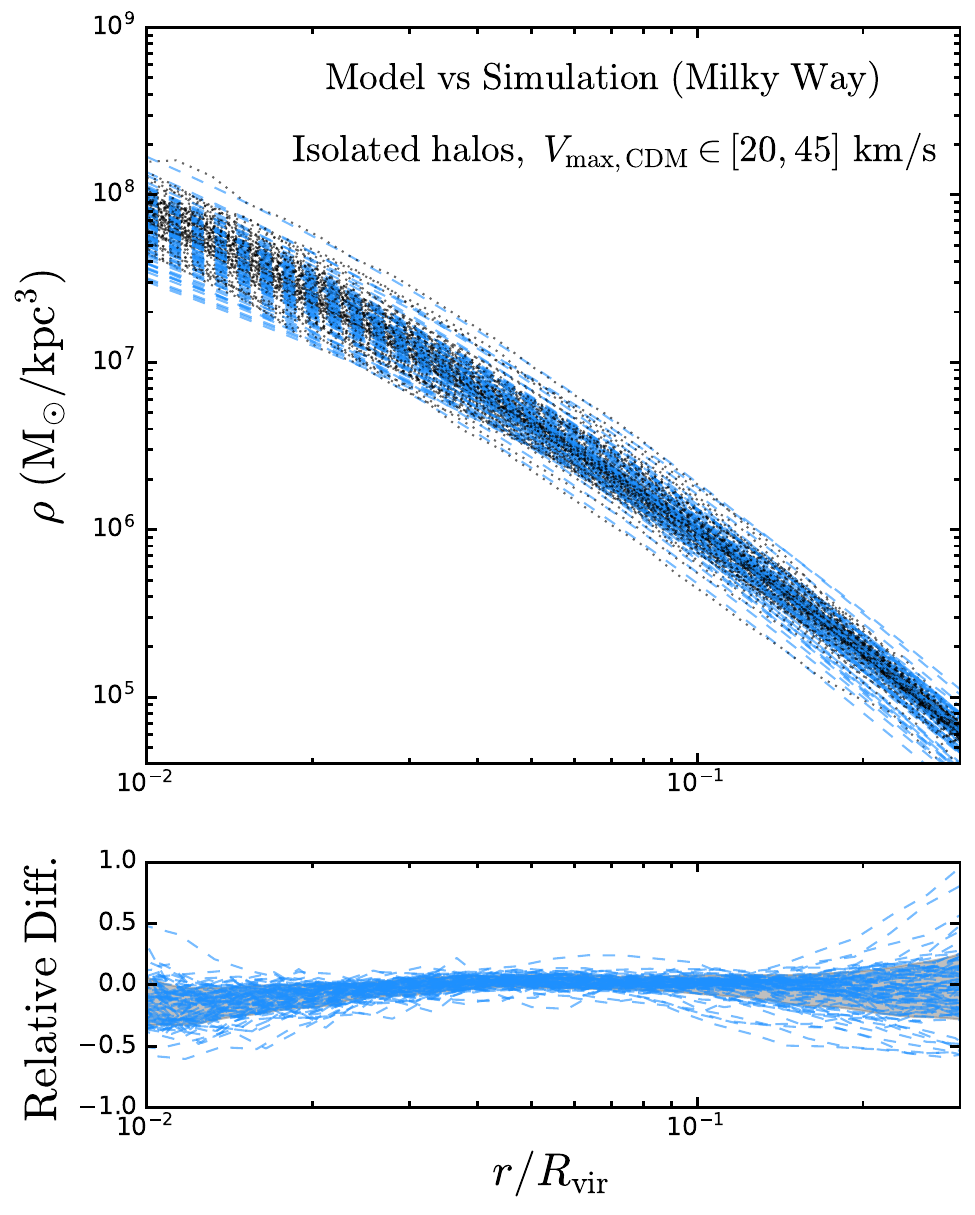}
  \includegraphics[width=5.6cm]{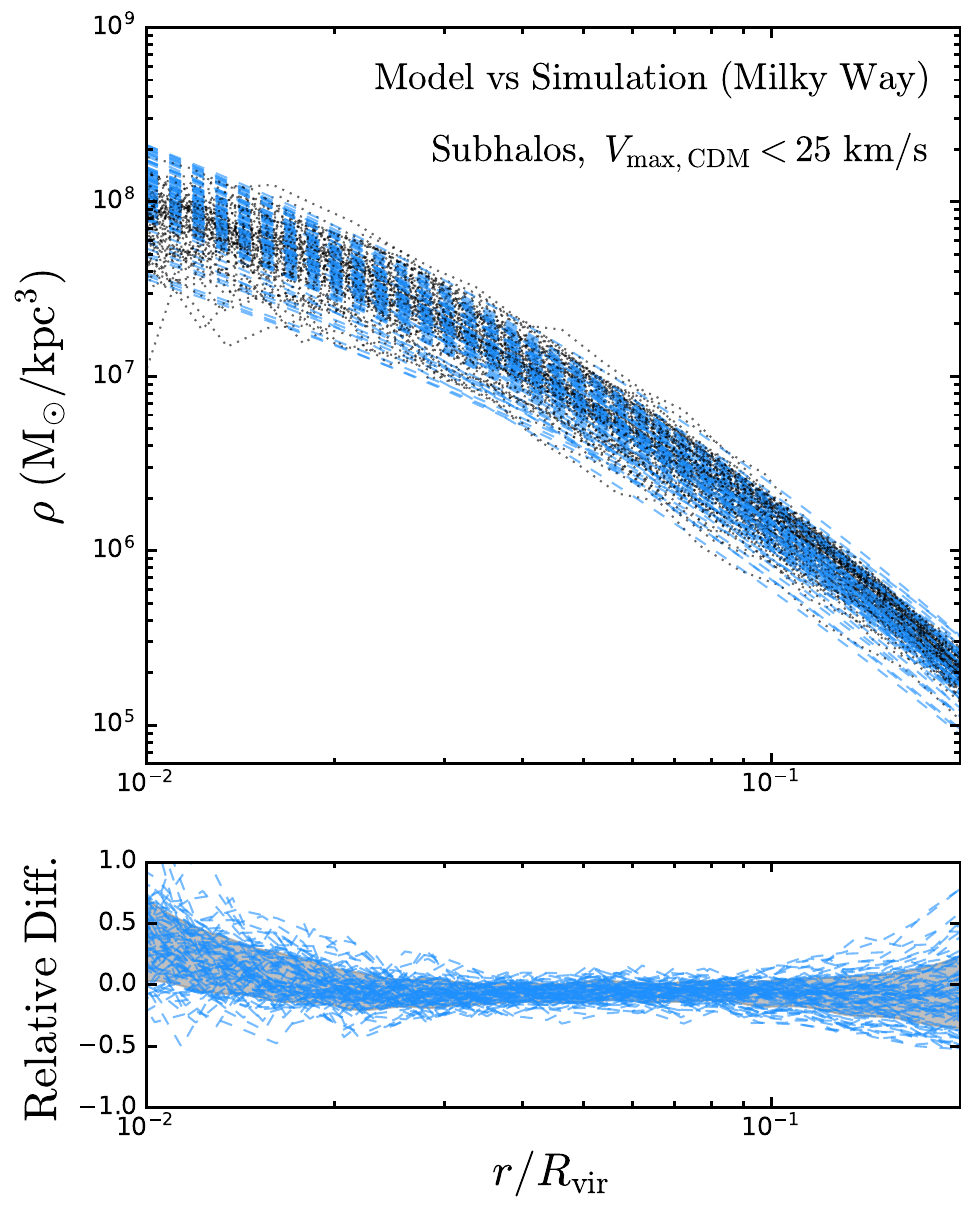}
  \includegraphics[width=5.6cm]{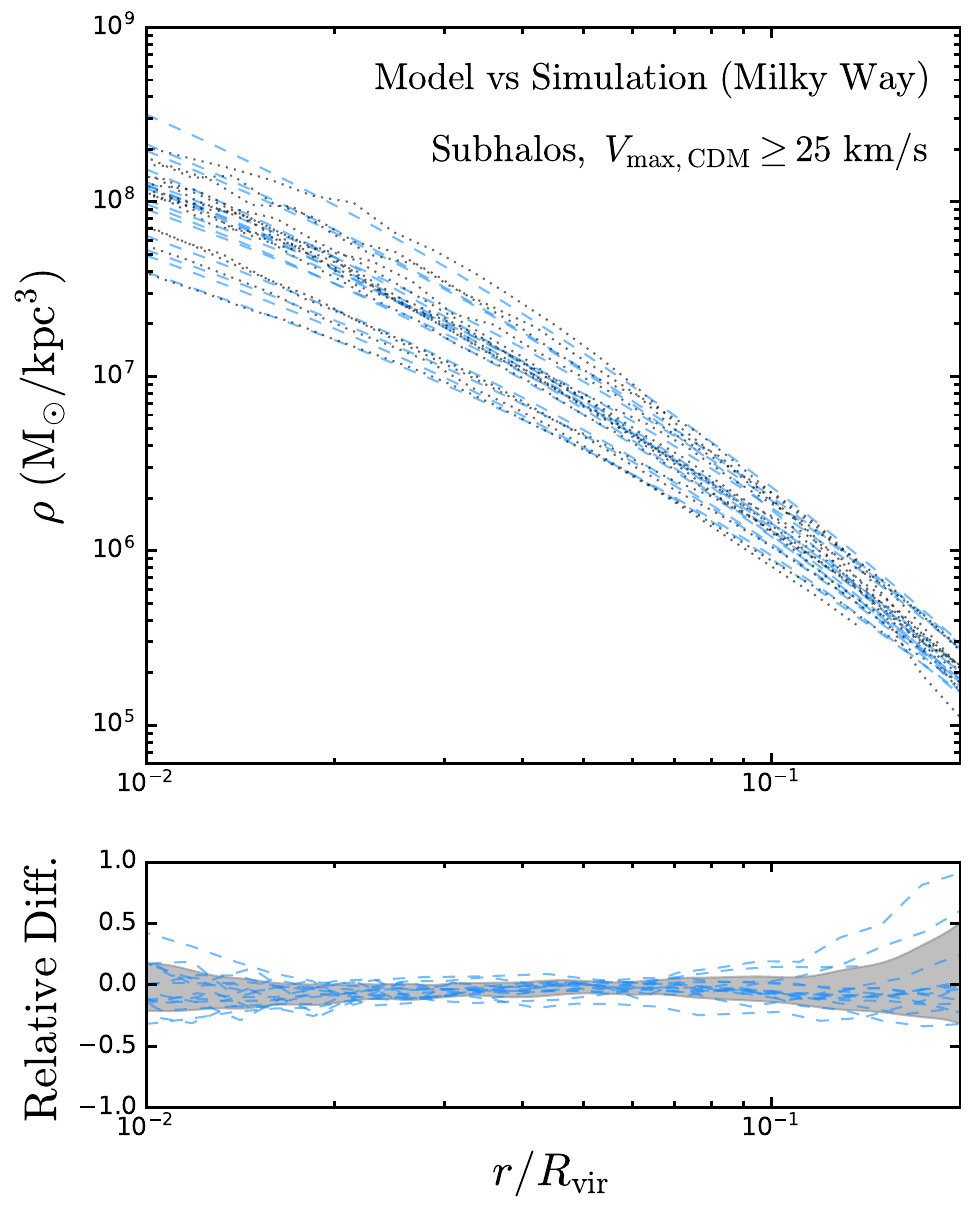}
  \caption{\label{fig:mwcdm}
The model-predicted (dashed) vs simulated (dotted) density profiles for isolated halos (left), subhalos with $V_{\rm max,CDM}$ smaller (middle) and larger (right) than $25~\rm km/s$, in the Milky Way CDM simulation of Ref.~\cite{Yang:2022mxl}.
The relative difference between each pair of simulated (Sim) and model-predicted (Mod) curves is measured as $\rm 2(Mod-Sim)/(Mod+Sim)$, with the $\pm 1\sigma$ band of the results shaded in gray.
}
\end{figure*}

\begin{figure*}[htbp]
  \centering
  \includegraphics[width=6cm]{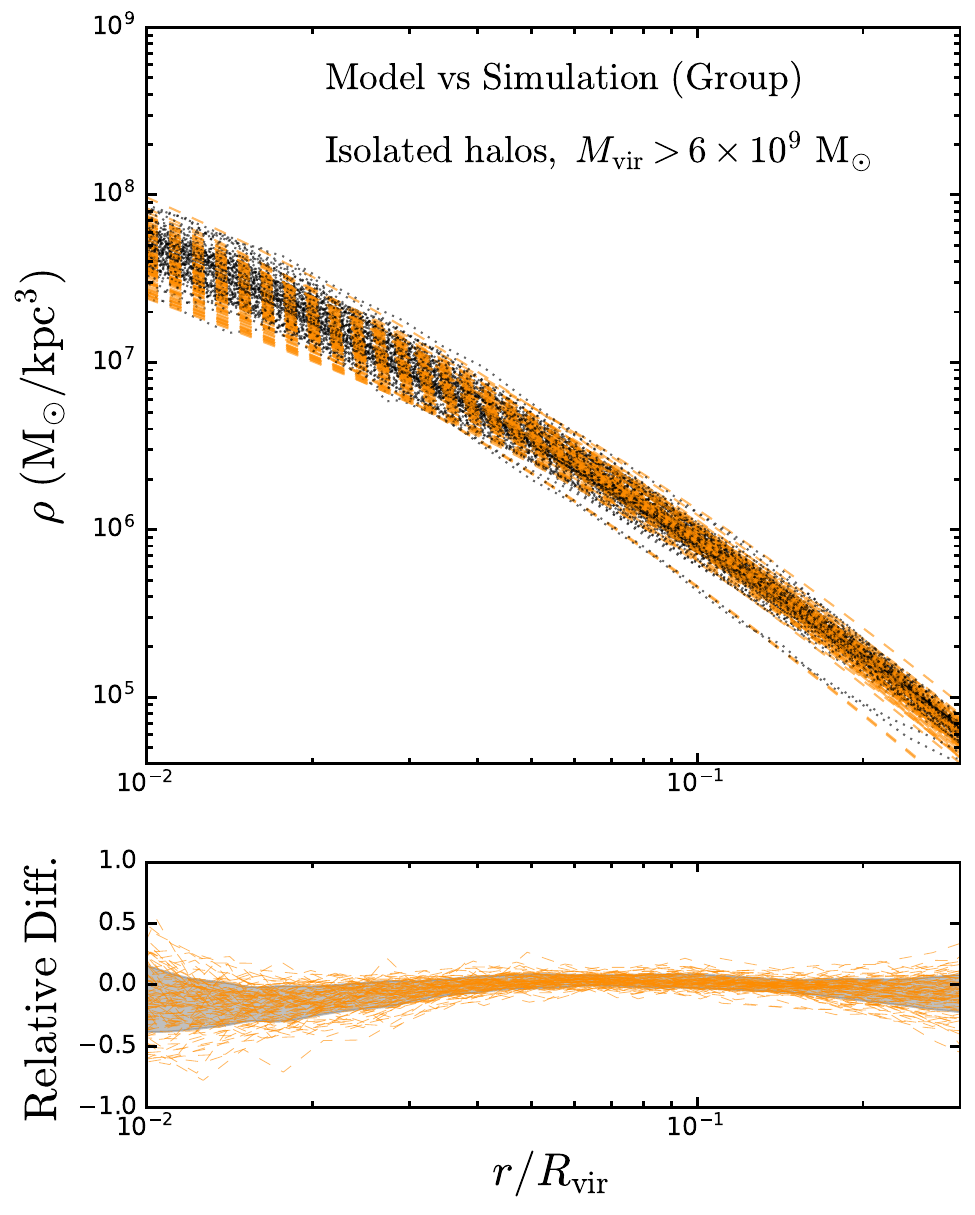}
  \includegraphics[width=6cm]{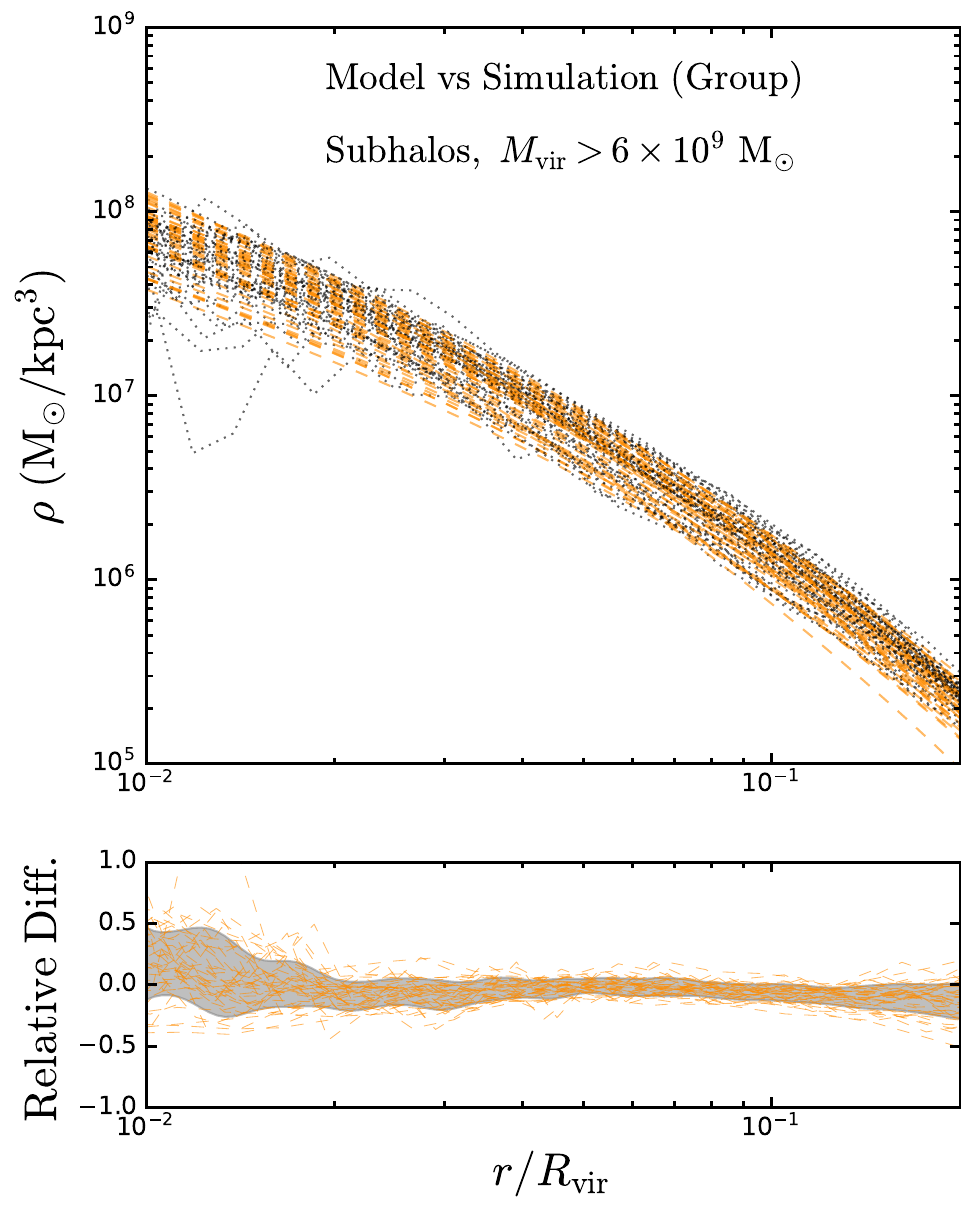}
  \caption{\label{fig:groupcdm} The model-predicted (dashed) vs simulated (dotted) density profiles for isolated halos (left) and subhalos (right) in the Group CDM simulation of Ref.~\cite{Nadler:2023nrd}.
The relative difference between each pair of simulated (Sim) and model-predicted (Mod) curves is measured as $\rm 2(Mod-Sim)/(Mod+Sim)$, with the $\pm 1\sigma$ band of the results shaded in gray.
}
\end{figure*}

For a given simulated SIDM halo, we find its CDM counterpart by examining their evolution trajectories and the success rate for matching the pair is 98\%. 
For subhalos of the main halo in the MilkyWaySIDM simulation, we consider halos of virial masses at $z=0$ higher than $1.43\times 10^8~\rm M_{\odot}$. 
The virial mass $M_{\rm vir}$ is defined according to \cite{Bryan_1998}, corresponding to a density contrast of $\Delta_{\mathrm{vir}}\approx 99$ times the critical density of the universe at $z=0$. 
This selection yields a sample of $102$ matched pairs of subhalos.
For isolated halos, we require them to reside at $0.3\textup{--}3~{\rm Mpc}$ from the Milky Way halo analog and find $620$ matched pairs. 
For the GroupSIDM simulation, we consider subhalos of the group main halo of masses higher than $10^9~\rm M_{\odot}/h$, and there are $50$ pairs.
We require isolated halos to reside at $0.8\textup{--}6~{\rm Mpc}$ from the group main and have masses at least $6\times 10^9~\rm M_{\odot}$, and find $317$ pairs. 

The parametric model in the CDM limit gives predictions for CDM halos. 
Before examining the performance for SIDM halos, we check the predictions of the parametric model in the CDM limit and ensure that they are consistent with the simulations.
Fig.~\ref{fig:mwcdm} illustrates the comparison between model predictions and simulations for CDM halos in the Milky Way simulation of Ref.~\cite{Yang:2022mxl}. 
We quantify the relative difference between simulated (Sim) and model-predicted (Mod) density profiles as $\rm 2(Mod-Sim)/(Mod+Sim)$.
It works even if the value in a bin is very small or even zero, which could occur due to simulation fluctuations. 
Additionally, the $\pm 1\sigma$ bands of these relative differences are indicated with gray shading.
We see that the model predictions align closely with the simulation results for the CDM isolated halos with $25~{\rm km/s}<V_{\rm max,CDM}<45~\rm km/s$ (left) and subhalos with $V_{\rm max,CDM} \geq 25~\rm km/s$ (right). 
However, for the subhalos with $V_{\rm max,CDM} < 25~\rm km/s$ (middle), we observe a systematic overestimation in the inner densities by up to 50\% at around $r/R_{\rm vir}=0.01$. 
The deviation is likely because the inner density profile of the subhalos is shallower than $\rho\propto r^{-1}$ due to tidal stripping. 
One may consider using the Einasto profile~\cite{2008MNRAS.387..536G,Navarro:2008kc,Dutton:2014xda,2016MNRAS.457.4340K,Fielder:2020lpk} for better fits, but it goes beyond our current parametric model framework.
Aside from this, the overall agreement is within 10\% around $r/R_{\rm vir}\approx 0.04$, with the relative difference widening towards both the inner and outer regions. At the innermost region shown ($r/R_{\rm vir}=0.01$), the $1\sigma$ difference reaches approximately 25\%.
At radii larger than those shown in the figure, the uncertainty continues to increase slightly. However, this uncertainty is tied to the accuracy of the NFW profile in modeling density distributions and is not related to SIDM.

Fig.~\ref{fig:groupcdm} shows comparisons for isolated (left) and subhalos (right) from the Group CDM simulation~\cite{Nadler:2023nrd}. 
The agreement is similar to that found for the halos in the Milky Way CDM simulation. 
However, in the subhalo case, the systematic shift is reduced, falling within the $\pm 1\sigma$ band and thus aligning with zero. 
In this test, we find two isolated CDM halos demonstrate exceptionally large deviations and exclude them from our study.

When applying the integral approach, several cases in the simulation undergo numerical instabilities and are excluded. 
These instabilities are largely due to insufficient accuracy in modeling the accretion history, and hence, it is an issue with the simulations (including the halo finding and merger tree algorithms) rather than the model.
For simplicity, we excluded such cases in this study. However, we note that excluding specific points with abrupt and significant changes in the accretion history of, e.g., $R_{\rm max,CDM}$, can bypass such instabilities.

\section{Model validation}
\label{sec:valid}

\begin{figure*}[htbp]
  \centering
  \includegraphics[width=5.3cm]{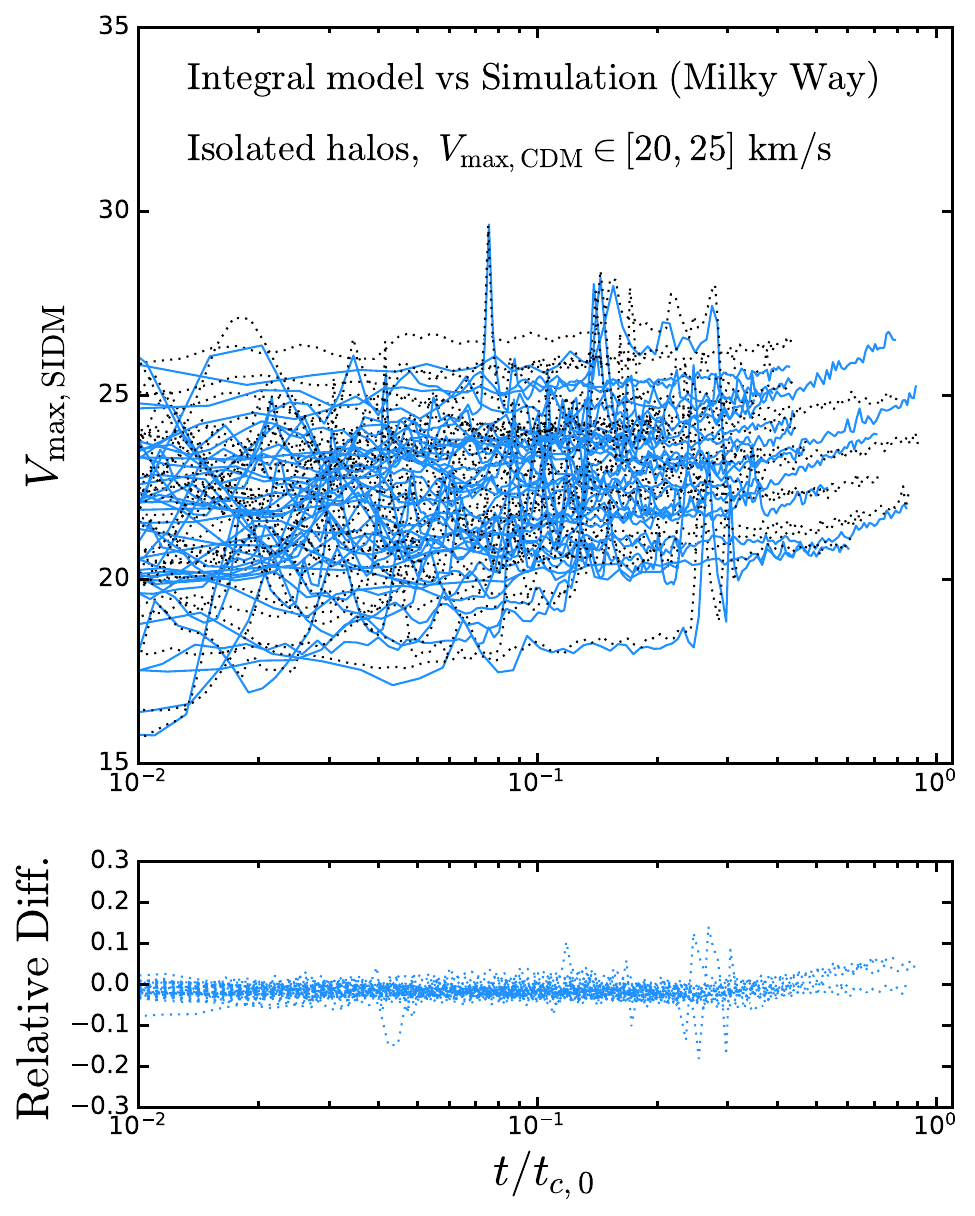}
  \includegraphics[width=5.3cm]{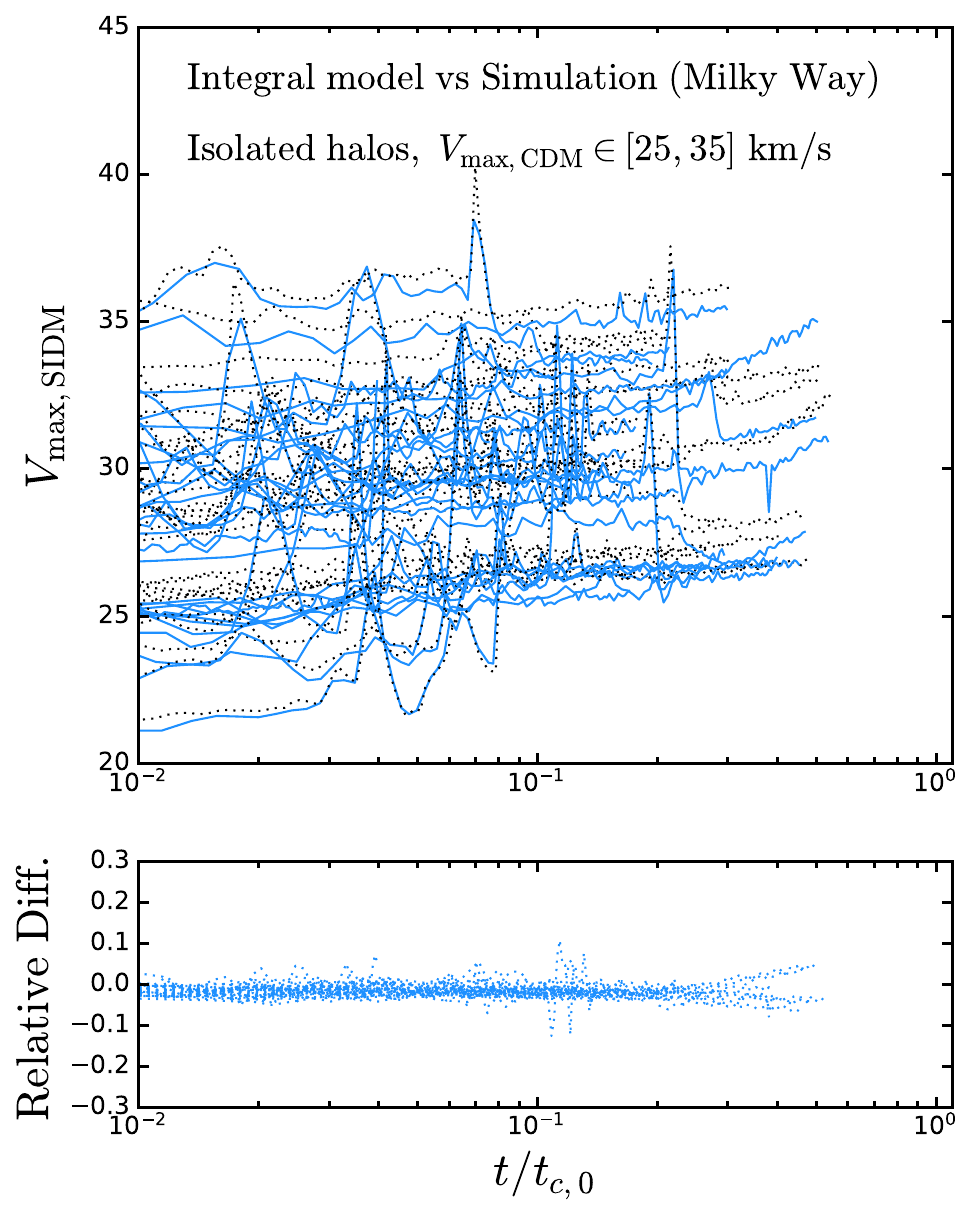}
  \includegraphics[width=5.3cm]{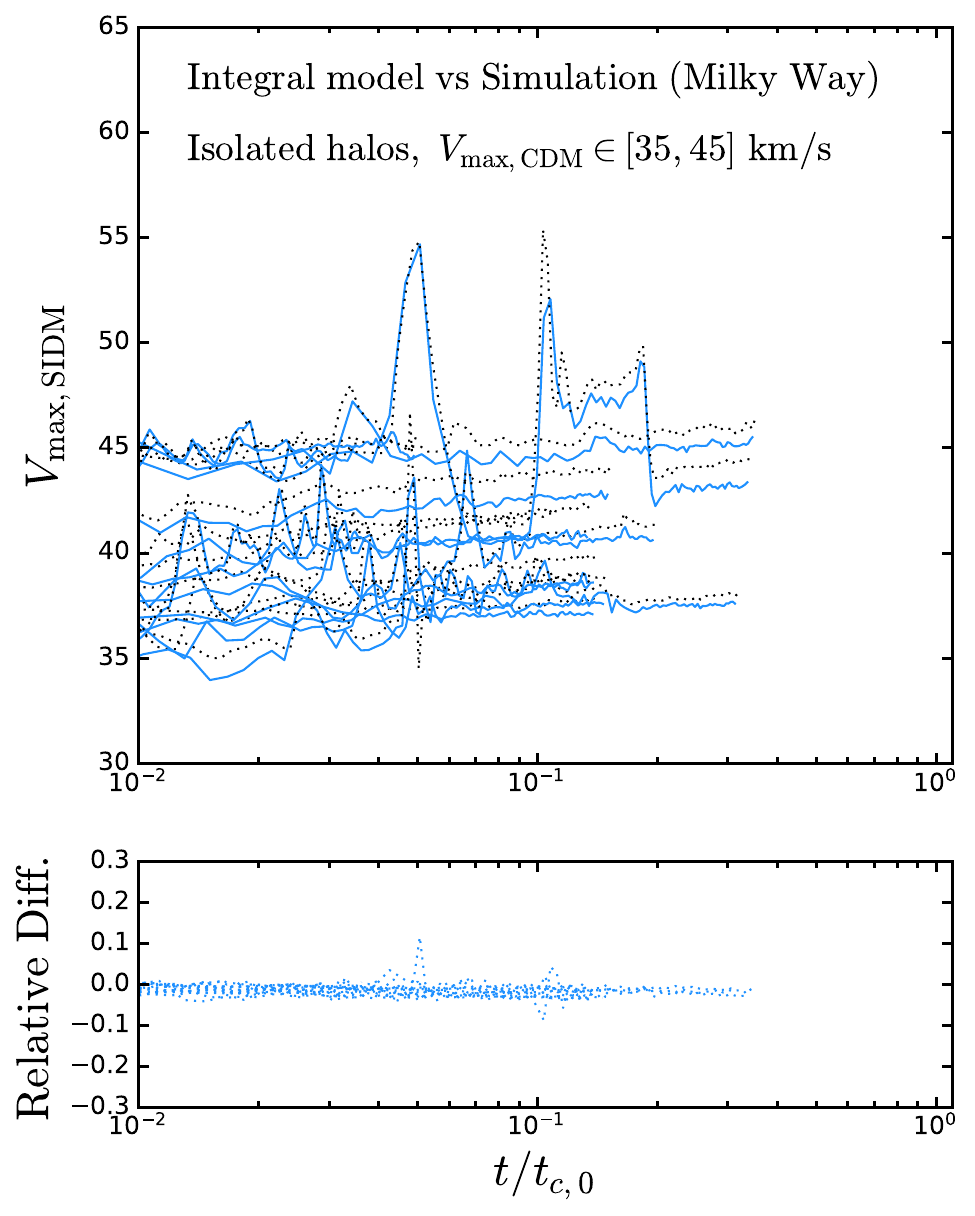}
  \caption{\label{fig:mahisos} The $V_{\rm max}$ as a function of normalized evolution time for isolated halos with $V_{\rm max,CDM} \in [20,25], ~ [25,35],$ and $[35,45]~\rm km/s$, in the MilkyWaySIDM simulation of Ref.~\cite{Yang:2022mxl}.
The solid and dotted curves denote the prediction of the parametric model and the simulation, respectively. 
In both cases, the evolution time starts at the estimated halo formation time and is normalized by the core collapse time estimated using $z=0$ halo properties.
The relative difference between each pair of simulated (Sim) and model-predicted (Mod) curves is measured as $\rm 2(Mod-Sim)/(Mod+Sim)$.
}
\end{figure*}

\begin{figure*}[htbp]
  \centering
  \includegraphics[width=5.3cm]{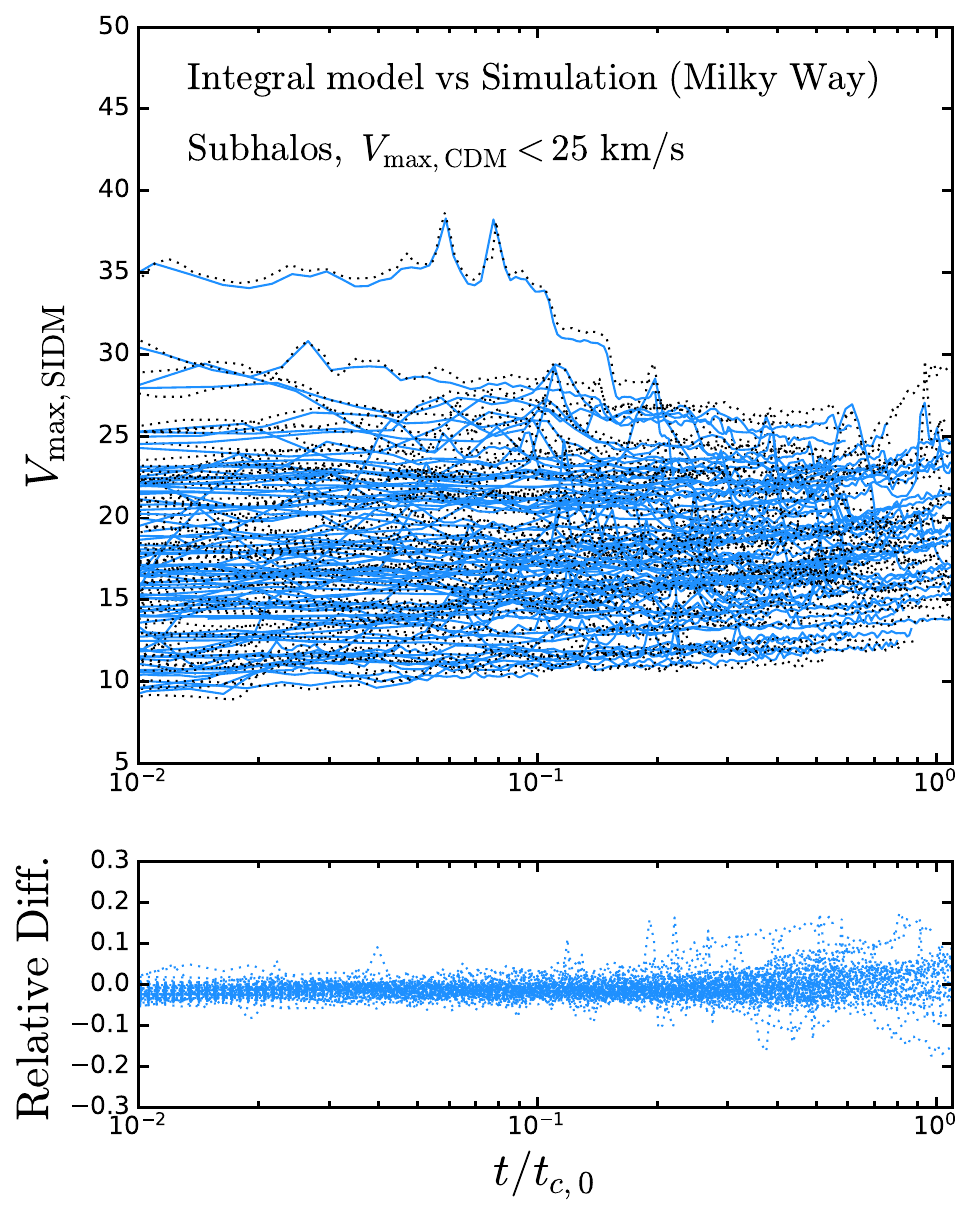}
  \includegraphics[width=5.3cm]{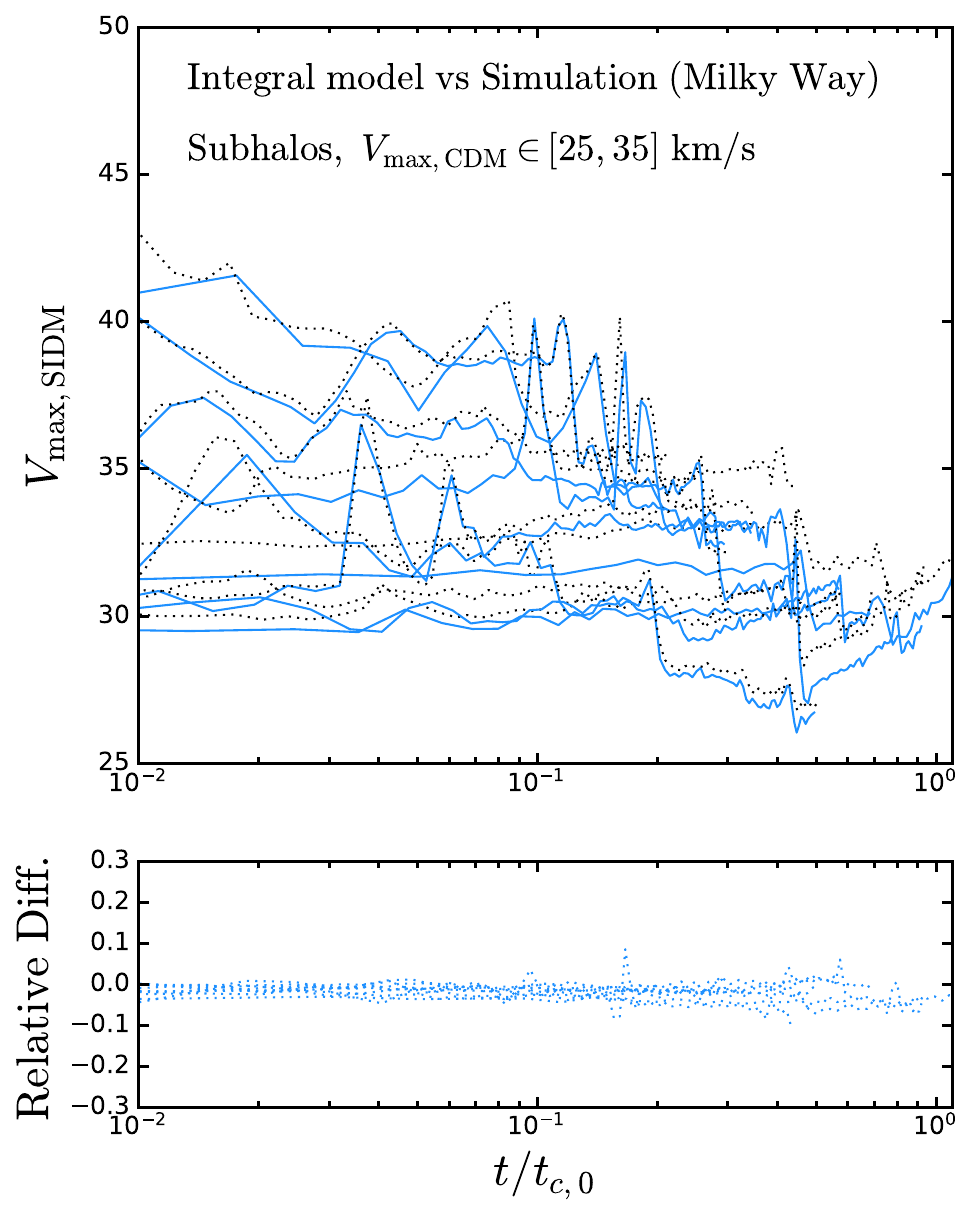}
  \includegraphics[width=5.3cm]{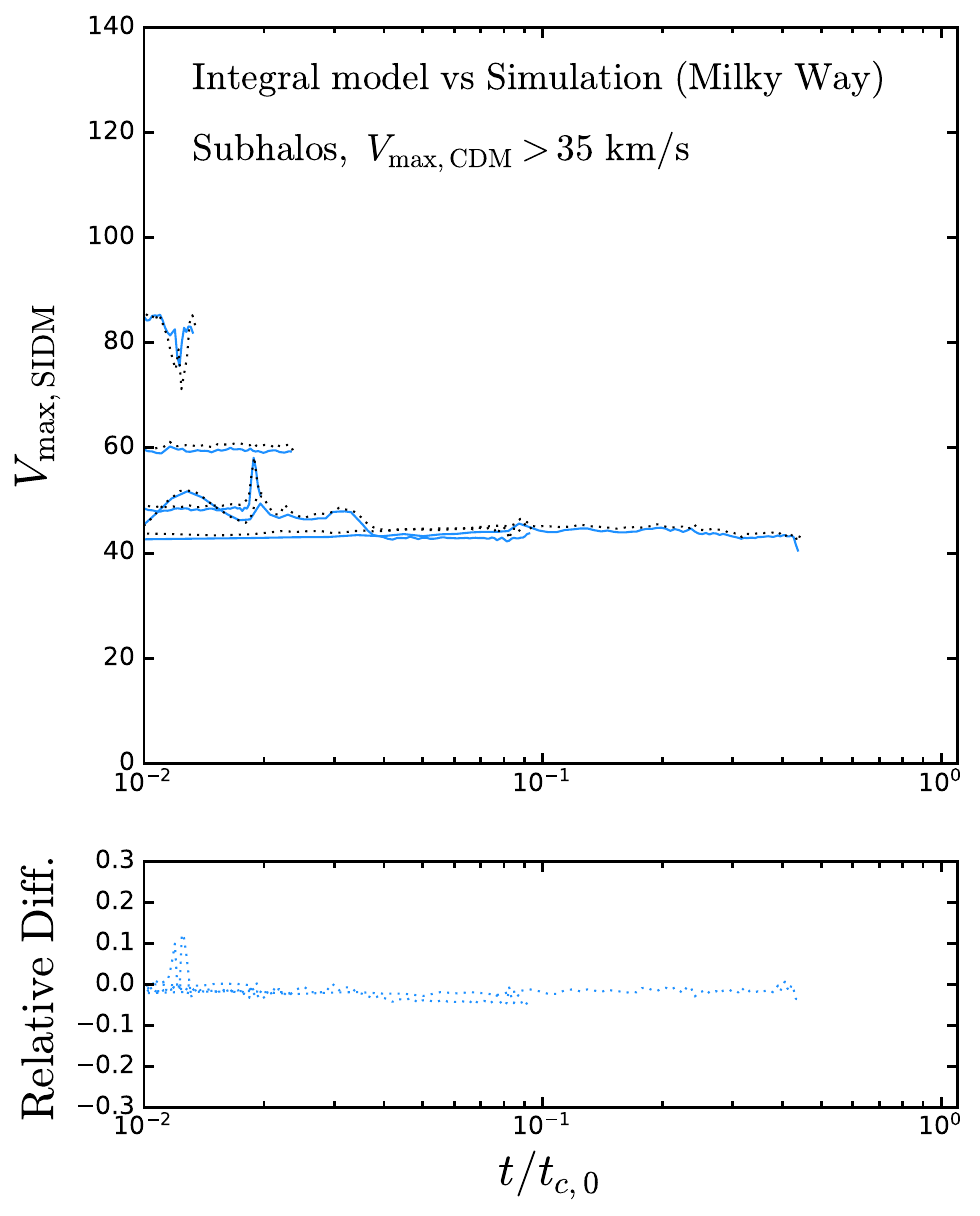}
  \caption{\label{fig:mahsubs} The $V_{\rm max}$ as a function of normalized evolution time for subhalos with $V_{\rm max,CDM} <25~\rm km/s$, $V_{\rm max,CDM}\in~ [25,35]~\rm km/s$, and $V_{\rm max,CDM}>35 ~\rm km/s$, in the MilkyWaySIDM simulation of Ref.~\cite{Yang:2022mxl}.
The solid and dotted curves denote the prediction of the parametric model and the simulation, respectively. 
In both cases, the evolution time starts at the estimated halo formation time and is normalized by the core collapse time estimated using $z=0$ halo properties.
The relative difference between each pair of simulated (Sim) and model-predicted (Mod) curves is measured as $\rm 2(Mod-Sim)/(Mod+Sim)$.
}
\end{figure*}

\begin{figure*}[htbp]
  \centering
  \includegraphics[width=5.3cm]{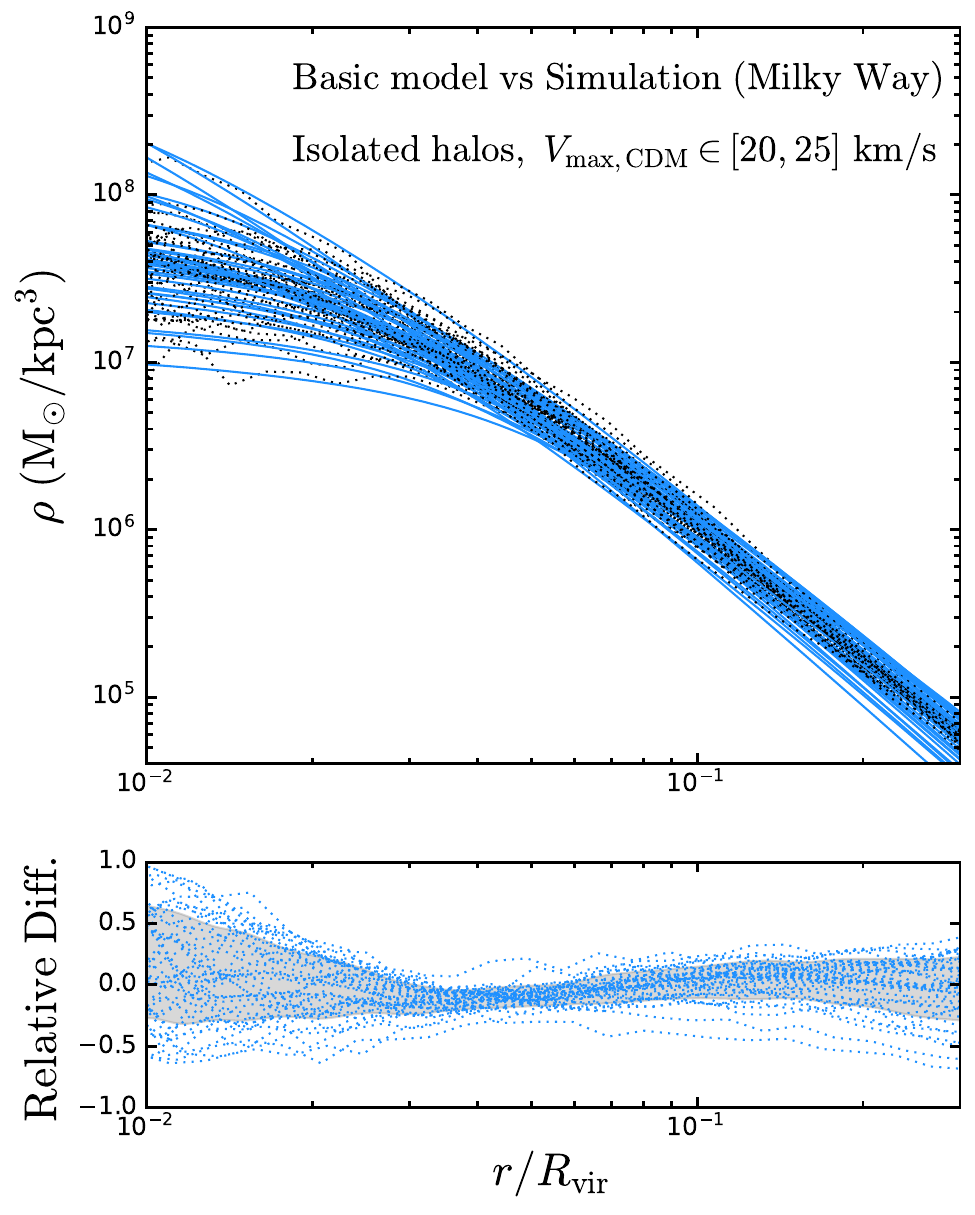}
  \includegraphics[width=5.3cm]{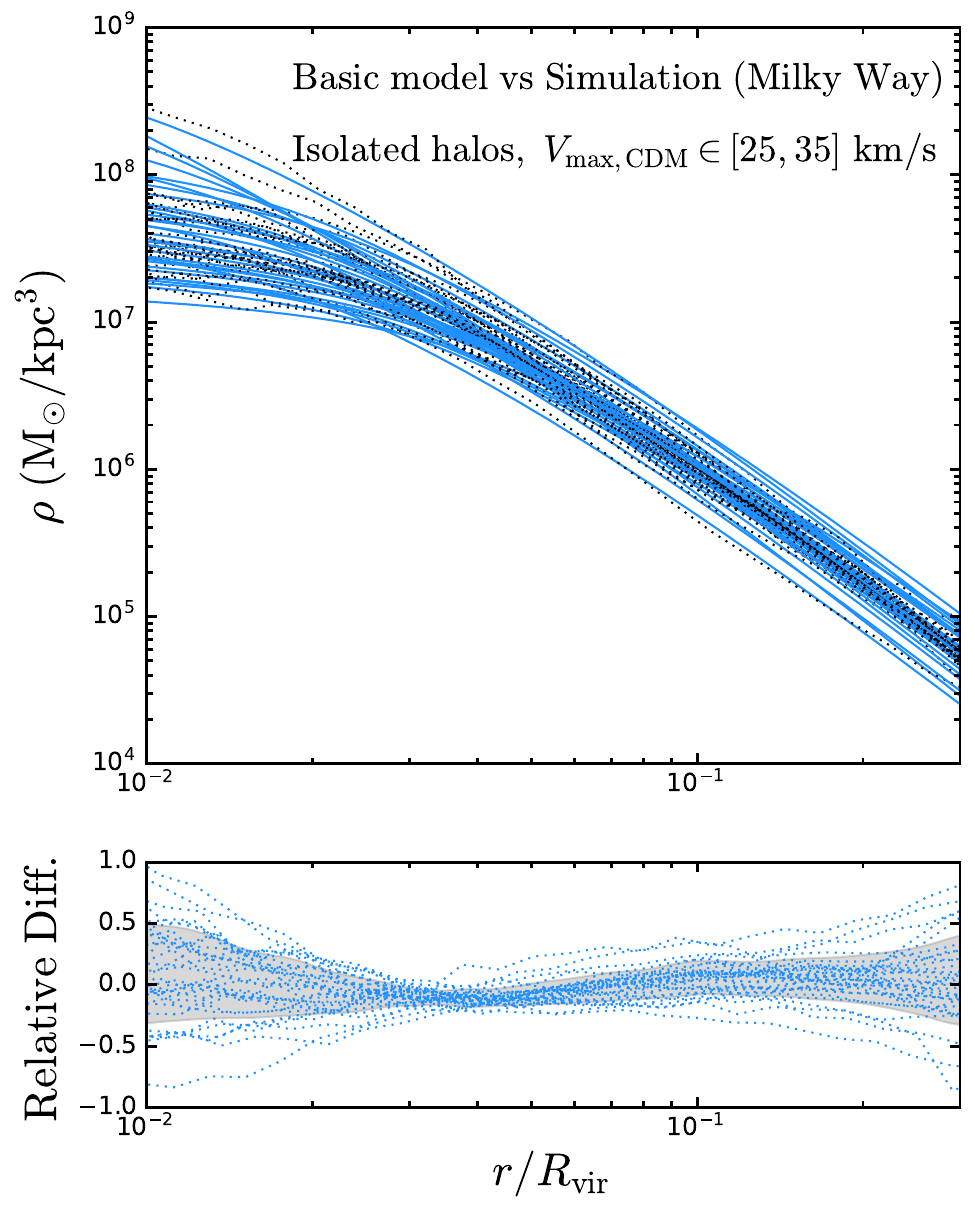}
  \includegraphics[width=5.3cm]{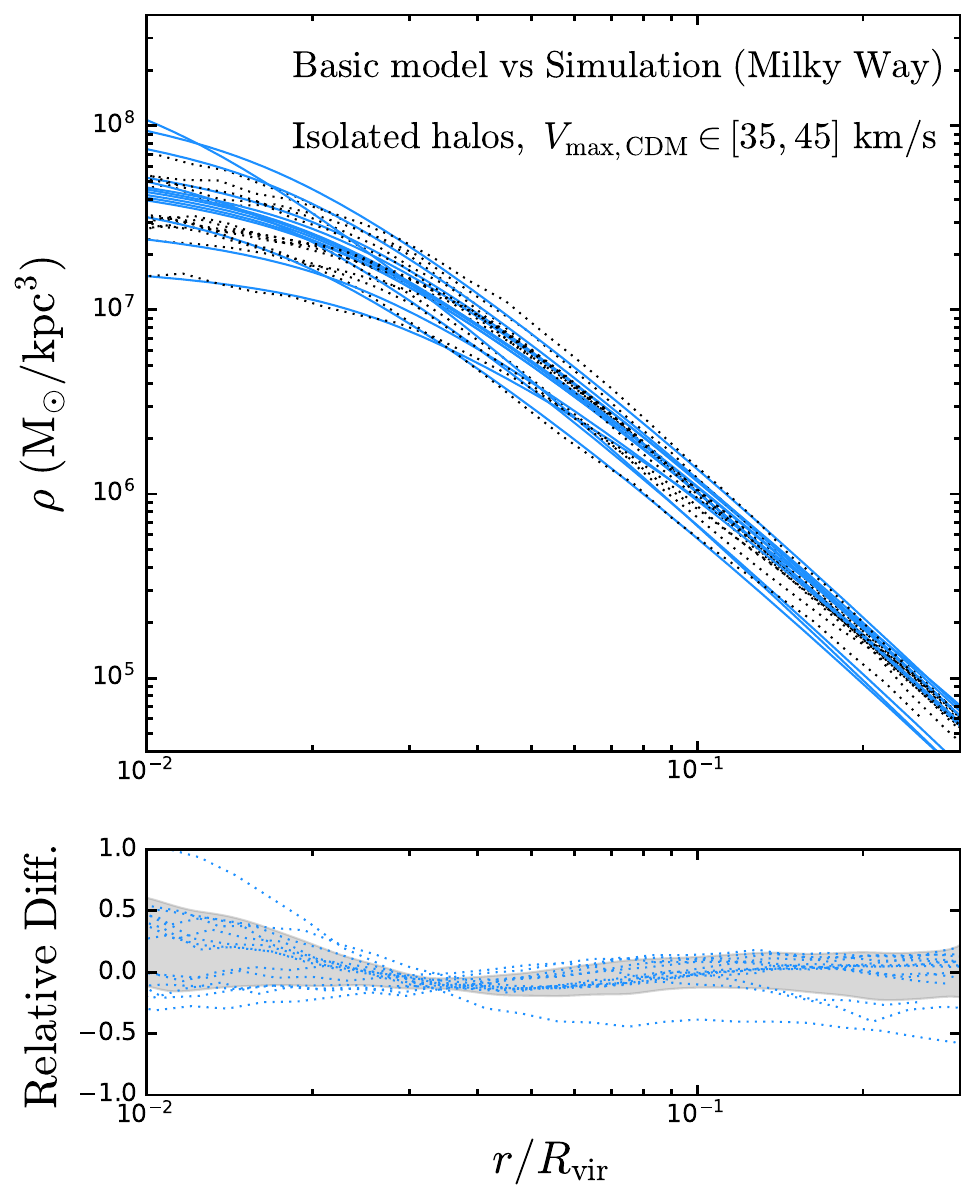}
  \caption{\label{fig:validisos1} 
The density profiles from the parametric model with the basic approach (solid) and the simulation (dotted) for isolated halos with $V_{\rm max,CDM} \in [20,25], ~ [25,35],$ and $[35,45]~\rm km/s$, in the MilkyWaySIDM simulation of Ref.~\cite{Yang:2022mxl}. 
In both cases, the evolution time starts at the estimated halo formation time and is normalized by the core collapse time estimated using $z=0$ halo properties.
The relative difference between each pair of simulated (Sim) and model-predicted (Mod) curves is measured as $\rm 2(Mod-Sim)/(Mod+Sim)$, with the $\pm 1\sigma$ band of the results shaded in gray.
}
\end{figure*}

In this section, we evaluate the performance of both basic and integral approaches of the parametric model, using the matched pairs of halos in the Milky Way zoom-in simulations of Ref.~\cite{Yang:2022mxl}. 
Building upon the initial validation performed in Ref.~\cite{Yang:2023jwn}, we extend the previous work by systematically matching and testing the majority of simulated halos.
We mainly focus on the evolution of $V_{\rm max,SIDM}$ and the density profile of halos at $z=0$. 
As in the CDM cases discussed in the previous section, we quantify the relative differences between each pair of matched simulated (Sim) and model-predicted (Mod) halos using the formula $\rm 2(Mod-Sim)/(Mod+Sim)$ and shade the $\pm 1\sigma$ bands of these curves in gray. 

Fig.~\ref{fig:mahisos} shows the evolution of $V_{\rm max,SIDM}$ for the isolated halos in $20~{\rm km/s}<V_{\rm max,CDM}<25~\rm km/s$ (left), $25~{\rm km/s}<V_{\rm max,CDM}<35~\rm km/s$ (middle), and $35~{\rm km/s}<V_{\rm max,CDM}<45~\rm km/s$ (right).
The $V_{\rm max,SIDM}$ evolution is shown as a function of normalized time $t/t_{c,0}$, where $t$ is the time since the halo formation and $t_{c,0}$ in Eq.~\ref{eq:tc0} is evaluated using the halo parameters at $z=0$.
We see that the relative differences remain small when $t/t_{c,0}\lesssim 0.2$, at the level of a few percent. 
After that, the differences increase. 
However, even for the cases with $t/t_{c,0}\sim 1$, where the halo enters the phase of deep collapse, the deviation is $\sim 10\%$.

Fig.~\ref{fig:mahsubs} shows the evolution of $V_{\rm max,SIDM}$ for the subhalos with $V_{\rm max,CDM}<25~\rm km/s$ (left), $25 {~\rm km/s}\leq V_{\rm max,CDM}<35~\rm km/s$ (middle), and $V_{\rm max,CDM}\geq 35~\rm km/s$ (right).
The performance of the model for subhalos is similar to that of isolated halos, but with more subhalos evolving to $t/t_{c,0}\sim 1$ in the $V_{\rm max,CDM} <25~\rm km/s$ case, where the relative difference could reach to $10\textup{--}20\%$. We observe tidal stripping to accelerate the gravothermal evolution in these cases.

\begin{figure*}[htbp]
  \centering
  \includegraphics[width=5.3cm]{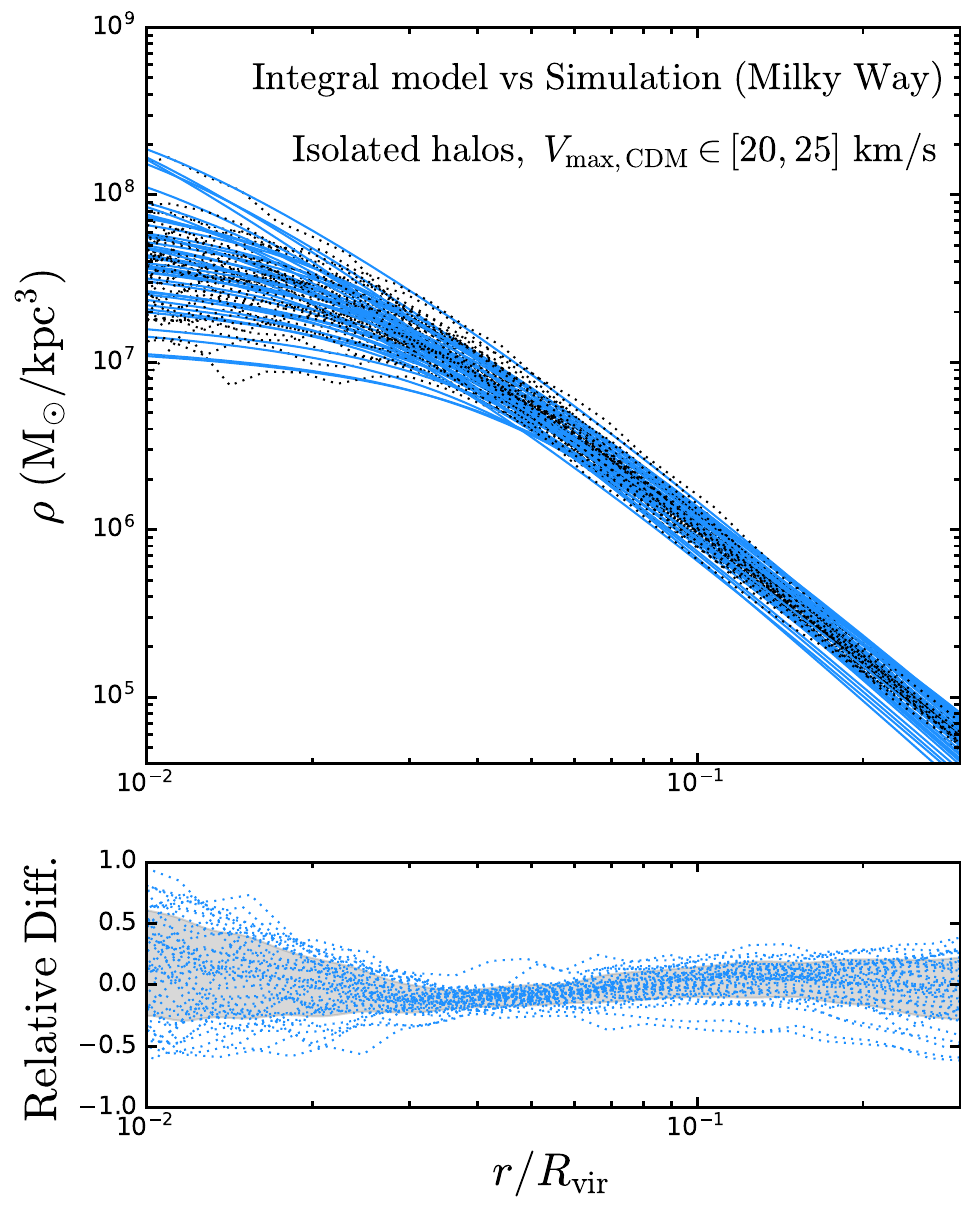}
  \includegraphics[width=5.3cm]{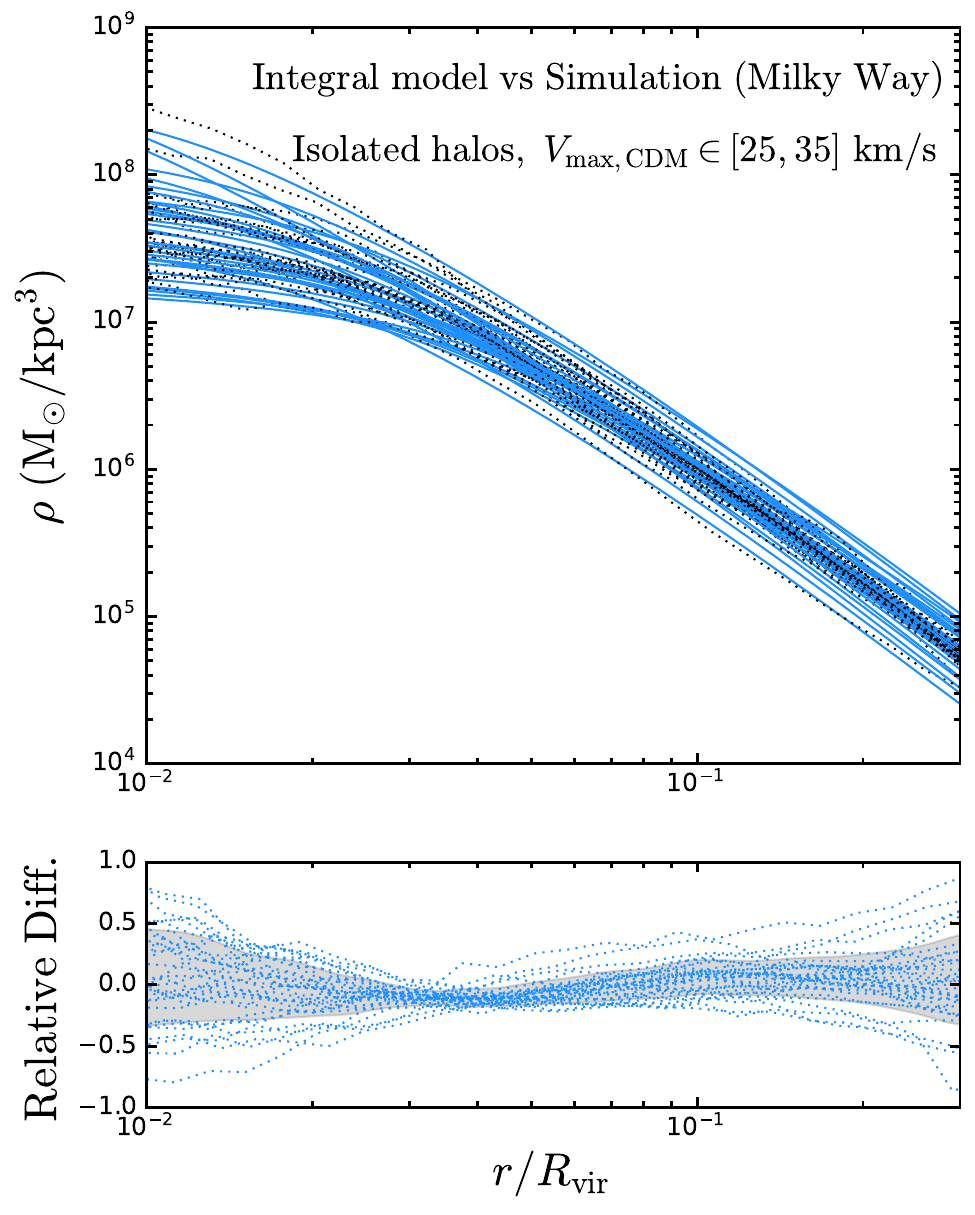}
  \includegraphics[width=5.3cm]{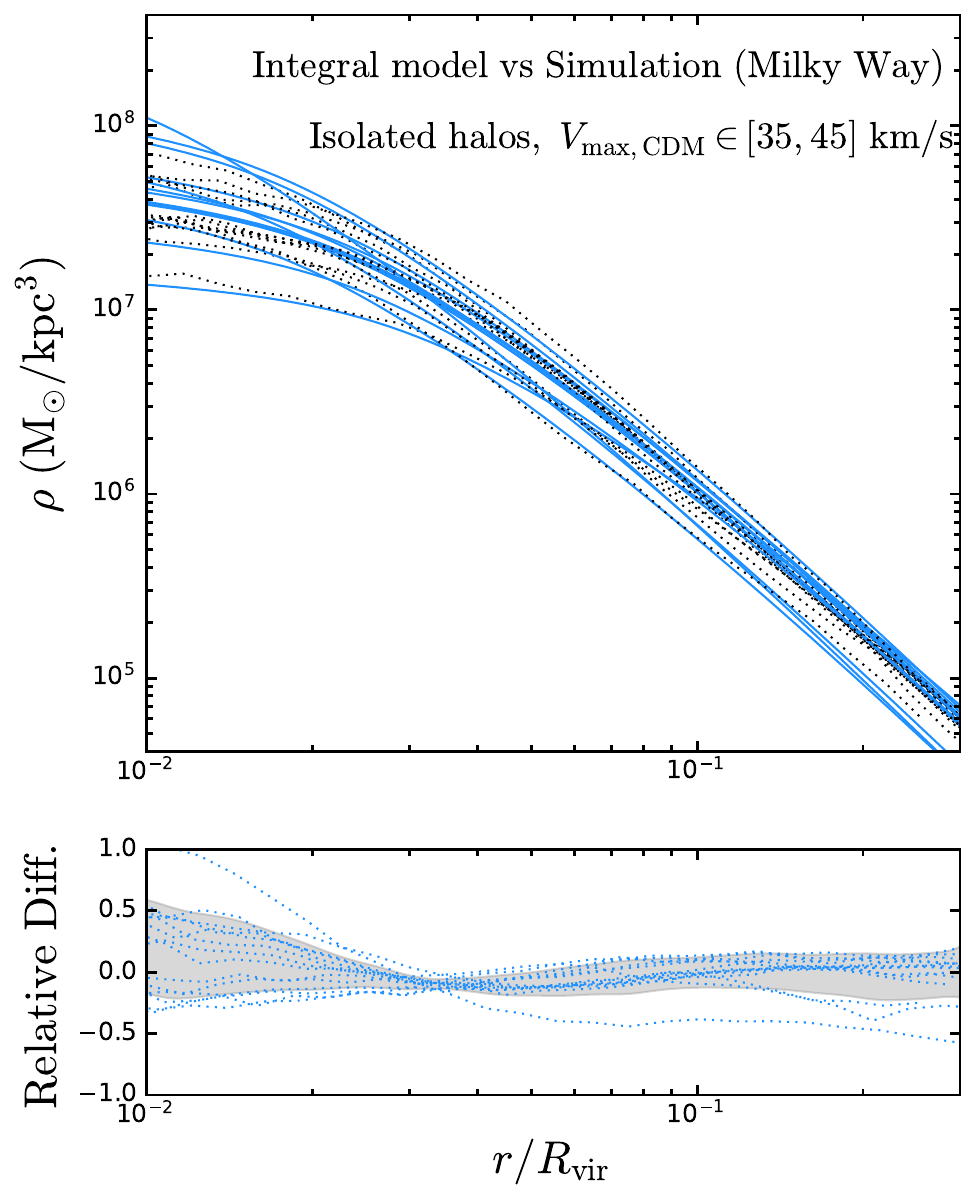}
  \caption{\label{fig:validisos2} The density profiles from the parametric model with the integral approach (solid) and the simulation (dotted) for isolated halos with $V_{\rm max,CDM} \in [20,25], ~ [25,35],$ and $[35,45]~\rm km/s$, in the MilkyWaySIDM simulation of Ref.~\cite{Yang:2022mxl}. 
The relative difference between each pair of simulated (Sim) and model-predicted (Mod) curves is measured as $\rm 2(Mod-Sim)/(Mod+Sim)$, with the $\pm 1\sigma$ band of the results shaded in gray.
}
\end{figure*}

\begin{figure*}[htbp]
  \centering
  \includegraphics[width=5.3cm]{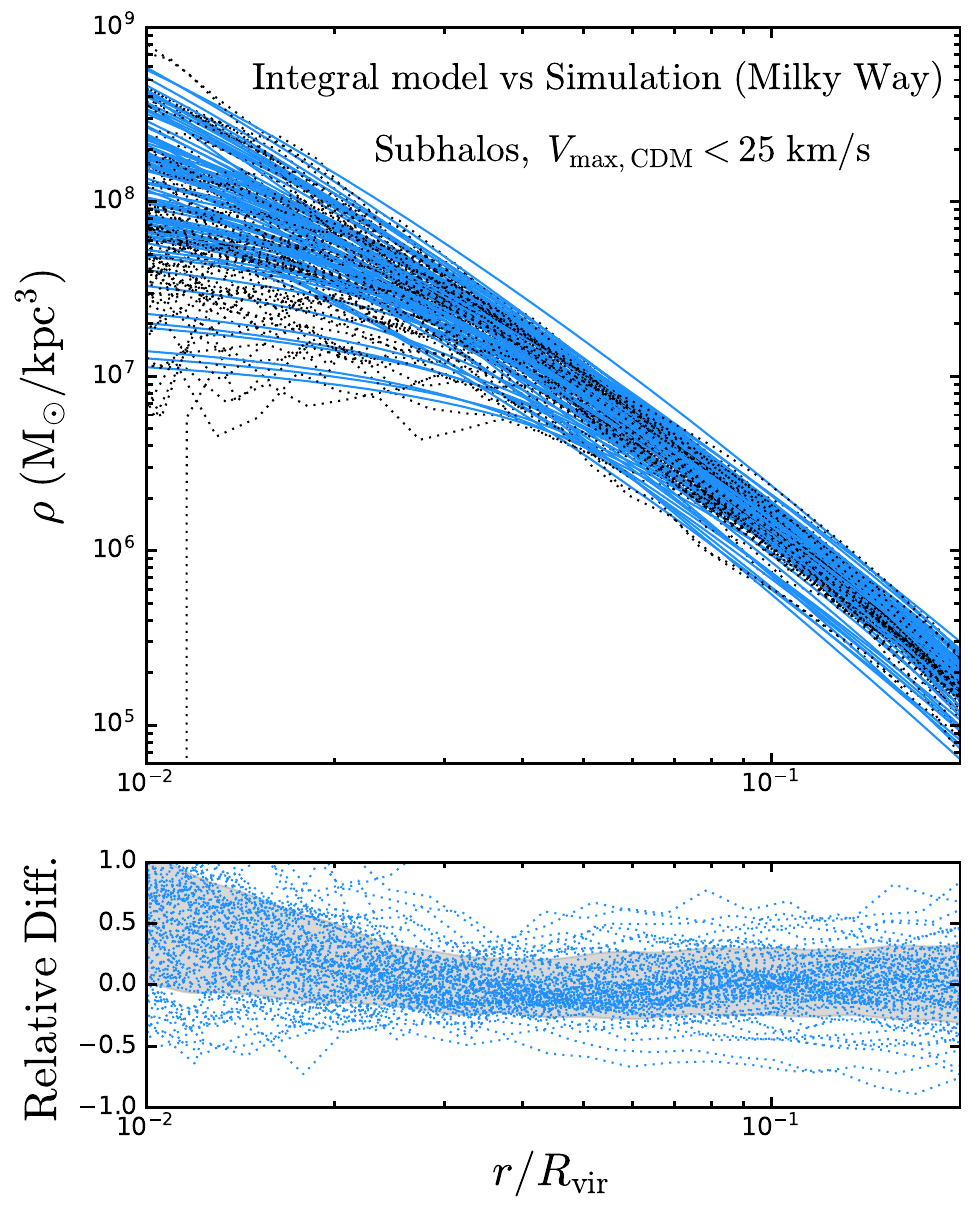}
  \includegraphics[width=5.3cm]{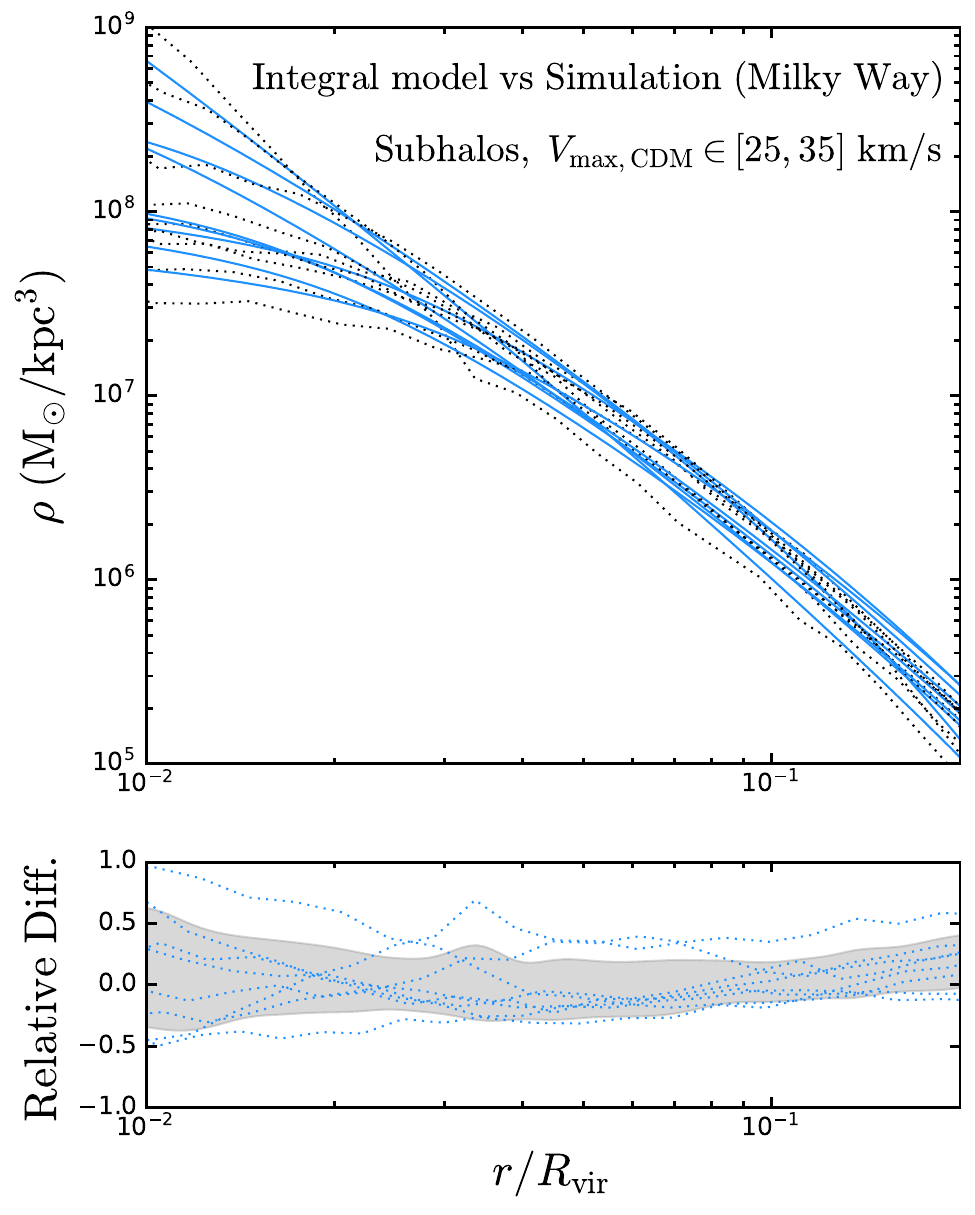}
  \includegraphics[width=5.3cm]{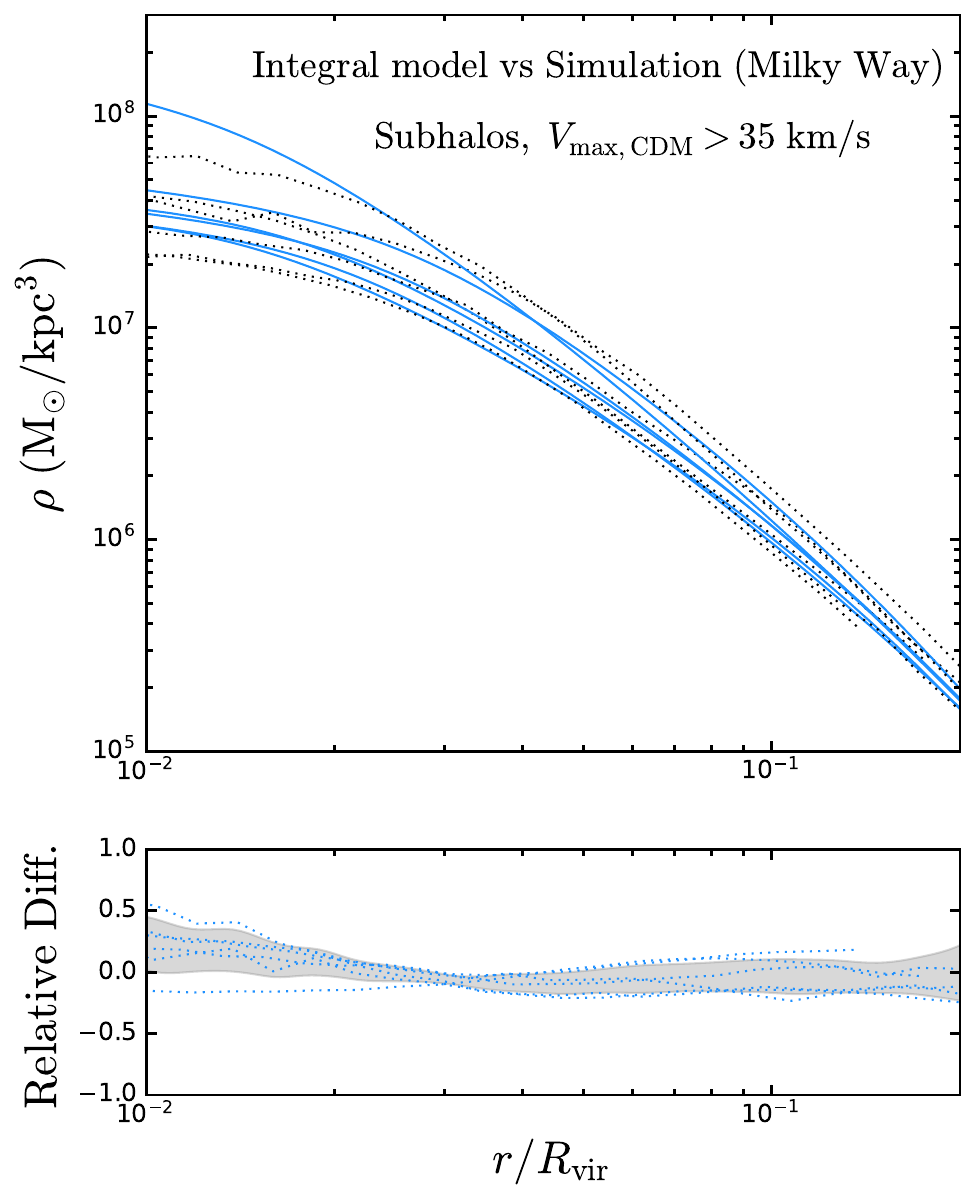}
  \caption{\label{fig:validsubs} The density profiles from the parametric model with the integral approach (solid) and the simulation (dotted) for subhalos with $V_{\rm max,CDM} <25~\rm km/s$, $V_{\rm max,CDM}\in~ [25,35]~\rm km/s$, and $V_{\rm max,CDM}>35 ~\rm km/s$, in the MilkyWaySIDM simulation of Ref.~\cite{Yang:2022mxl}.
The relative difference between each pair of simulated (Sim) and model-predicted (Mod) curves is measured as $\rm 2(Mod-Sim)/(Mod+Sim)$, with the $\pm 1\sigma$ band of the results shaded in gray.
}
\end{figure*}

In Figs.~\ref{fig:validisos1} and \ref{fig:validisos2}, we show the density profiles from the model prediction (solid) and the simulation (dotted), based on the basic and integral approaches, respectively.
The halos are divided into three $V_{\rm max,CDM}$ bins as in Fig.~\ref{fig:mahisos}. 
We find that the relative differences in both approaches are comparable, averaging around 10\% near the scale radii but increasing to about 50\% in the inner and outer regions. 
Both approaches do not exhibit any noticeable systematic shifts across all radii.

Fig.~\ref{fig:validsubs} shows the comparison for the density profiles of the subhalos. 
The agreement level with the model decreases slightly compared to the isolated halos, with a notable $V_{\rm max}$ dependency. For $V_{\rm max,CDM}<25~\rm km/s$, the relative difference within about 25\% for $r/R_{\rm vir} > 0.05$ but overall increase towards smaller radii. 
A similar upward systematic shift appears in the CDM case, as shown in Fig.~\ref{fig:mwcdm} (middle).
This suggests that the analytical density profile in Eq.~\ref{eq:cnfw} may overestimate the inner density of the CDM subhalos in the low $V_{\rm max}$ bin, which could also result in an overestimation of their SIDM counterparts.
For larger $V_{\rm max}$, the agreement becomes better, especially in the $V_{\rm max,CDM}>35~\rm km/s$ case, where performance is comparable to that of isolated halos. 
This trend may be related to the fact that the most massive subhalos often accreted into the host recently (see the left panel of Fig.~\ref{fig:evp100} in Appendix~\ref{sec:ram}), such that the inner halo profiles have not been significantly affected by tidal stripping.

We have provided a comprehensive analysis of the evolution histories and density profiles of both isolated halos and subhalos, from the Milk Way CDM and SIDM simulations in Ref.~\cite{Yang:2022mxl}. From Figs.~\ref{fig:mahisos} to \ref{fig:validsubs}, we have demonstrated that the parametric model can effectively capture the key features of the SIDM halos.
For the density profiles of the isolated halos, the agreement between the model prediction and the simulation is well within $50\%$ for all radii, and the small deviation is not systematic.  
For the subhalos, while the overall trends are similar, the model accuracy decreases somewhat, especially in the lowest \(V_{\rm max,CDM}\) bin, suggesting areas for further improvement. 
For example, the systematic upward shift in the lowest $V_{\rm max,CDM}$ bin could be fixed by adjusting the analytical density profile. With this analysis, we have quantified the relative difference between the model and the simulation, which should be taken into account in practical applications.

\section{Matched halos with an extreme cross section}
\label{sec:group}

\begin{figure*}[htbp]
  \centering
  \includegraphics[width=5.3cm]{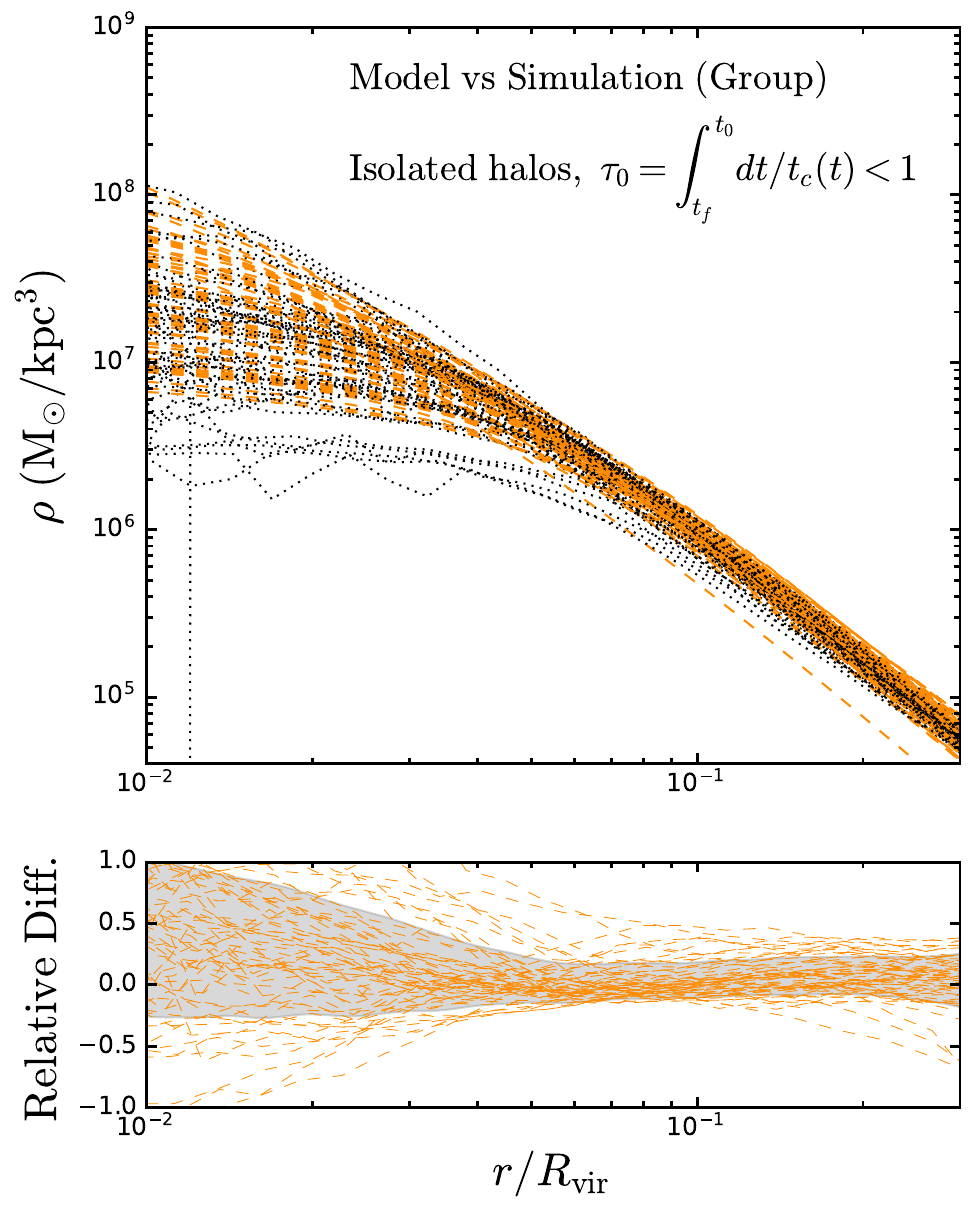}
  \includegraphics[width=5.3cm]{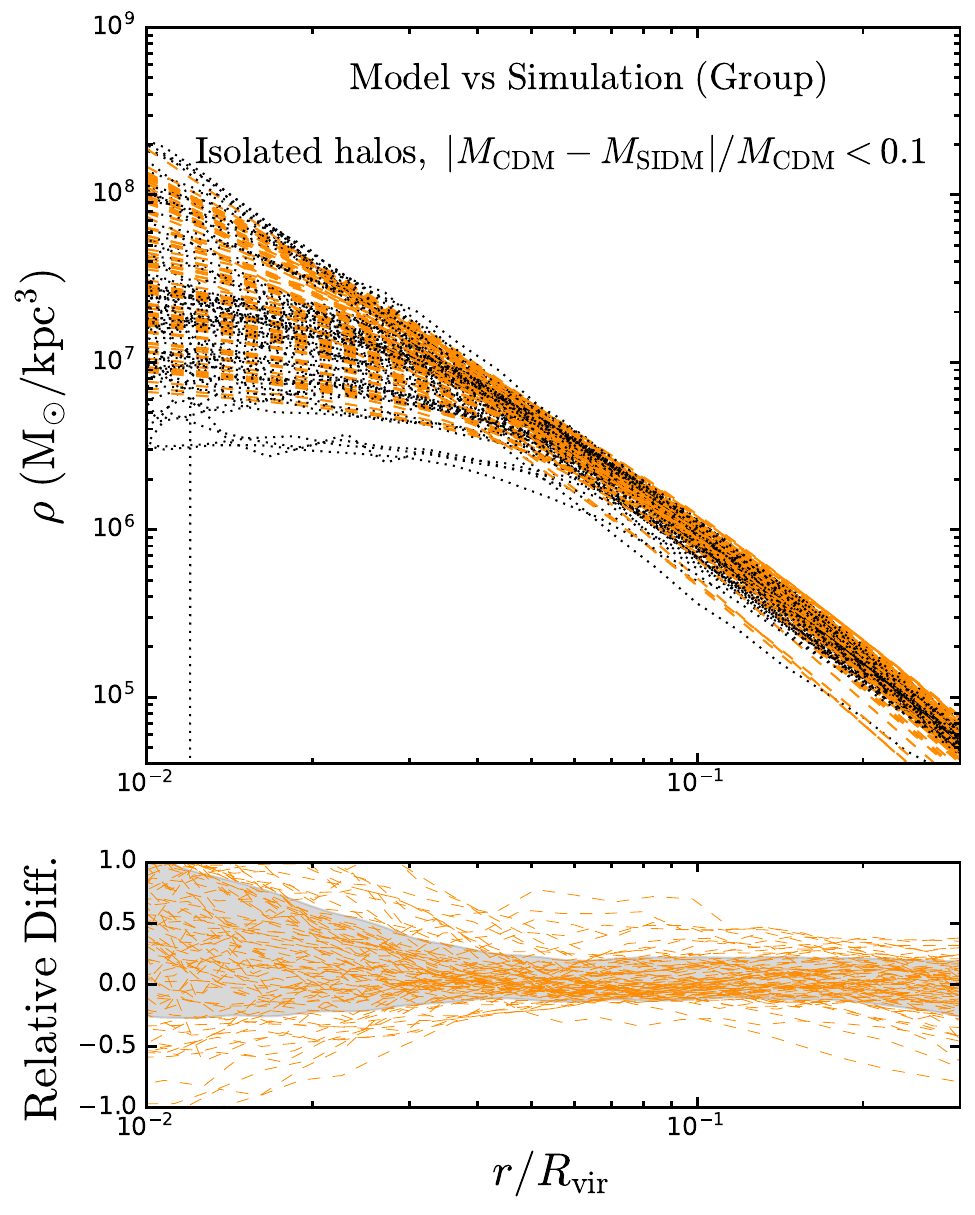}
  \includegraphics[width=5.3cm]{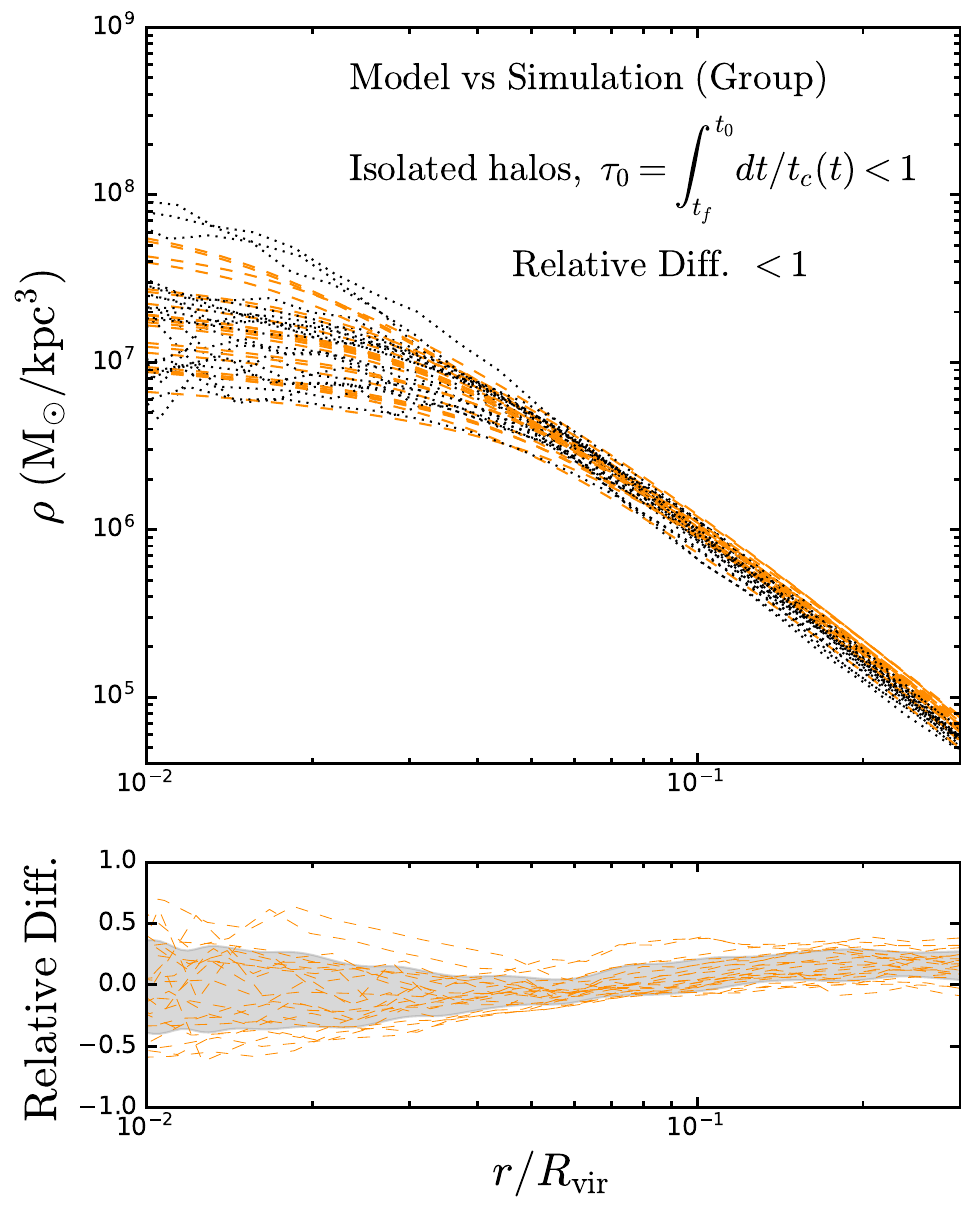}
  \caption{\label{fig:group1} The density profiles from the parametric model with the integral approach (dashed) and the simulation (dotted) for isolated halos in the GroupSIDM simulation in Ref.~\cite{Nadler:2023nrd}. 
From left to right, the results are presented considering three conditions: 
cases with $\tau_0 = \int_{t_h}^{t_0} dt /t_c(t) <1$ (left), cases with the relative mass difference $|M_{\rm CDM} - M_{\rm SIDM}|/M_{\rm CDM}$ within 10\% (middle), and cases with $\tau_0 = \int_{t_h}^{t_0} dt /t_c(t) <1$ with the additional requirement that the relative differences be smaller than one (right).
The relative difference between each pair of simulated (Sim) and model-predicted (Mod) curves is measured as $\rm 2(Mod-Sim)/(Mod+Sim)$, with the $\pm 1\sigma$ band of the results shaded in gray.
}
\end{figure*}

\begin{figure*}[htbp]
  \centering
  \includegraphics[width=5.3cm]{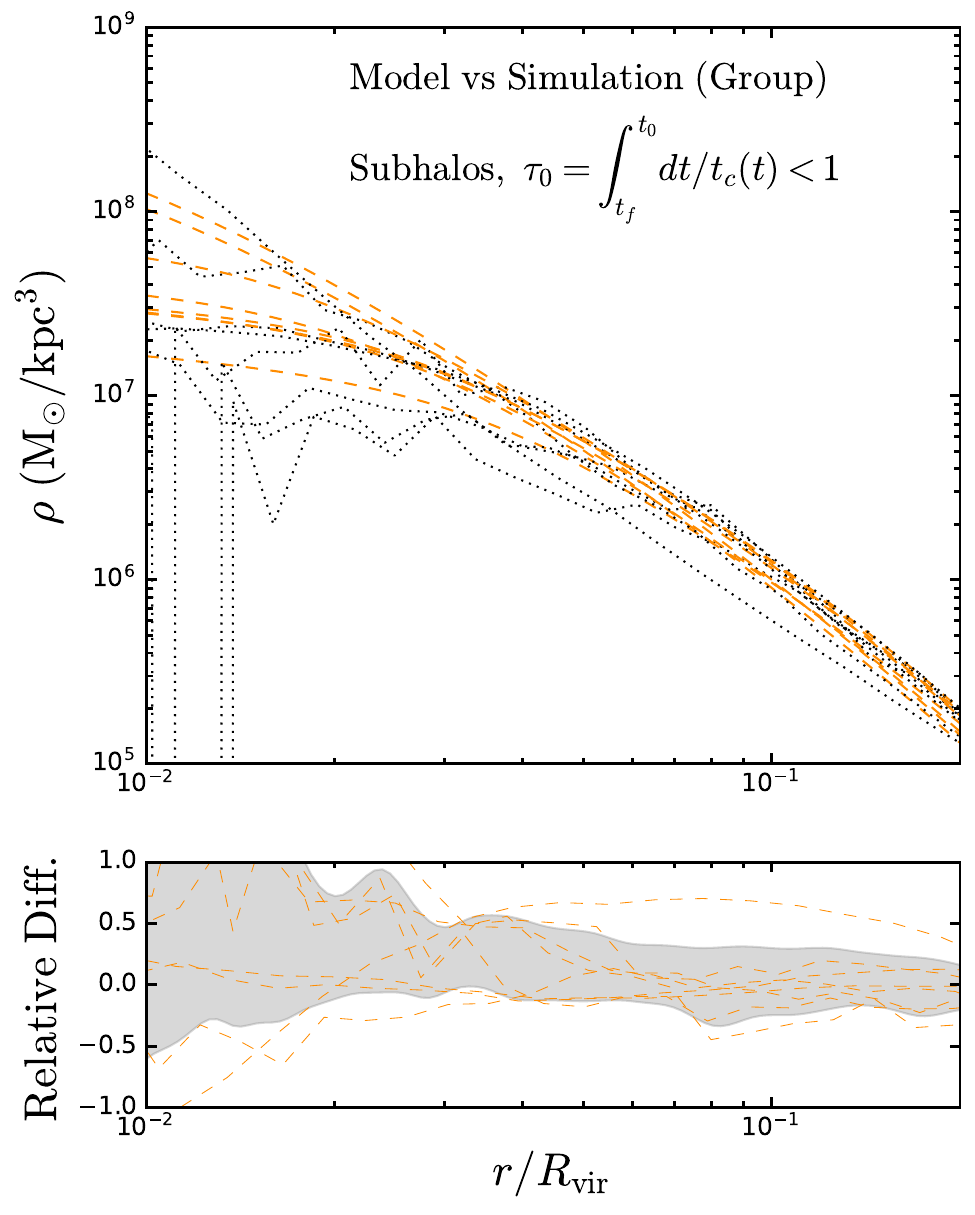}
  \includegraphics[width=5.3cm]{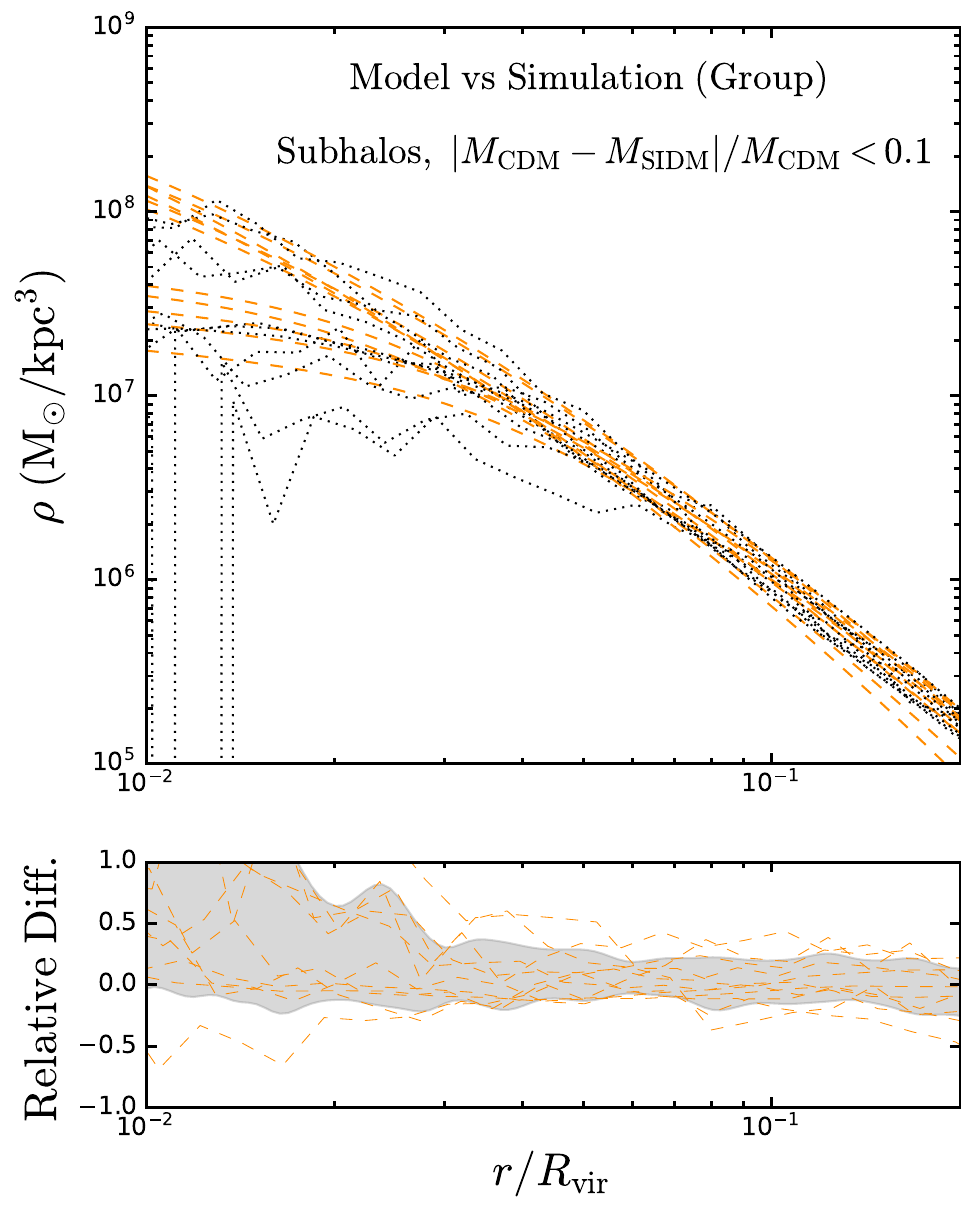}
  \includegraphics[width=5.3cm]{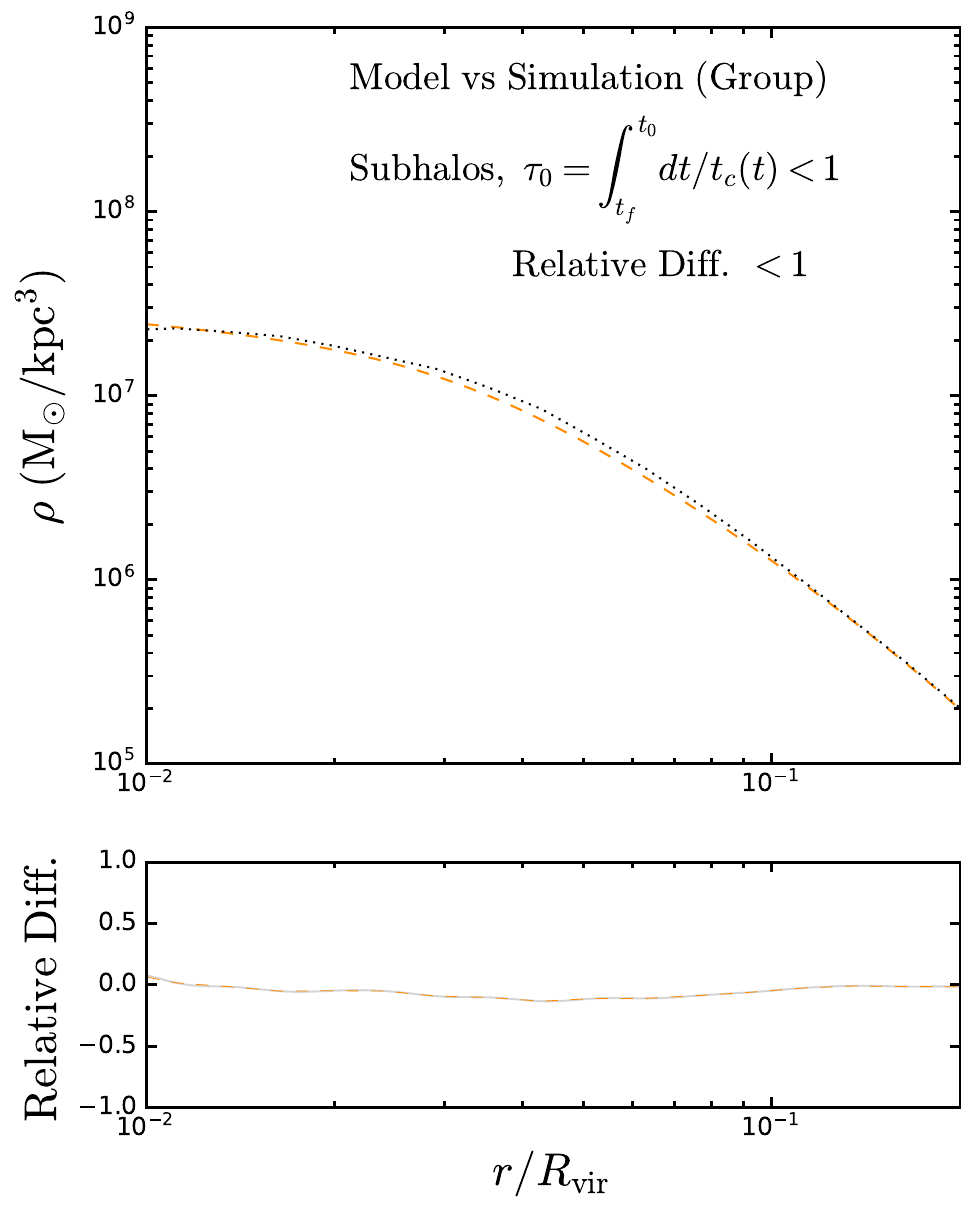}
  \caption{\label{fig:group2} The density profiles from the parametric model with the integral approach (dashed) and the simulation (dotted) for subhalos (bottom) in the GroupSIDM simulation in Ref.~\cite{Nadler:2023nrd}.
From left to right, the results are presented considering three conditions:
cases with $\tau_0 = \int_{t_h}^{t_0} dt /t_c(t) <1$ (left), cases with the relative mass difference $|M_{\rm CDM} - M_{\rm SIDM}|/M_{\rm CDM}$ within 10\% (middle), and cases with $\tau_0 = \int_{t_h}^{t_0} dt /t_c(t) <1$ with the additional requirement that the relative differences be smaller than one (right).
The relative difference between each pair of simulated (Sim) and model-predicted (Mod) curves is measured as $\rm 2(Mod-Sim)/(Mod+Sim)$, with the $\pm 1\sigma$ band of the results shaded in gray.
}
\end{figure*}

\begin{figure*}[htbp]
  \centering
  \includegraphics[width=16.cm]{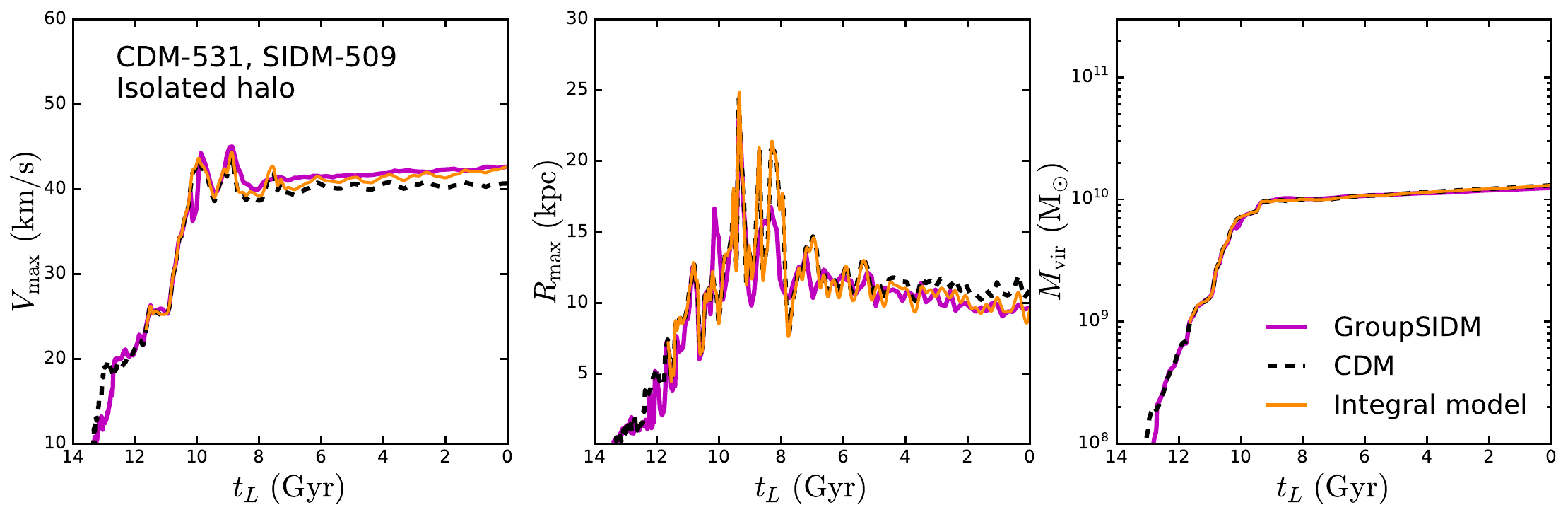} \\
  \includegraphics[width=16.cm]{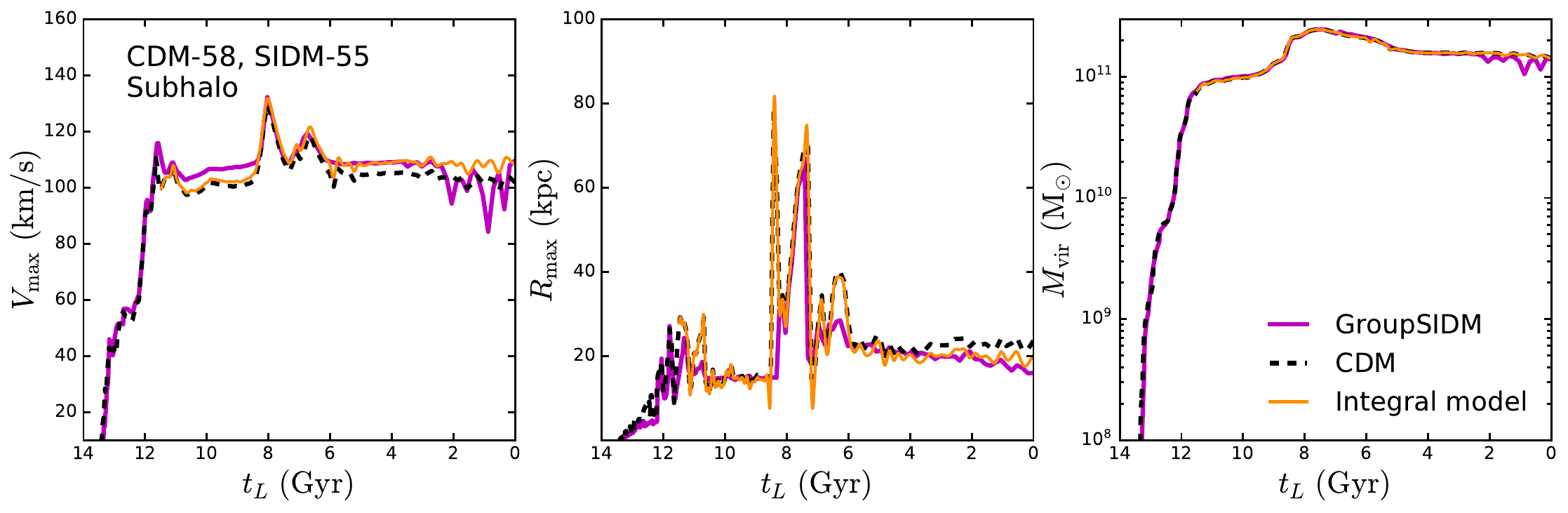}
  \caption{\label{fig:good1}
Evolution of $V_{\rm max}$ (left), $R_{\rm max}$ (middle), and $M_{\rm vir}$ (right) for halos in the GroupSIDM simulation (magenta) and as predicted by the integral model (orange), compared to the curves for CDM (dashed black). The top panels feature a core-collapsing halo from the right panel of Fig.~\ref{fig:group1}, whose relative difference is smaller than one. The bottom panels show the core-collapsing subhalo from the right panel of Fig.~\ref{fig:group2}, also with a relative difference smaller than one.
}
\end{figure*}

\begin{figure*}[htbp]
  \centering
  \includegraphics[width=16.cm]{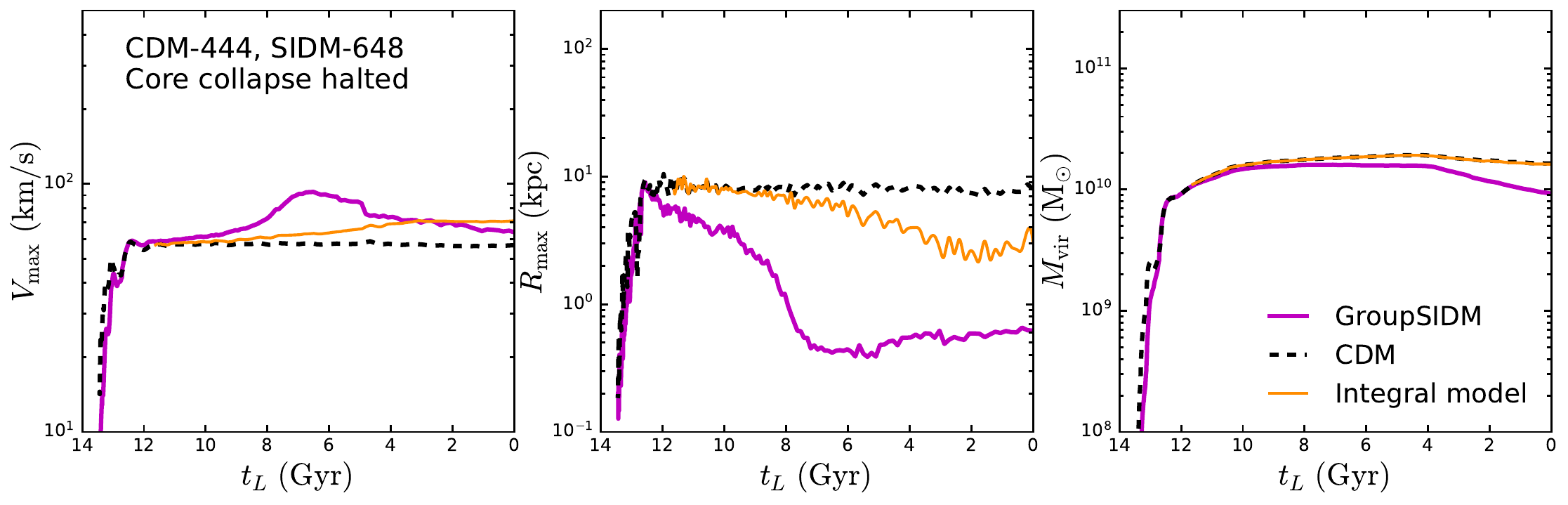} \\
  \includegraphics[width=16.cm]{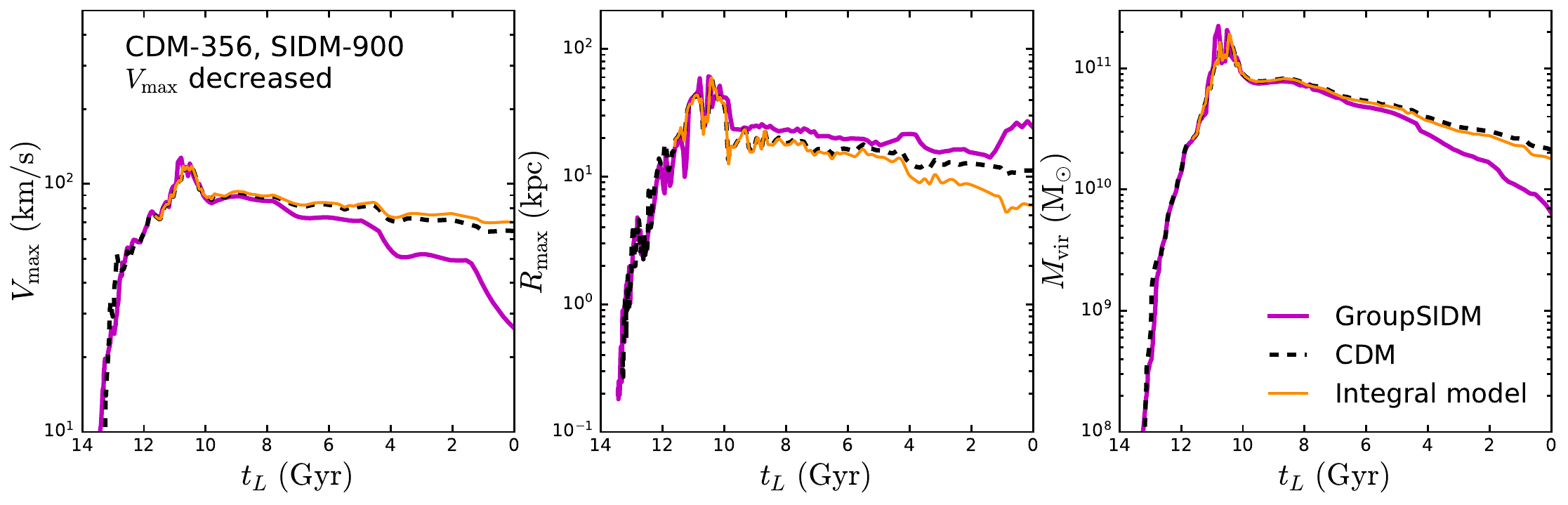} 
  \caption{\label{fig:sepcial} 
Evolution of $V_{\rm max}$ (left), $R_{\rm max}$ (middle), and $M_{\rm vir}$ (right) for halos in simulated (magenta) and model-predicted (orange; integral approach) SIDM, alongside CDM evolution curves (dashed black).
The case corresponding to the top panels has a halted core collapse, the case corresponding to the bottom panels demonstrates an enhanced decrease in $V_{\rm max}$. 
}
\end{figure*}

The GroupSIDM simulation presented in Ref.~\cite{Nadler:2023nrd} explores an extreme SIDM scenario at group scales, $10^{13}~\rm M_{\odot}$. The cross section has the same normalization $\sigma_0/m = 147.1~\rm cm^2/g$ as in MilkyWaySIDM, but with a larger transition velocity of $w = 120~\rm km/s$. 
This larger $w$ shifts the suppression of the cross section due to the velocity dependence to more massive halos, and significantly increases the effective cross section for halos with typical velocity scales greater than $w = 24.33~\rm km/s$ in the MilkyWaySIDM case.
In MilkyWaySIDM, there are $9$ subhalos with $\tau_0 > 1$ in the matched samples. 
No isolated halos have $\tau_0 > 1$. In GroupSIDM, there are many such cases: $45$ out of $106$ for isolated halos; $38$ out of $46$ for subhalos.
Since the parametric model is calibrated with the SIDM simulation within $\tau_0 < 1$, and we will select halos from the GroupsSIDM simulation with $\tau_0 < 1$.

We also incorporate the ram-pressure evaporation (RPe) effect following the procedures in Refs.~\cite{Kummer:2017bhr,Shirasaki:2022ttb}. 
Generally, the RPe is insignificant for halos in both our SIDM simulations because the large relative velocities between subhalo and host halo particles suppress the cross section to the $1~\rm cm^2/g$ level. However, certain cases with long evolutionary trajectories in the host halo can have a non-negligible effect, which we will illustrate with examples in Appendix~\ref{sec:ram}.

In Fig.~\ref{fig:group1}, we compare the integral model's predictions (dashed) with the simulated (dotted) density profiles of isolated halos from the GroupSIDM simulation.
We consider three selection criteria: halos with $\tau_0 = \int_{t_h}^{t_0} dt / t_c(t) < 1$ (left), $|M_{\rm CDM} - M_{\rm SIDM}|/M_{\rm CDM}<10\%$ (middle), and halos with $\rm 2(Mod-Sim)/(Mod+Sim)<1$ and $\tau_0 < 1$ (right). 
Overall, the model predictions align less well with the simulation results compared to the Milky Way simulation.
As shown in the left panel, in the innermost regions ($r/R_{\rm vir}=0.01$), the relative differences can be as large as 100\%.
Aside from the $\tau_0<1$ requirement, we also test the $|M_{\rm CDM} - M_{\rm SIDM}|/M_{\rm CDM}<0.1$ criterion, given that the parametric model conserves the halo mass. Interestingly, the performance after applying this mass difference requirement is similar to that obtained from requiring $\tau_0<1$.
In the right panel, we enforce $\rm 2(Mod-Sim)/(Mod+Sim) < 1$ and find a subgroup of candidates with significantly better agreement, displaying relative differences well within a $\pm50$\% band, see Fig.~\ref{fig:group1} (right) for an illustration. These cases have more continuous and simpler accretion histories than others.

Fig.~\ref{fig:group2} shows similar results for subhalos within the group main halo, following the same criteria as in Fig.~\ref{fig:group1}. 
For the $\tau_0 < 1$ and $|M_{\rm CDM} - M_{\rm SIDM}|/M_{\rm CDM}<0.1$, the model performance is comparable to the corresponding isolated halo cases.
Additionally, when the condition $\rm 2(Mod-Sim)/(Mod+Sim)<1$ and $\tau_0 < 1$ are combined, only one subhalo meets these criteria, exhibiting remarkable agreement between simulations and model predictions.

We pick two halos satisfying the combined criteria $\rm 2(Mod-Sim)/(Mod+Sim)<1$ and $\tau_0 < 1$, shown in the right panels of Figs.~\ref{fig:group1} and \ref{fig:group2}. 
Fig.~\ref{fig:good1} (top) shows the evolution of $V_{\rm max}$ (left), $R_{\rm max}$ (middle), and $M_{\rm vir}$ (right) of isolated core-collapsing halos from the integral approach (orange) and the Group CDM (black) and SIDM (magenta) simulations. 
We see that the model prediction agrees with the GroupSIDM simulation at almost all times in the evolution history of this halo. 
Fig.~\ref{fig:good1} (bottom) shows a similar case, but for a subhalo. 
These examples highlight the successful application of the integral model in these cases. Aside from the $z=0$ density profiles in the right panels of Fig.~\ref{fig:group1} and Fig.~\ref{fig:group2}, their evolution histories also match the SIDM simulation results quite well.

We further present the evolutionary histories of two example cases in Fig.~\ref{fig:sepcial} to illustrate the challenges. 
We depict the evolution of $V_{\rm max}$ (left), $R_{\rm max}$ (middle), and $M_{\rm vir}$ (right) for both simulated (magenta) and integral model-predicted (orange) SIDM, together with the evolution of its CDM counterpart (dashed black). 
The top panels illustrate a boosted evolution in the SIDM simulation at early times ($t_L > 8~$Gyr). 
This leads to the highest density being reached around $t_L = 7~$Gyr. 
After this point, core collapse ceases, and $V_{\rm max}$ begins to decrease. 
This is likely due to the numerical issues associated with N-body simulations in the deeply collapsed regime and the energy conservation condition is violated~\cite{Zhong:2023yzk,Mace:2024uze,Palubski:2024ibb,Fischer:2024eaz}.
Interestingly, the halo masses in CDM and SIDM start to diverge from each other early on at $t_L > 10~$Gyr, potentially causing the parametric model prediction to become inaccurate. 
The bottom panels correspond to a subhalo with multiple pericenter passages, indicated by steep decreases in the $V_{\rm max}$ evolutions. In SIDM, when the core size becomes comparable to the tidal radius, especially at pericenter passages, tidal stripping is enhanced, causing the subhalo mass to be smaller than its CDM counterpart. This decrease occurred in this halo early on at $t_L \approx 8~$Gyr. Subsequent pericenter passages further accelerate the mass loss, amplifying the difference between the SIDM simulation and the model prediction.

These examples illustrate that applying the parametric model to halos with large effective cross sections and complex late mergers requires caution. 
In the GroupSIDM simulation, the accretion histories of halos are generally more noisy than those in the MilkyWaySIDM simulation, as more massive halos tend to form later. Additionally, merger events may induce secondary effects beyond the scope of our current parametric model, altering the gravothermal state of an SIDM halo. 
Therefore, it is crucial to inspect the CDM and predicted evolutionary trajectories to rule out numerical inaccuracies and explore potential secondary effects through matched SIDM halos in the simulation. 
Additionally, the $\tau_0 > 1$ regime deserves dedicated study, and it would be interesting to extend the parametric model for $\tau_0>1$.
In future work, we plan to address these issues explicitly.

\section{Implications for observing strong lensing perturbers}
\label{sec:application}

We now demonstrate the utility of our model by applying it to predict the inner density profiles of subhalos relevant for strong gravitational lensing analyses. Several strong-lensing perturbers have been detected through the gravitational imaging technique~\cite{Vegetti:2009cz,Vegetti:2012mc}. Intriguingly, some of these systems are significantly denser than standard CDM predictions~\cite{Minor:2020hic,Zhang:2023wda}. Ref.~\cite{Nadler:2023nrd} demonstrated that the GroupSIDM model produces substructure in excellent agreement with the properties of the SDSSJ0946+1006 perturber; however, this study relied on the Group zoom-in simulation described above, which is numerically expensive and evaluated for a single SIDM cross section. 
The ability to rapidly generate such predictions for a range of SIDM models is therefore timely, as upcoming facilities will drastically increase the strong lens sample sizes \cite{LSSTDarkMatterGroup:2019mwo,Gezari:2022rml,ORiordan:2022qds}.

\begin{figure*}[htbp]
  \centering
  \includegraphics[width=5.3cm]{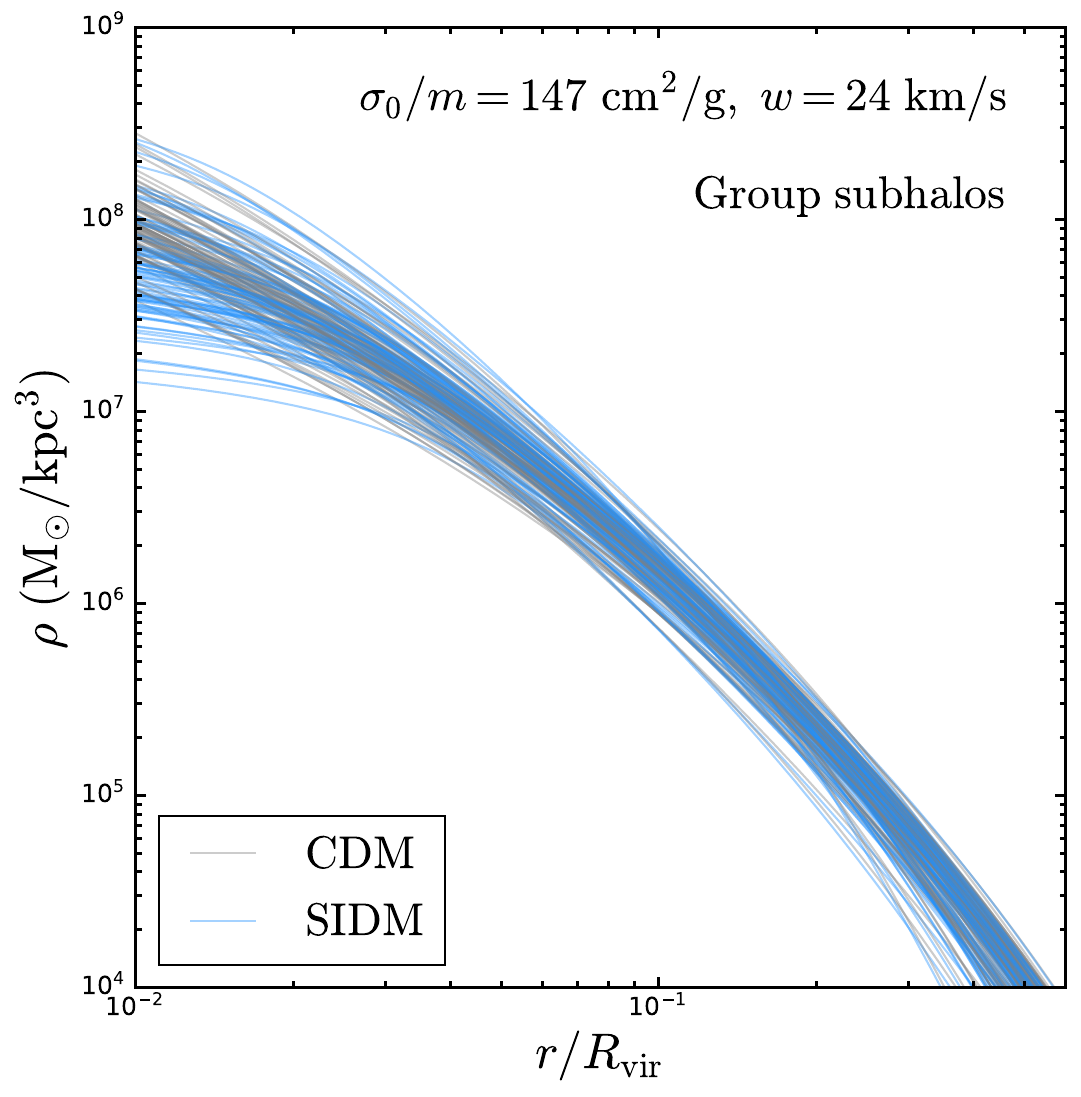}
  \includegraphics[width=5.3cm]{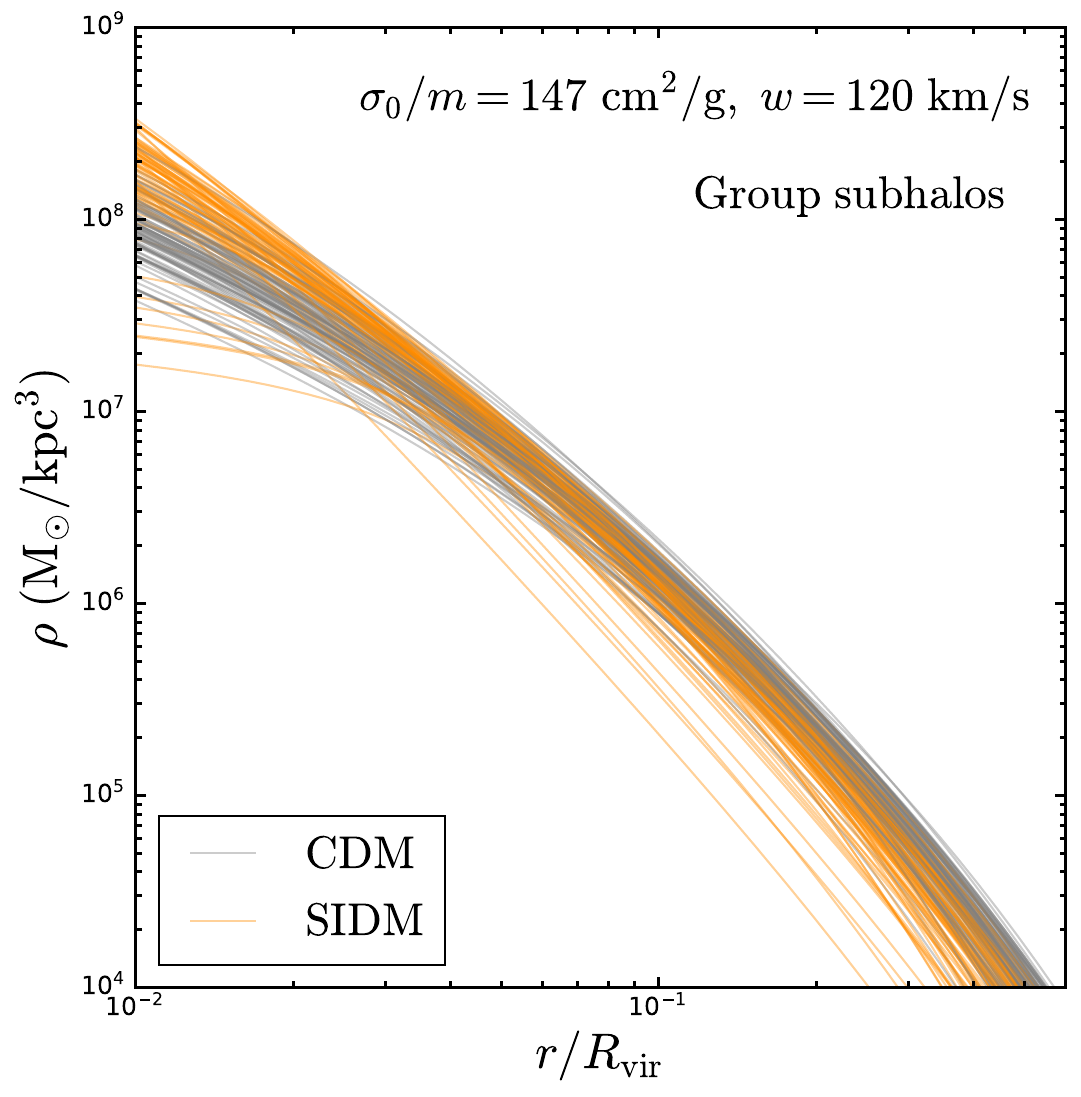} 
  \includegraphics[width=5.3cm]{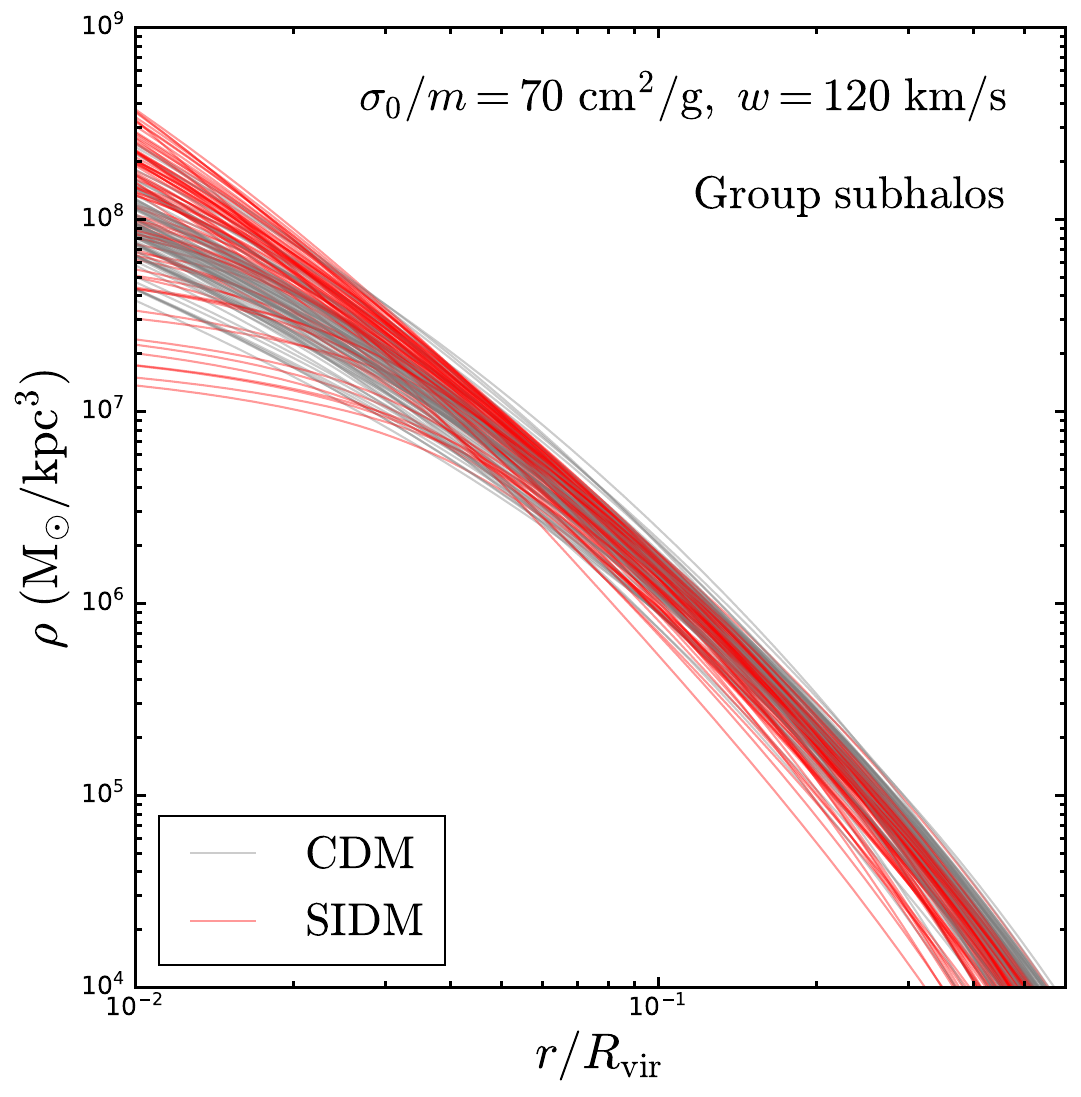} \\
  \includegraphics[width=5.3cm]{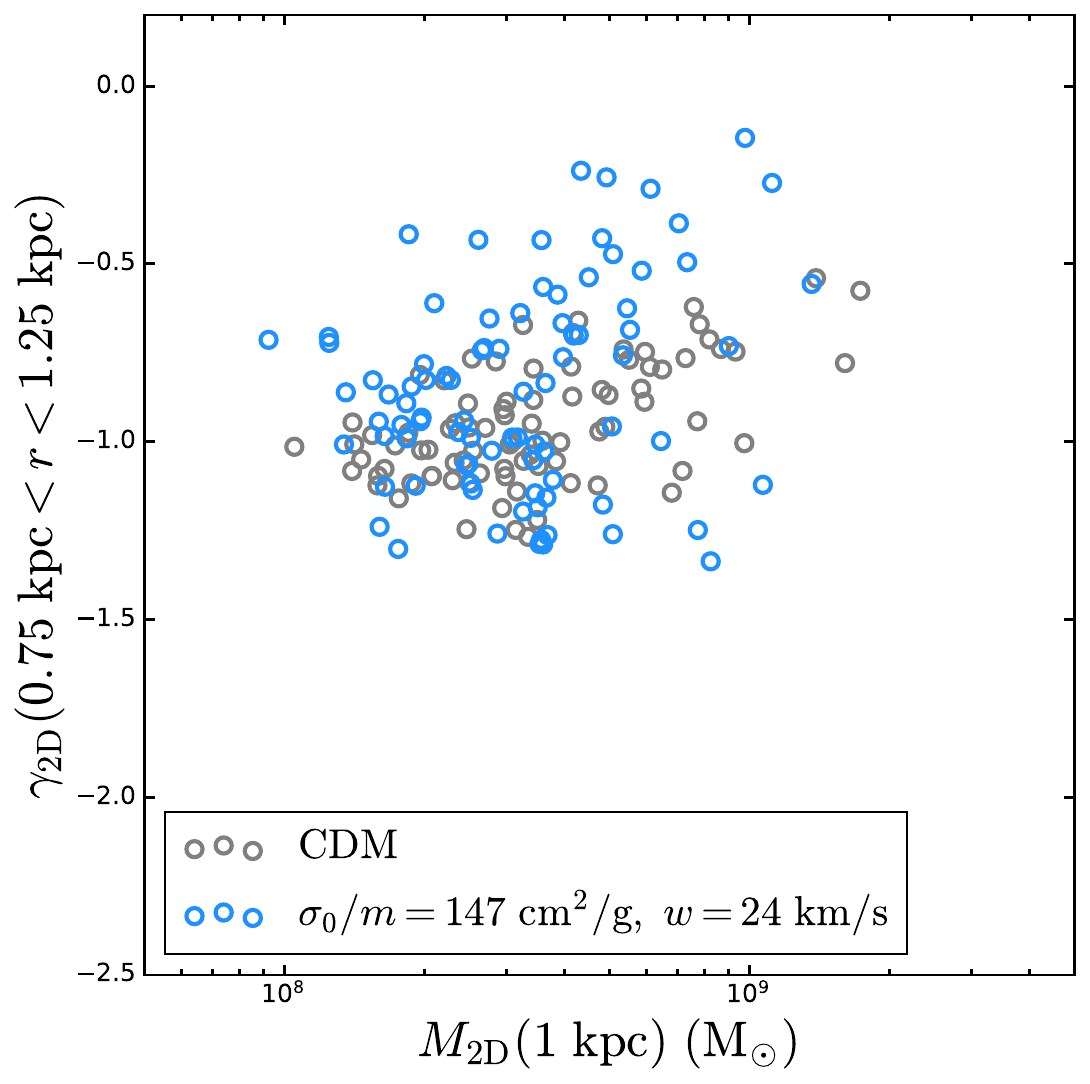}
  \includegraphics[width=5.3cm]{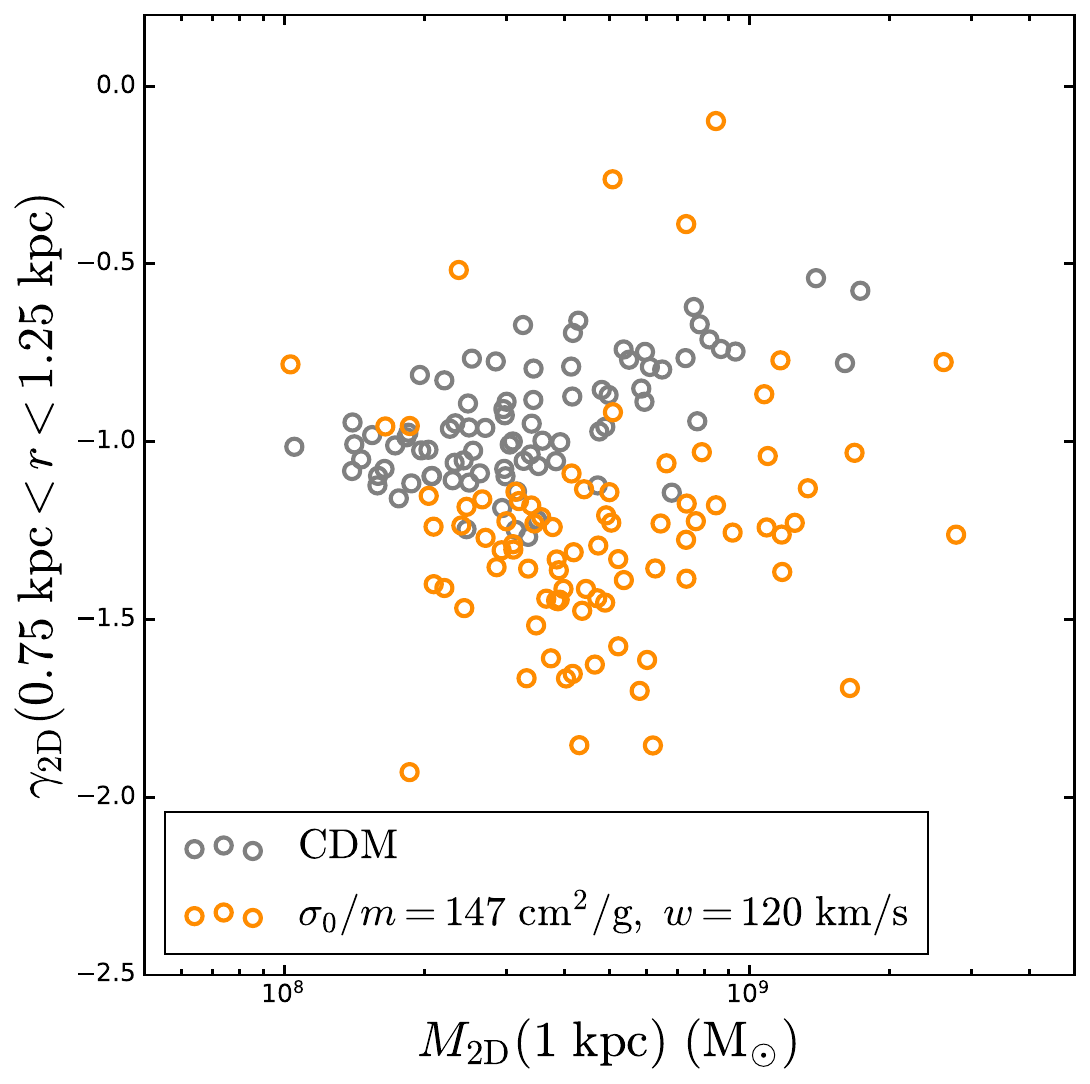}
  \includegraphics[width=5.3cm]{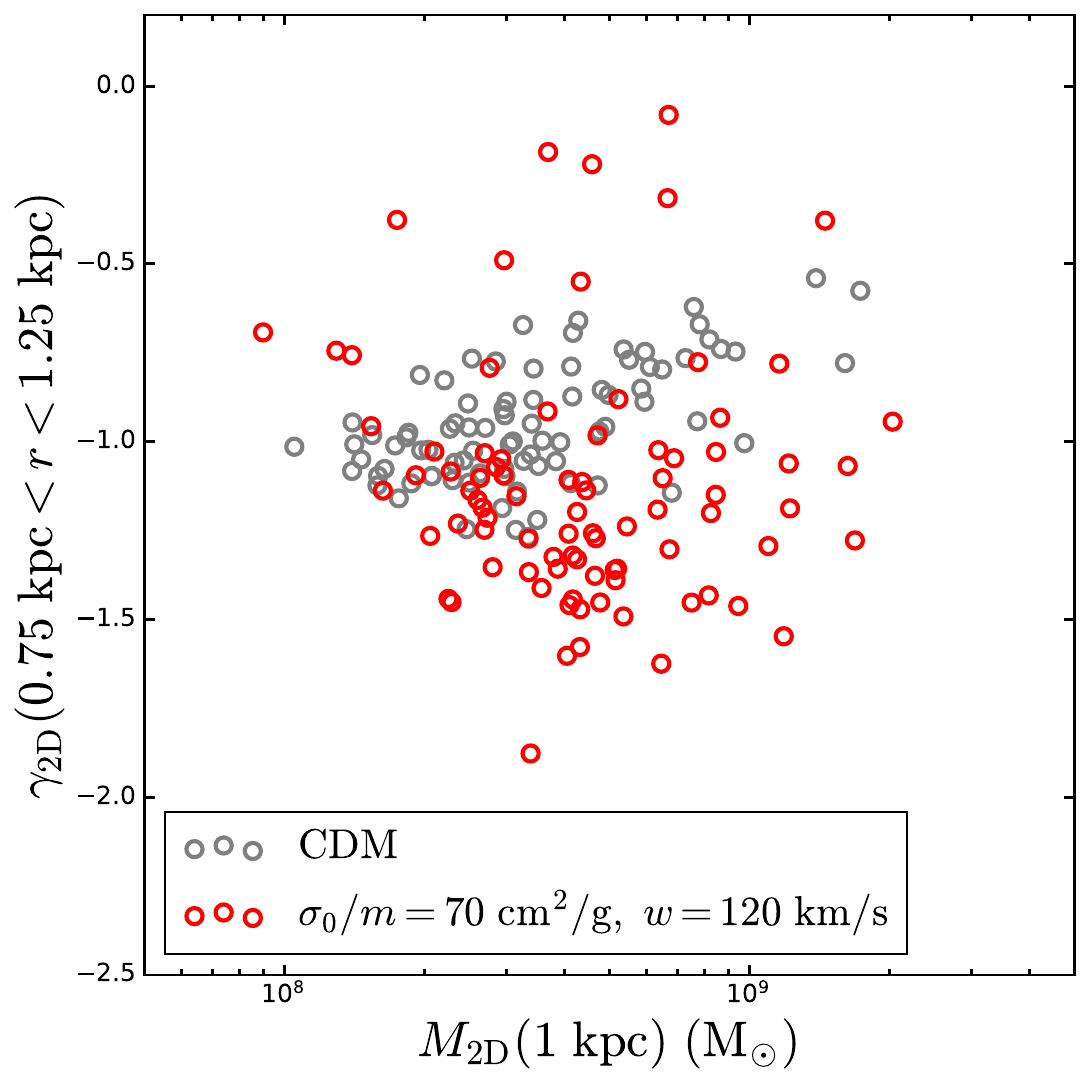}
  \caption{\label{fig:gamma2D} 
{\it Top: } Density profiles of CDM (gray) and SIDM (colored) of subhalos of the group host halo. 
{\it Bottom: } The projected logarithmic density profile slope $\gamma_{\rm 2D}$, averaged over $0.75~{\rm kpc} < r < 1.25~\rm kpc$, vs enclosed projected mass within $1~$kpc for subhalos of the group host halo. 
We consider the following three SIDM scenarios. The MilkyWaySIDM model~\cite{Yang:2022mxl} has $\sigma_0/m = 147~\rm cm^2/g$ and $w=24~\rm km/s$ (left). 
The GroupSIDM model~\cite{Nadler:2023nrd} features $\sigma_0/m$ and $w=120~\rm km/s$ (middle). 
An intermediate model is set at $\sigma_0/m = 70~\rm cm^2/g$ and $w=120~\rm km/s$ (right).
}
\end{figure*}

As an example application of the parametric model, we explore features of the density profiles under the three SIDM scenarios as characterized by Eq.~\ref{eq:xsr}:
$\sigma_0/m=147.1~\rm cm^2/g$, $w=24.33~\rm km/s$ (MilkyWaySIDM); $\sigma_0/m=147.1~\rm cm^2/g$, $w=120~\rm km/s$ (GroupSIDM); and $\sigma_0/m=70~\rm cm^2/g$, $w=120~\rm km/s$. 
We apply the integral approach described in Sec.~\ref{sec:model} to the CDM subhalos of the most massive main halo in the Group simulation and obtain their SIDM counterparts for the three SIDM scenarios.
The top panels of Fig.~\ref{fig:gamma2D} show the SIDM (colored) and CDM (gray) density profiles, illustrating how SIDM diversifies inner halo structure by producing both core-forming and -collapsing subhalos.  
In the MilkyWaySIDM scenario, there are more core-forming subhalos than collapsing ones, and the overall density profiles are shallower compared to CDM. In GroupSIDM, the trend is opposite and most subhalos are collapsed. 
The third SIDM scenario has smaller $\sigma_0/m$, while the same $w$ compared to GroupSIDM, and hence the number of collapsed subhalos decreases, but is still higher than that in MilkyWaySIDM.  

We further compute the projected logarithmic density profile slope $\gamma_{\rm 2D}\equiv d\ln\Sigma/d\ln R$, averaged over $0.75~{\rm kpc} < r < 1.25~\rm kpc$, and the enclosed projected mass within $1~$kpc $M_{\rm 2D}(1~{\rm kpc})$. The projected surface mass densities are computed as~\cite{Minor201110627,Nadler:2023nrd}
\begin{eqnarray}
\Sigma(R) = \int_{- R_{\rm vir}}^{R_{\rm vir}} \rho(\sqrt{R^2+z^2}) dz.
\end{eqnarray}

Fig.~\ref{fig:gamma2D} (bottom) shows the $M_{\rm 2D}(1~{\rm kpc}) \textup{--} \gamma_{\rm 2D}$ distribution for SIDM (red) and CDM (blue) subhalos in the group host across the three SIDM scenarios. 
In the MilkyWaySIDM scenario, the number of core-forming subhalos dominates, and the overall $\gamma_{\rm 2D}$ value shifts upwards; only a few SIDM subhalos have a steeper density slope than their CDM counterparts. 
Conversely, GroupSIDM is characterized by a large population of core-collapsing subhalos, which have lower $\gamma_{\rm 2D}$ values than their CDM counterparts, leading to a systematic downward shift in the $\gamma_{\rm 2D}$ distribution. 
The third SIDM scenario exhibits a trend of the $\gamma_{\rm 2D}$ distribution in between MilkWaySIDM and GroupSIDM, as expected. 
The comparison demonstrates that dark matter self-interactions diversify the inner density profiles of subhalos, and the significance is strongly correlated with the size of the cross section. 

More specifically, the CDM subhalos have $-1.5\lesssim\gamma_{2D}\lesssim-0.5$. In contrast, for $\sigma_0/m=70~{\rm cm^2/g}\textup{--}147~{\rm cm^2/g}$ and $w=120~{\rm km/s}$, a considerable number of SIDM subhalos have $\gamma_{\rm 2D}\lesssim-1.5$, and some of them reach $\gamma_{\rm 2D}\approx -2$. 
This lower limit is due to the analytical density profile we assume in Eq.~\ref{eq:cnfw}, where the density scales as $\propto r^{-3}$ in the steepest limit, with both $r_c$ and $r_s$ shrinking to small values. Thus, the projected surface mass density is
$$
\Sigma(R) \propto \int \left( R^2 + z^2 \right)^{-3/2} d z \propto R^{-2}.  
$$
We see that $\gamma_{2D}$ cannot fall below $-2$.
Compared with the simulated SIDM halos, we found that some core-collapsing halo profiles can exhibit distortions at the 10\% level in the density profile, which are not currently accounted for in our density profile models. For instance, the core-collapsing halo with the highest inner density in the left panel of Fig.~\ref{fig:group2} shows an inner density slope steeper than the parametric model prediction. Incorporating these distortions could enhance the precision of the predicted $\gamma_{\rm 2D}$ values. 
Nevertheless, the parametric model successfully reproduces the major trend in the $M_{\rm 2D}(1{\rm kpc}) \textup{--} \gamma_{\rm 2D}$ distribution of the GroupSIDM subhalos illustrated in Fig.~2 of \cite{Nadler:2023nrd}. 
With the current model, we can still differentiate between different SIDM scenarios and estimate the spread in inner densities. The parametric model provides a conservative estimate of the downward shifts, which is useful for exploring strong lensing systems.

\section{Conclusions}
\label{sec:sum}

In this study, we conducted a comprehensive evaluation of the parametric SIDM model introduced in Ref.~\cite{Yang:2023jwn}, utilizing matched halo pairs from CDM and SIDM simulations. Our analysis primarily focused on the SIDM simulation of the Milky Way analog (Ref.~\cite{Yang:2022mxl}), examining both isolated halos and subhalos. These halos, predominantly in the stages of gravothermal evolution with $t/t_c<1$, exhibit mass evolution histories akin to those in CDM simulations, making them ideal cases for testing the parametric model. 

We showed the $V_{\rm max}$ evolution from the parametric model is well consistent with that from the N-body simulation.  
In examining the density profiles at redshift $z=0$, we noted that relative differences are predominantly within 50\% in the inner regions, decreasing to below 10\% around $r/R_{\rm vir}\sim 0.04$, and then moderately rising again. 
Overall, we observed no systematic shifts in density profiles of isolated halos from either the basic or integral approaches. 
Thus, the parametric model accurately predicts the full diversity of (sub)halo density profiles in our simulations, with a spread that strictly exceeds CDM.

Modeling halos that are deeply collapsed is challenging. 
As these halos evolve, their $t/t_c$ approaches unity, it is necessary to use finer timesteps to accurately resolve the escalating inner densities. 
This requirement substantially slows down the simulation process. 
In the GroupSIDM simulation of Ref.~\cite{Nadler:2023nrd}, the effective cross section exceeds $100~\rm cm^2/g$ for typical dwarf galaxy halos, causing many of them to enter the deeply core collapsed phase. Nevertheless, the model performs well for halos with $\tau_0 < 1$. By freezing the SIDM-induced gravothermal evolution once $\tau_0=1$, the model can still provide reasonable SIDM predictions for comparing model predictions and identifying potential signatures. We presented an example application for lensing perturber systems in three SIDM scenarios, demonstrating that they produce distinguishable signatures on the $M_{\rm 2D}(1{\rm kpc})\textup{--}\gamma_{\rm 2D}$ plane.

In conclusion, the parametric SIDM halo model provides an efficient tool for making predictions for given SIDM scenarios and CDM halos. 
It has recently been implemented into the semi-analytic model program {\sc SASHIMI-SIDM}, for phenomenological SIDM studies down to very low masses, relevant for, e.g., stellar stream perturbations, indirect detection, and direct detection.
It has also been extended to incorporate the effect of baryons~\cite{Yang:2024tba}, enabling more realistic theoretical predictions for halos hosting massive galaxies. 
Moreover, the model's efficiency and flexibility make it suitable for predicting galaxy rotation curves, based on which one can explore a wider parameter space and different SIDM scenarios, see, e.g., Ref.~\cite{ManceraPina:2024ybj}.
Additionally, using the parametric model's results as leading-order predictions provides a starting point for investigating new SIDM signatures across cosmic environments. 
We leave these investigations for future work. 

We provide example scripts for applying the parametric model at: \url{https://github.com/DanengYang/parametricSIDM}

\acknowledgments
We thank Yi-Ming Zhong, Shin'ichiro Ando, Andrew Benson, Siddhesh Raut for valuable comments on the parametric model, and Fangzhou Jiang, Zhichao Zeng for helpful discussion about the ram-pressure evaporation effect and the numerical convergence in SIDM. D.Y. and H.-B. Y were supported by the John Templeton Foundation under grant ID \#61884 and the U.S. Department of Energy under grant No.\ de-sc0008541. 
This research was supported in part by grant NSF PHY-2309135 to the Kavli Institute for Theoretical Physics (KITP).
D.Y. was also supported in part by the National Key Research and Development Program of China (No. 2022YFF0503304), the Project for Young Scientists in Basic Research of the Chinese Academy of Sciences (No. YSBR-092).


\appendix

\section{Performance of the Read profile compared with the $\beta4$ profile}
\label{app:compProfs}

\begin{figure*}[htbp]
  \centering
  \includegraphics[width=7.2cm]{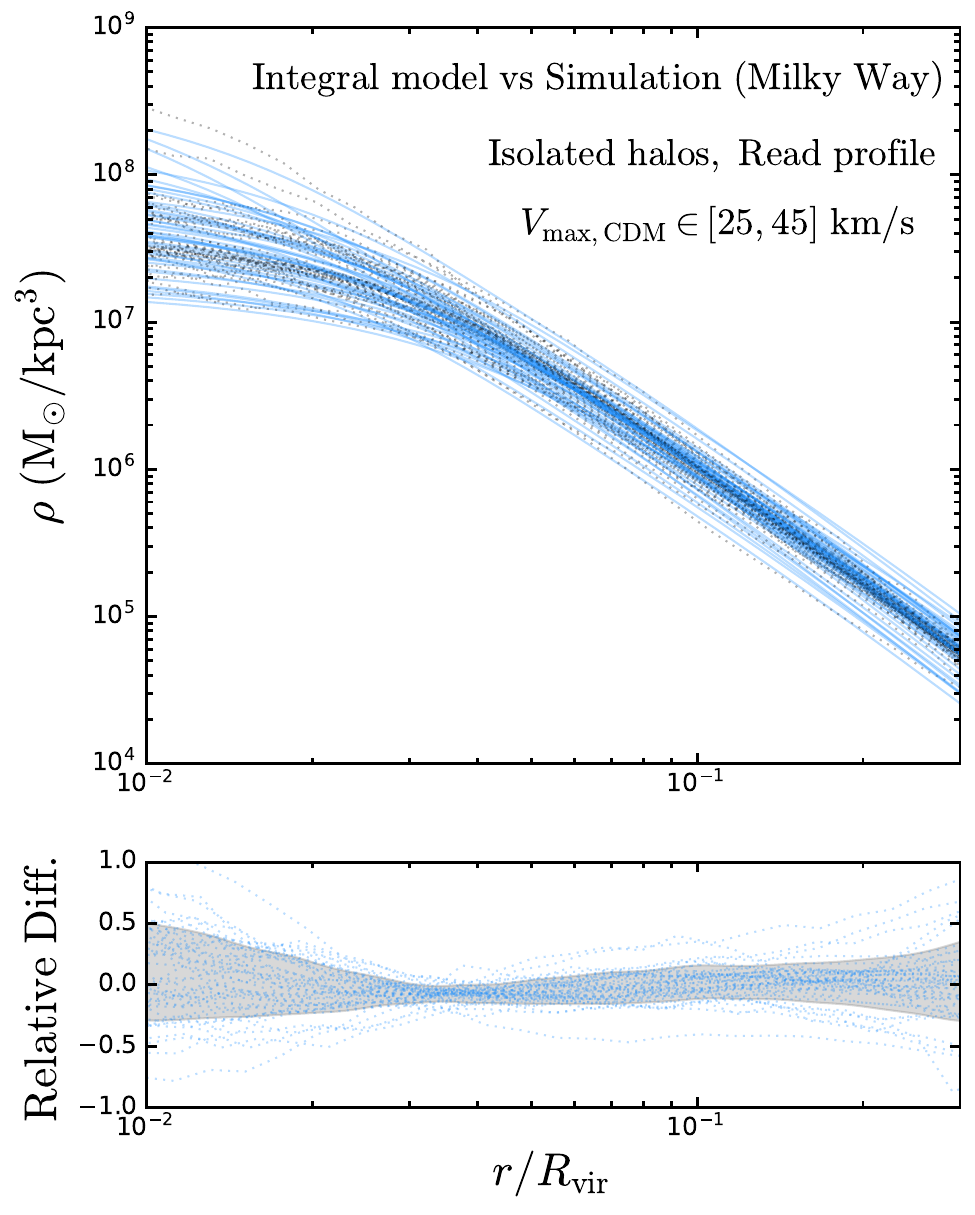}
  \includegraphics[width=7.2cm]{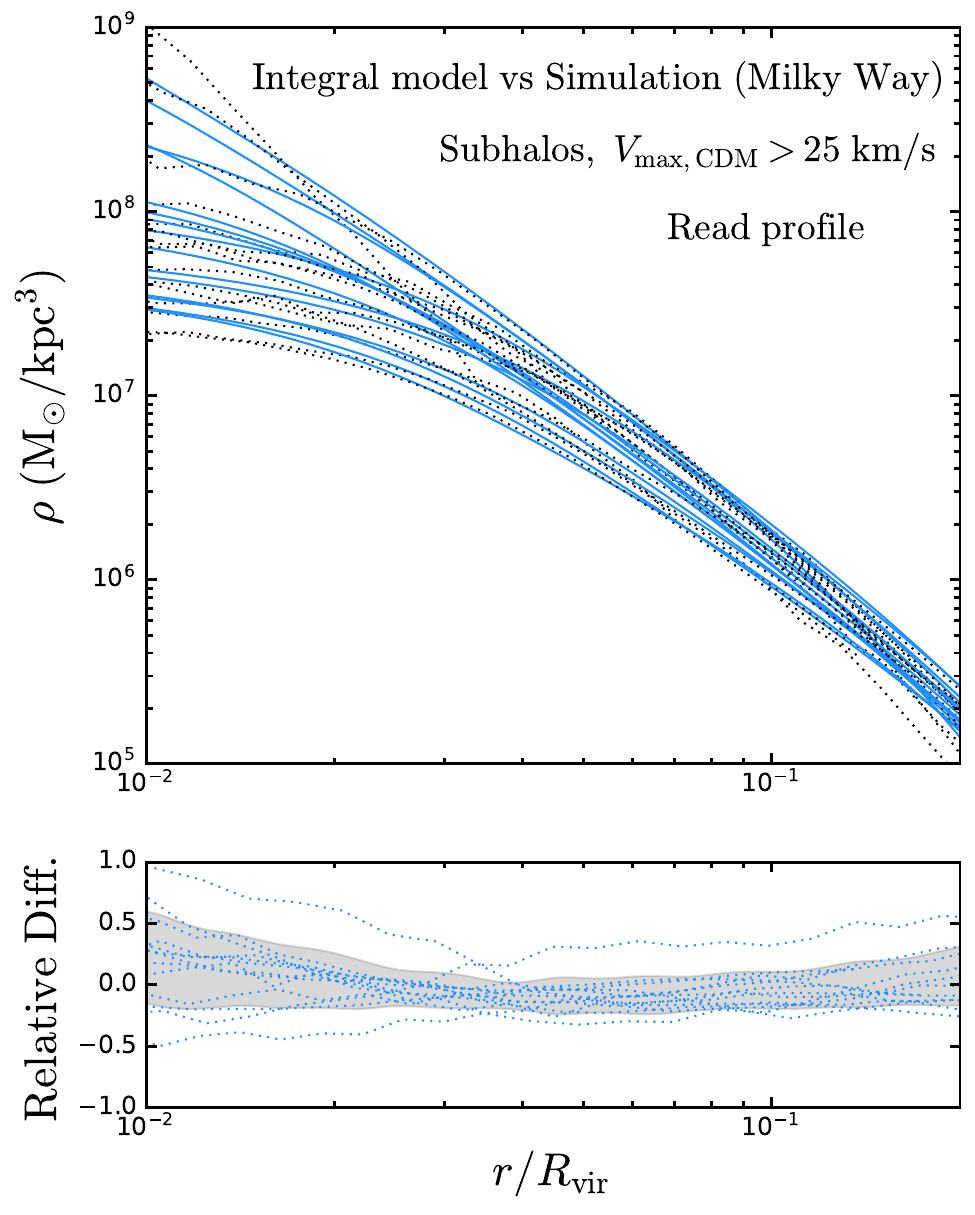}
  \caption{\label{fig:app1} Testing the Read profile based model predictions for isolated halos (left) with $V_{\rm max,CDM}\in [25,45]~\rm km/s$ and subhalos (right) with $V_{\rm max,CDM} > 25~\rm km/s$.
The density profiles from the parametric model using the integral approach (solid) and from the simulation (dotted) are plotted.
The relative difference between each pair of simulated (Sim) and model-predicted (Mod) curves is measured as $\rm 2(Mod-Sim)/(Mod+Sim)$, with the $\pm 1\sigma$ band of the results shaded in gray.
}
\end{figure*}

The Read profile proposed in Refs.~\cite{2016MNRAS.462.3628R,2016MNRAS.459.2573R} has the same performance as the $\beta 4$ one in obtaining the model predictions. 
The Read profile is formulated based on the NFW profile,
\begin{eqnarray}
\label{eq:read}
\rho_{\rm Read}(r) = f^n \rho_{\rm NFW} + \frac{nf^{n-1}(1-f^2)}{4\pi r^2 r_c } M_{\rm NFW},
\end{eqnarray}
where $r_c$ is the core radius, $f(r) = \tanh(r/r_c)$, and $n$ is a parameter in the range $0<n\leq1$. The NFW density and mass profiles are
\begin{widetext}
\begin{eqnarray}
\rho_{\rm NFW}(r) = \frac{\rho_s}{\frac{r}{r_s} \left(1 +\frac{r}{r_s} \right)^2},~
M_{\rm NFW}(r) = 4\pi \rho_s r_s^3 \left[ \ln\left(1+\frac{r}{r_s}\right) -\frac{r}{r+r_s} \right]. 
\end{eqnarray}
\end{widetext}
The enclosed mass follows the relation $M_{\rm Read} = f^n(r) M_{\rm NFW}(r)$. 
For $r\gg r_c$, $f^{n}(r)=1$, and $M_{\rm Read} = M_{\rm NFW}(r)$. 
In the limit of $r_c\rightarrow0$, $f\rightarrow1$ and $\rho_{\rm Read}(r)\rightarrow\rho(r)_{\rm NFW}$. 
In this work, we fix $n=1$.

The evolution trajectories of the parameters $\rho_s$, $r_s$, and $r_c$ in the Read profile are described by the following equations~\cite{Yang:2023jwn},
\begin{widetext}
\begin{eqnarray}
\label{eq:m0read}
\frac{\rho_s}{\rho_{s,0}} &=& 1.335 + 0.7746 \tau + 8.042 \tau^5 -13.89 \tau^7  + 10.18 \tau^9 + (1-1.335) (\ln 0.001)^{-1} \ln \left( \tau + 0.001 \right), \nonumber \\
\frac{r_s}{r_{s,0}} &=& 0.8771 - 0.2372 \tau +  0.2216 \tau^2 -0.3868 \tau^3 + (1-0.8771) (\ln 0.001)^{-1} \ln \left( \tau + 0.001 \right), \nonumber \\
\frac{r_c}{r_{s,0}} &=& 3.324 \sqrt{\tau} -4.897 \tau + 3.367 \tau^2 -2.512 \tau^3 + 0.8699 \tau^4,  
\end{eqnarray}
\end{widetext}
where the subscript ``$0$'' denotes the corresponding value of the initial NFW profile.
We found the same functional forms of Eq.~\ref{eq:m0} work well for the Read profile, with the adjustment of the coefficients.
Note that the fitted values of $\rho_s$, $r_s$, and $r_c$ of the Read profile are different from those of our cored profile in Eq.~\ref{eq:cnfw}. 

For applying the integral model, we present the fitting functions for the $V_{\rm max}$ and $R_{\rm max}$ evolution, based on the Read profile modeled by Eq.~\ref{eq:m0read}. 
\begin{widetext}
\begin{eqnarray}
\label{eq:appm1}
\nonumber
\frac{V_{\rm max}}{V_{\rm max,0}} &=& 1+ 0.2289 \tau -5.018 \tau^3 + 17.75 \tau^4 - 19.35 \tau^5 + 8.953 \tau^7 - 2.364 \tau^9  \\ 
\frac{R_{\rm max}}{R_{\rm max,0}} &=& 1 -0.6026 \tau + 1.043 \tau^2 - 1.484 \tau^3 + 0.5263 \tau^4,  
\end{eqnarray}
\end{widetext}
where $\tau=t/t_c$.

We apply the integral approach with the Read profile to obtain predictions for the halos in the Milky Way simulation (Ref.\cite{Yang:2022mxl}).
In Fig.~\ref{fig:app1}, we show the obtained results for isolated halos with $V_{\rm max,CDM}$ values between $25$ and $45~\rm km/s$ (left),
and subhalos with $V_{\rm max,CDM}$ greater than $25~\rm km/s$ (right).
The results closely mirror those obtained using the $\beta4$ profile.

\section{Incorporating the ram-pressure evaporation}
\label{sec:ram}

We follow the procedures in Refs.~\cite{Kummer:2017bhr,Shirasaki:2022ttb}, with minor adjustments, to model the ram-pressure evaporation (RPe) effect induced by the collisions between subhalo and host halo particles.
First, we estimate an escape velocity assuming an NFW potential at $r_s$,
$v_{\rm esc} = \sqrt{-2 \Phi_{\rm NFW}(r_s)} = \sqrt{8\pi (\ln 2) G \rho_s r_s^2}$.
We also compute the 1D velocity dispersion $\sigma_{\rm 1D,host}(d)$ of the host halo assuming an NFW halo at the position of the subhalo, which has a distance $d$ from the host halo.
The 1D velocity dispersion $\sigma_{\rm NFW}(r)$ can be obtained analytically as~\cite{Lokas:2000mu}.
$$
\sigma_{\rm NFW}(r) = \sqrt{4\pi G \rho_s r_s^2 F(r/r_s)}, 
$$
with
\begin{widetext}
\begin{eqnarray}
F(x) &=& \frac{1}{2} x (x+1)^2 \left(6
   \text{Li}_2(-x)+\left(\frac{1}{x^2}-\frac{2}{x+1} -\frac{4}{x}+1\right) \log(x+1) \right. \\ \nonumber
 && \left. -\frac{1}{x}-\frac{6}{x+1}-\frac{1}{(x+1)^2}+3 \log ^2(x+1)-\log (x)+\pi ^2\right). 
\end{eqnarray}
\end{widetext}
Based on the $v_{\rm esc}$ and $\sigma_{\rm 1D,host}(d)$, we compute the evaporation fraction $\chi_e$ as
\begin{eqnarray}
\chi_e = \frac{1-y^2}{1+y^2},
\end{eqnarray}
where $y=v_{\rm esc}/\sqrt{v_{\rm sub}^2+\sigma_{\rm 1D,host}^2(d)}$ and $v_{\rm sub}$ refers to the magnitude of the subhalo's velocity.
In a given timestep, the mass change rate due to ram-pressure evaporation can be computed as
\begin{eqnarray}
\left( \frac{d\ln M_{\rm sub}}{dt} \right)_{\rm RPe} = - \chi_e \frac{\sigma_V(v_r)}{m} v_{\rm sub} \rho_h(d),
\end{eqnarray}
where $\sigma_V(v_r)/m$ refers to the viscosity cross section evaluated at $v_r= \sqrt{v_{\rm sub}^2+\sigma_{\rm 1D,host}^2(d)}$ and $\rho_h(d)$ is the host halo density at radius equals $d$.
The viscosity cross section averages over the angular distribution of a differential cross section as~\cite{Tulin:2013teo}
\begin{equation}
\label{eq:xsrD}
\sigma_V = \frac{3}{2} \int d\cos\theta \sin^2\theta \frac{d\sigma}{d\cos\theta}. 
\end{equation}

We convert the mass change rate into that of $V_{\rm max}$ and $R_{\rm max}$ assuming $M_{\rm sub}\propto V_{\rm max}^3$ and $M_{\rm sub}\propto R_{\rm max}^2$ which we have tested using the simulated CDM halos.
It follows that
\begin{eqnarray}
\left( \frac{d V_{\rm max} }{dt} \right)_{\rm RPe} &=& \frac{V_{\rm max,CDM}}{3} \left( \frac{d\ln M_{\rm sub}}{dt} \right)_{\rm RPe}  \\ \nonumber
\left( \frac{d R_{\rm max} }{dt} \right)_{\rm RPe} &=& \frac{R_{\rm max,CDM}}{2} \left( \frac{d\ln M_{\rm sub}}{dt} \right)_{\rm RPe}.  \\
\end{eqnarray}

\begin{figure*}[htbp]
  \centering
  \includegraphics[height=5.3cm]{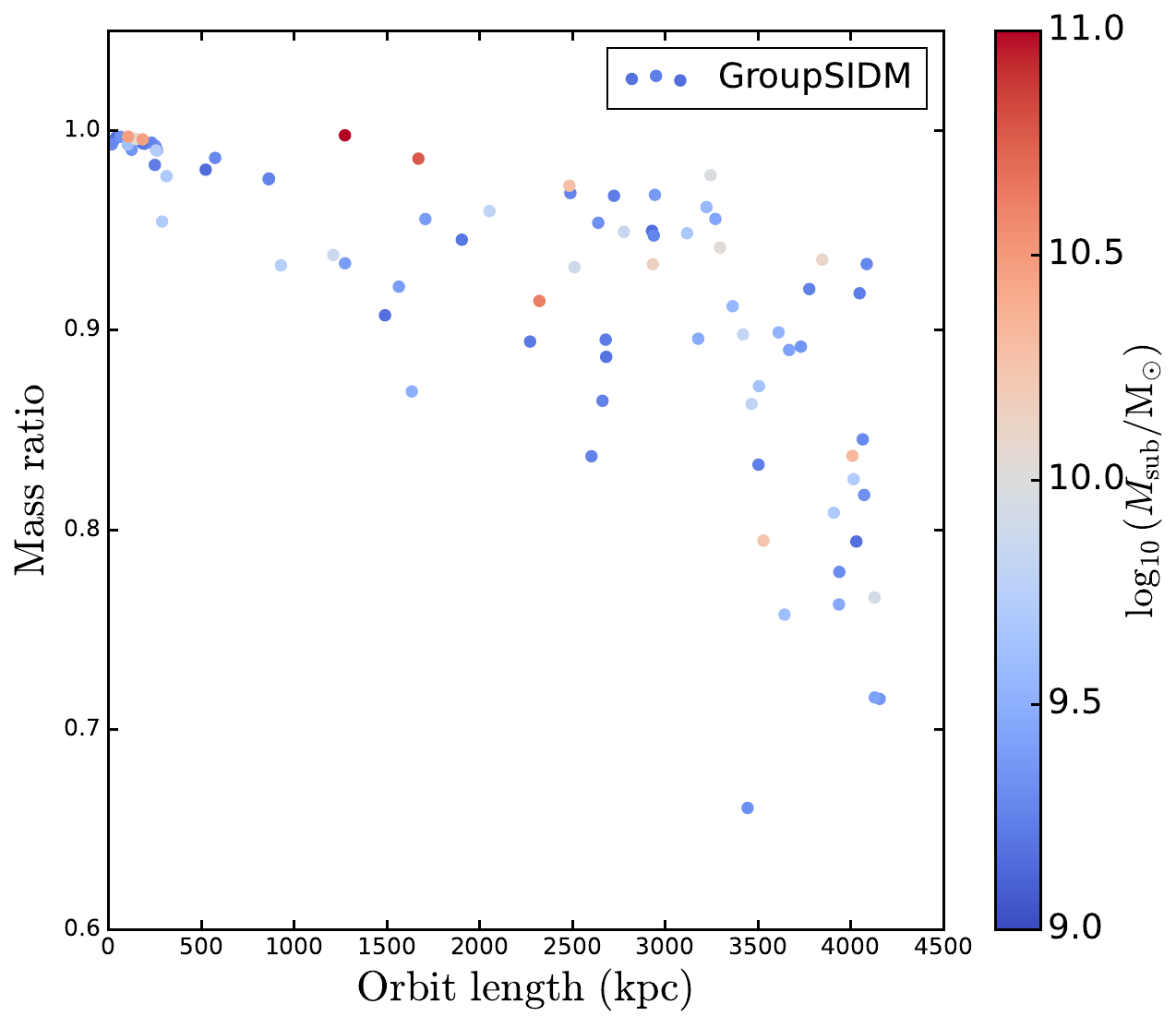} 
  \includegraphics[height=5.3cm]{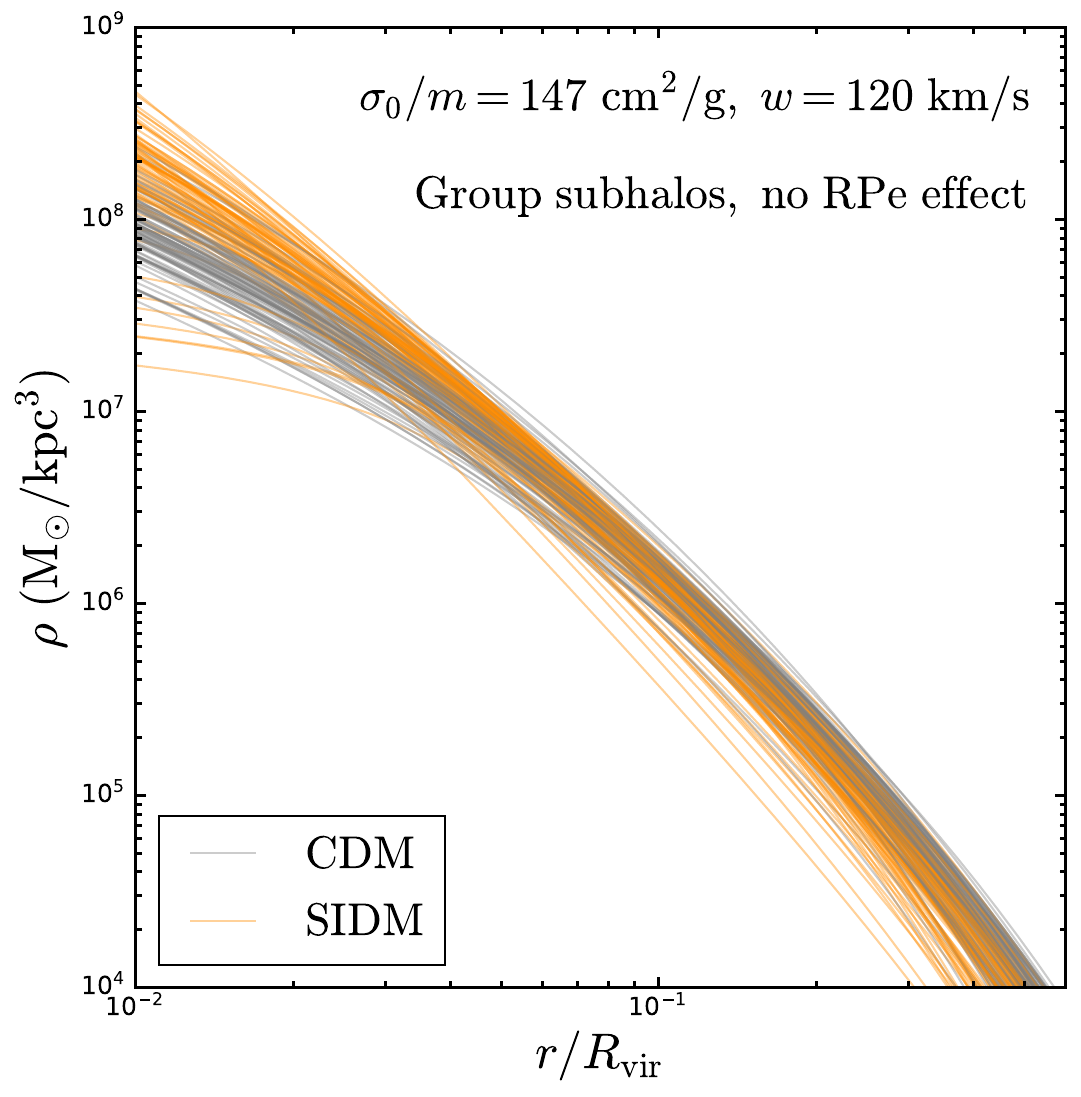}
  \includegraphics[height=5.3cm]{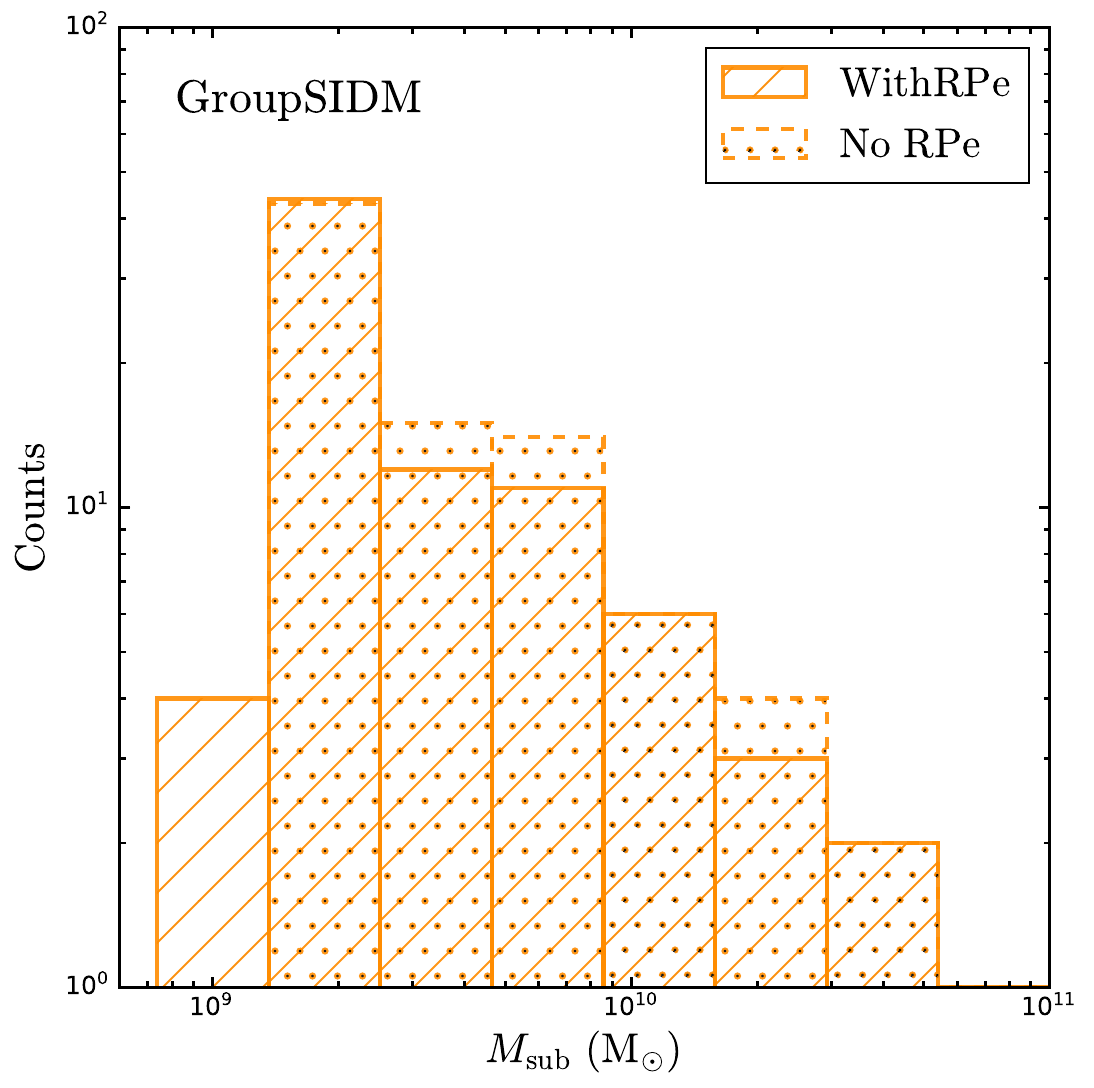}
  \caption{\label{fig:evp100} {\it Left:} Scatter for subhalo orbital length in the host halo vs ratio of masses with and without the RPe effect.
{\it Middle:} The same as Fig.~\ref{fig:gamma2D} (upper right), except that the subhalo density profiles are obtained without incorporating the RPe effect.
{\it Right:} Subhalo mass distribution with (shaded lines) and without (shaded dotts) the RPe effect.
}
\end{figure*}

To better illustrate the RPe effect, we apply the parametric model to subhalos within a group host, considering the GroupSIDM cross section. 
In the left panel of Fig.~\ref{fig:evp100}, we show the ratio of masses with and without the RPe effect vs the orbital length of a subhalo as it evolves in the host halo. The data points are color-coded by the mass at $z=0$, revealing that more massive subhalos tend to have lower evaporated masses and shorter orbital lengths. This observation can be attributed to the tidal mass loss of subhalo progenitors: extensive orbital evolution removes most of their mass, rendering them lighter.
In the middle panel, we plot density profiles of CDM (blue) and SIDM (red) subhalos of the group host halo, {\it without} incorporating the RPe effect.
Compared with the results that include the RPe effect shown in the upper right panel of Fig.~\ref{fig:gamma2D}, we find the profiles of some lowest density subhalos at around $r/R_{\rm vir}\approx 0.1$ are shifted to higher values.
In the right panel, we compare the subhalo mass distribution with and without the RPe effect. 
We find that approximately 10\% of halos with masses greater than $10^9~\rm M_{\odot}/h$ without the RPe effect are shifted below this threshold when the RPe effect is included. 
We have checked the reduction in the mass function found in the GroupSIDM simulation~\cite{Nadler:2023nrd} is more significant than that estimated using the RPe model. This suggests that the effect of enhanced tidal striping due to SIDM core formation may have a stronger impact than the RPe effect, and we will leave it for future study.

\section{A hybrid method for approximating integral approach results}
\label{app:phase}

\begin{figure*}[htbp]
  \centering
  \includegraphics[width=7.2cm]{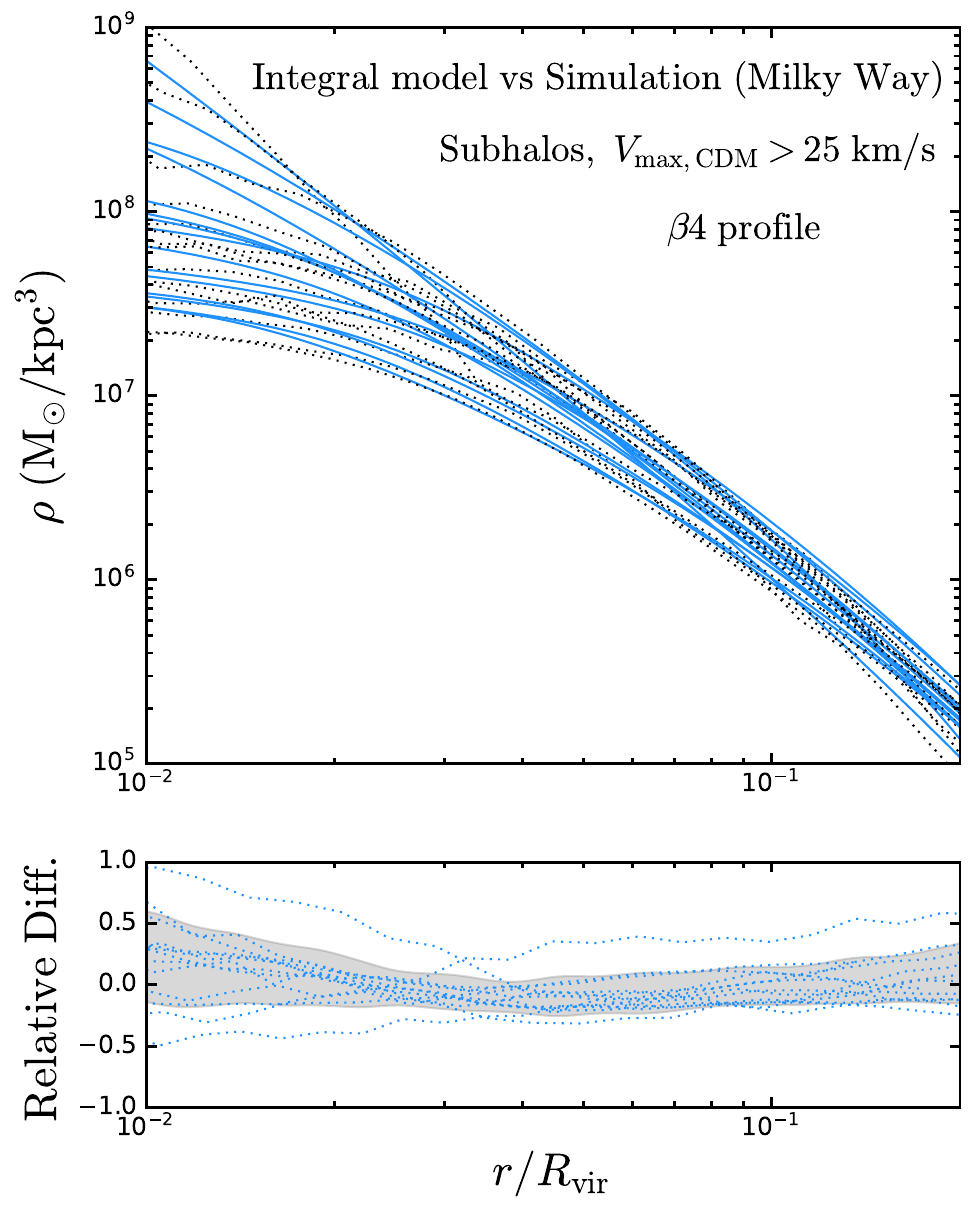}
  \includegraphics[width=7.2cm]{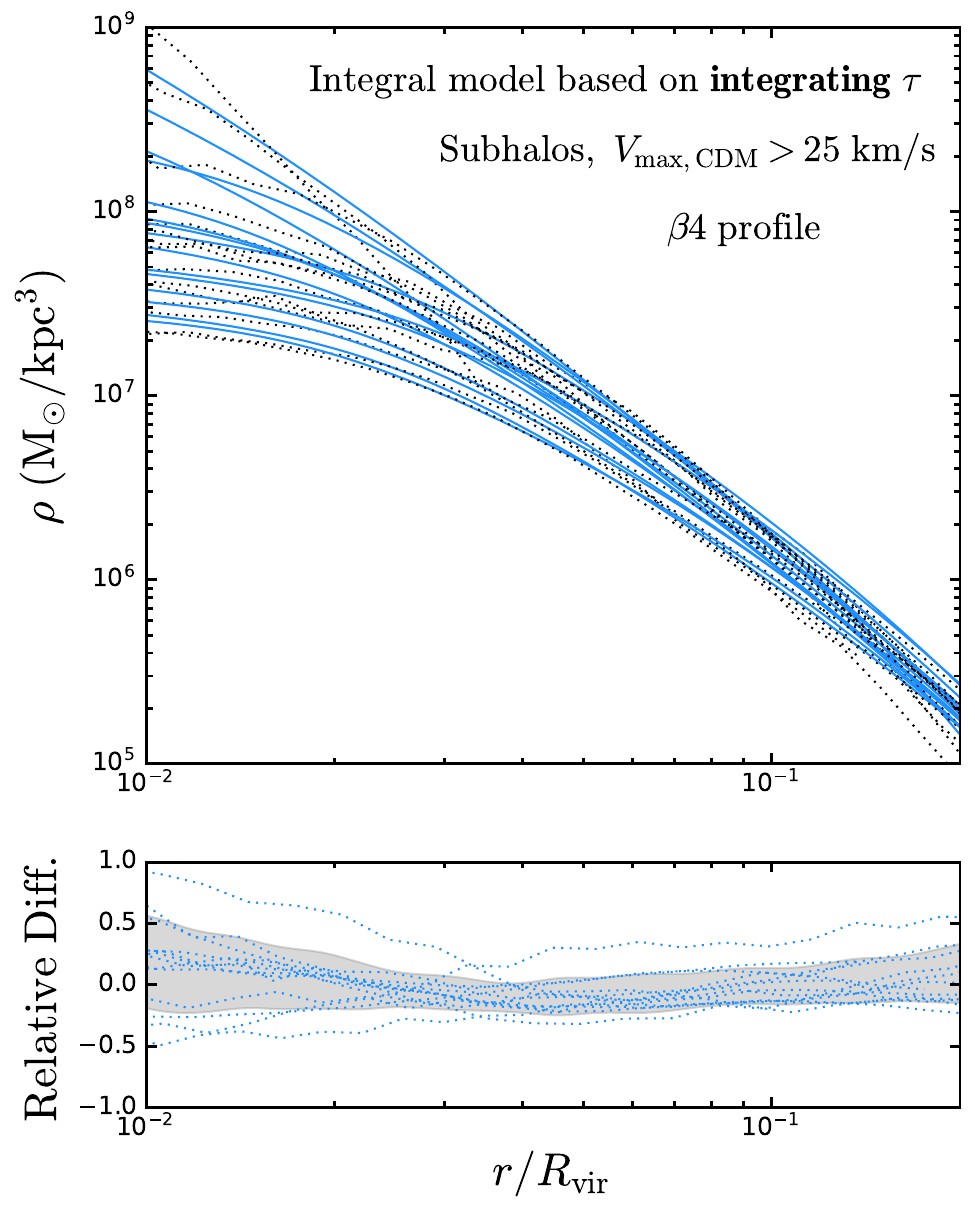}
  \caption{\label{fig:phase1}
Comparison of the integral approach results, obtained considering Eq.~\ref{eq:int} (left) and Eq.~\ref{eq:int2} (right), respectively.
We consider subhalos with $V_{\rm max,CDM} > 25~\rm km/s$ and adopt the $\beta4$ profile for obtaining the predictions.
The density profiles from the parametric model using the integral approach (solid) and from the simulation (dotted) are plotted. 
The relative difference between each pair of simulated (Sim) and model-predicted (Mod) curves is measured as $\rm 2(Mod-Sim)/(Mod+Sim)$, with the $\pm 1\sigma$ band of the results shaded in gray.
}
\end{figure*}

\begin{figure*}[htbp]
  \centering
  \includegraphics[width=7.2cm]{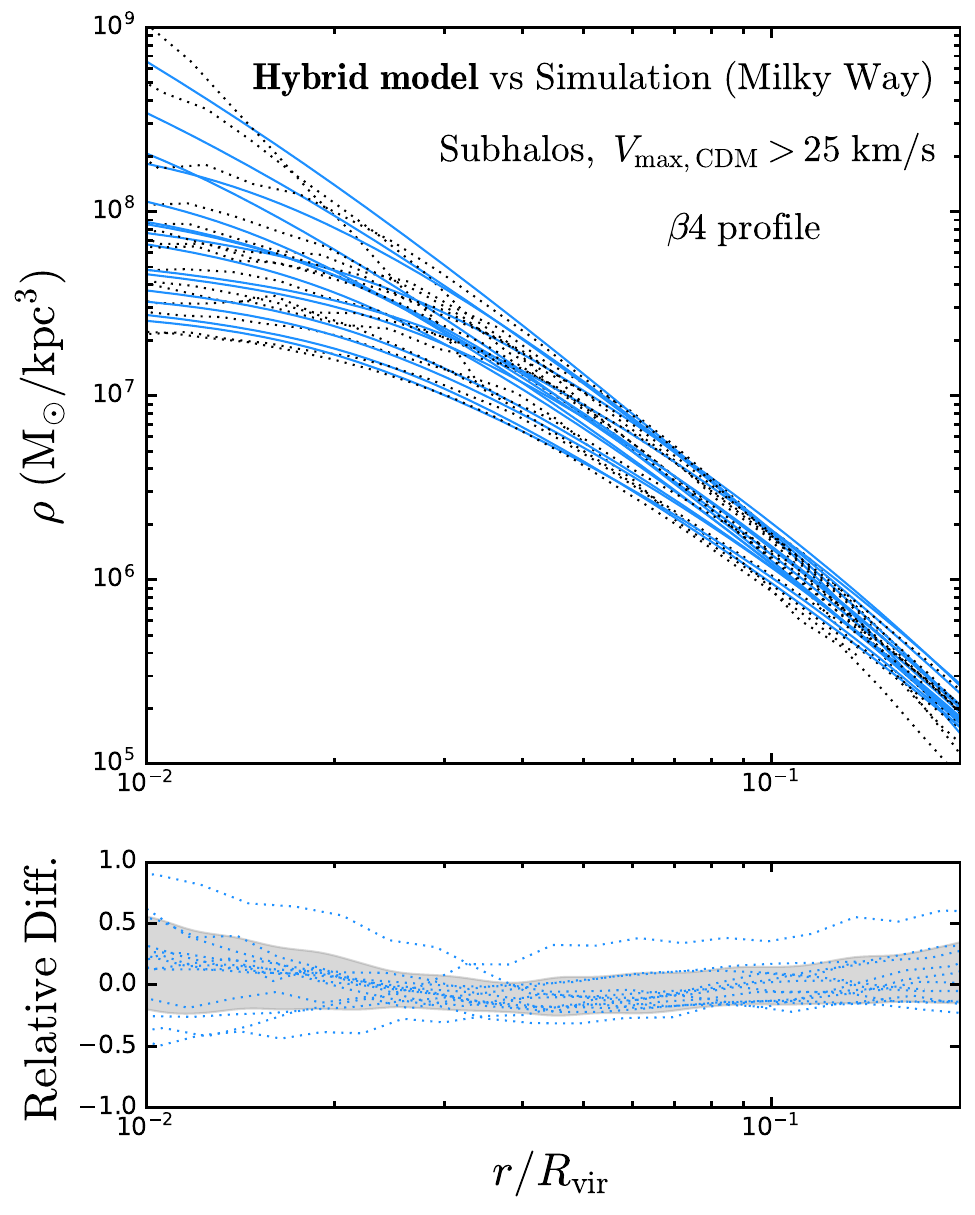}
  \includegraphics[width=7.2cm]{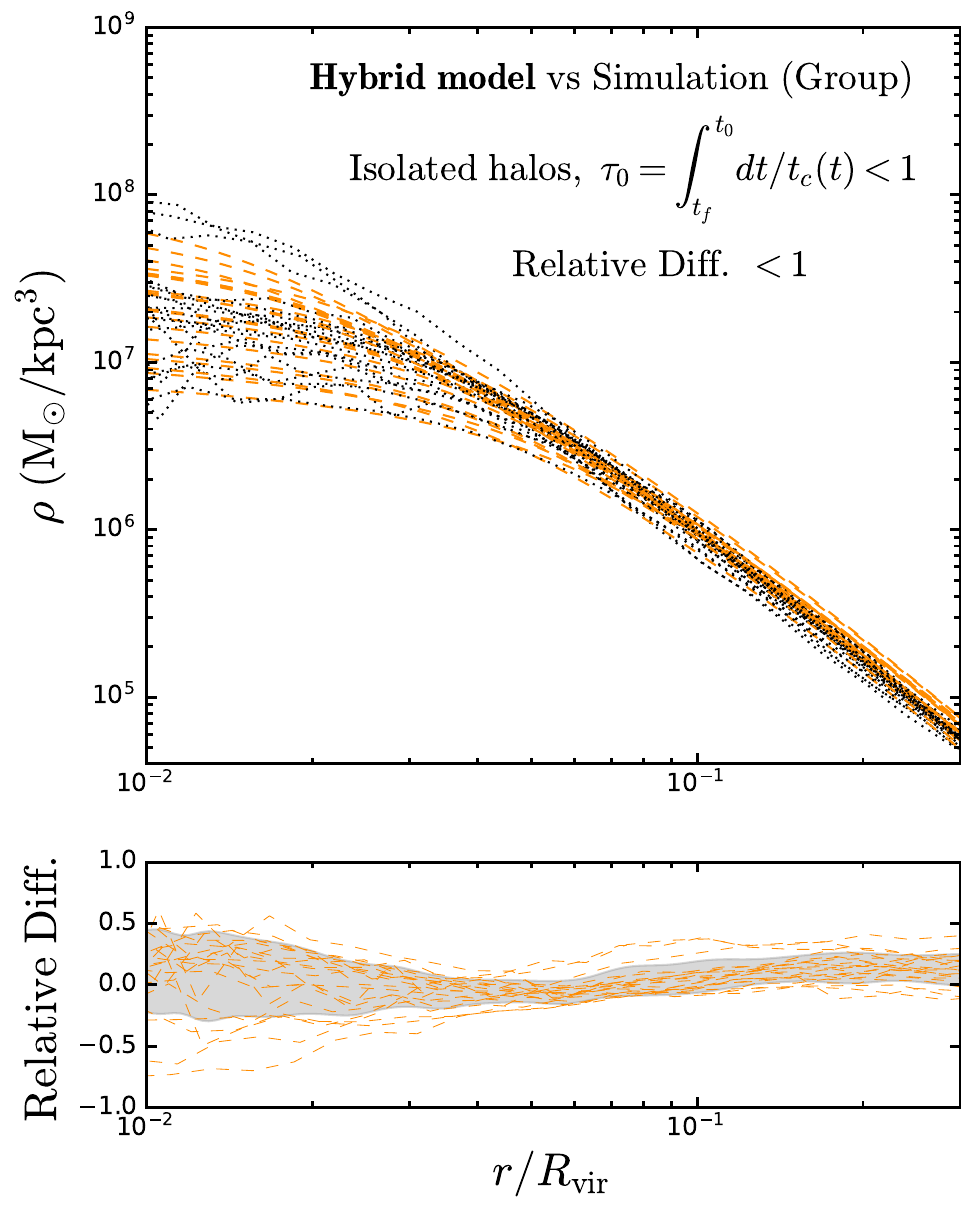}
  \caption{\label{fig:phase2} 
Model-predicted density profiles (solid) obtained using the hybrid approach, for subhalos in the Milky Way simulation (left) and isolated halos in the Group simulation (right). 
Their counterparts from the N-body SIDM simulations (dotted) are shown for comparison.
The relative difference between each pair of simulated (Sim) and model-predicted (Mod) curves is measured as $\rm 2(Mod-Sim)/(Mod+Sim)$, with the $\pm 1\sigma$ band of the results shaded in gray.
}
\end{figure*}

In this section, we explore a subtle aspect of the integral approach and introduce a hybrid method that approximates its results.
We propose using the following equation to replace Eq.~\ref{eq:int} in the integral approach 
\begin{widetext}
\begin{eqnarray}
\label{eq:int2}
\nonumber 
V_{\rm max}(t)    &=&  V_{\rm max, CDM}(t_h) + \int_{t_h}^{t} d t'  \frac{d V_{\rm max,CDM}(t')}{d t'} +  \int_{0}^{\tau(t)} d\tau' \frac{d V_{\rm max, Model} (\tau')}{d \tau'} \\  
R_{\rm max}(t)    &=&  R_{\rm max, CDM}(t_h) + \int_{t_h}^{t} d t'  \frac{d R_{\rm max,CDM}(t')}{d t'} +  \int_{0}^{\tau(t)} d\tau' \frac{d R_{\rm max, Model} (\tau')}{d \tau'}, 
\end{eqnarray}
\end{widetext}
where $\tau(t)=\int_0^t t'/t_c(t')$ can be computed independently. 
The primary distinction relative to Eq.~\ref{eq:int} lies in the interpretation of the integral terms for SIDM. For the cases without accretion histories, these measures converge; 
however, they are different conceptually: $t'$ indicates the duration of halo evolution, whereas $\tau'$ denotes the halo's gravothermal state at time $t'$. This difference reflects a potential ``memory effect,'' representing how much a halo's current state depends on its accretion history.
Theoretically, the differences between the fictitious CDM halo and the simulated CDM halo stem from the mass changes in the progenitor halos during the integration over $\tau$. Consequently, only halos that have experienced significant mass and gravothermal evolution exhibit a clear differentiation in the dark matter-only case. 
By exploring this subtle effect, we could better understand the influence of accretion histories on gravothermal evolution.

We perform such a comparison for the subhalos in the Milky Way simulation~\cite{Yang:2022mxl}.
As illustrated in Fig.~\ref{fig:phase1}, both methods produce almost identical results. This similarity is anticipated, given that the differences are suppressed by the rate of change of $t_c$ and an additional multiplicative factor involving $t_c$. 

Moreover, we evaluate the accuracy of approximating the results of the integral approach by disregarding the differences between the fictitious CDM halo and the simulated CDM halo. 
This simplification reduces the integrals in Eq.~(\ref{eq:int2}) to a simple evaluation of the gravothermal phase~\cite{Yang:2023jwn,Yang:2024tba} 
$$
\tau(t) = \int_{t_h}^{t} \frac{d t'}{t_{c}[\sigma_{\rm eff}(t')/m,\rho_s(t'),r_s(t')]},
$$
where $t_h$ refers to the halo formation time and $t_c(t')$ is the core collapse time computed at time $t'$ as in the original integral approach. 
The SIDM information is assumed to be entirely encoded by the gravothermal phase $\tau$.
Due to its position as an intermediary between the integral and basic approaches, we refer to this as the {\it hybrid} approach.

In Fig.~\ref{fig:phase2}, we test the hybrid approach for subhalos in the MilkyWaySIDM simulation (left) and isolated halos in the GroupSIDM simulation (right), respectively. 
In both cases, the performances are close to the original integral approach, supporting the effectiveness of the hybrid approach in dark matter-only simulations.

It is important to note, however, that in scenarios with a growing baryonic content, the differences between these two approaches may become more pronounced. In such cases, the integral approach continues to provide the most accurate theoretical predictions. For a detailed comparison of these effects, see Appendix B of Ref.~\cite{Yang:2024tba}.


\bibliography{reference}

\end{document}